\documentclass[a4paper,10pt,twoside]{cpc-hepnp}

\usepackage{lineno}
\usepackage{multirow}
\usepackage{color}
\usepackage{url}

\usepackage{multicol}
\usepackage{graphicx}
\usepackage{booktabs}
\usepackage{amssymb,bm,mathrsfs,bbm,amscd}
\usepackage[tbtags]{amsmath}
\usepackage{lastpage}
\usepackage{hyperref}

\graphicspath{{figures/}}

\begin{document}

\fancyhead[c]{\small Chinese Physics C~~~Vol. XX, No. X (201X)
XXXXXX} \fancyfoot[C]{\small 010201-\thepage}


\title{Improved Measurement of the Reactor Antineutrino Flux and Spectrum at Daya
Bay}


\newcommand{\ECUST}{1}
\newcommand{\Wisconsin}{2}
\newcommand{\Yale}{3}
\newcommand{\BNL}{4}
\newcommand{\NTU}{5}
\newcommand{\NUU}{6}
\newcommand{\NJU}{7}
\newcommand{\IHEP}{8}
\newcommand{\CUHK}{9}
\newcommand{\NCTU}{10}
\newcommand{\SDU}{11}
\newcommand{\TsingHua}{12}
\newcommand{\NCEPU}{13}
\newcommand{\SZU}{14}
\newcommand{\ZSU}{15}
\newcommand{\Dubna}{16}
\newcommand{\Siena}{17}
\newcommand{\IIT}{18}
\newcommand{\UIUC}{19}
\newcommand{\LBNL}{20}
\newcommand{\SJTU}{21}
\newcommand{\BNU}{22}
\newcommand{\UH}{23}
\newcommand{\VirginiaTech}{24}
\newcommand{\CIAE}{25}
\newcommand{\USTC}{26}
\newcommand{\NanKai}{27}
\newcommand{\UC}{28}
\newcommand{\TempleUniversity}{29}
\newcommand{\DGUT}{30}
\newcommand{\UCB}{31}
\newcommand{\HKU}{32}
\newcommand{\Charles}{33}
\newcommand{\Princeton}{34}
\newcommand{\XJTU}{35}
\newcommand{\CUC}{36}
\newcommand{\CalTech}{37}
\newcommand{\WM}{38}
\newcommand{\RPI}{39}
\newcommand{\CGNPG}{40}
\newcommand{\NUDT}{41}
\newcommand{\IowaState}{42}
\newcommand{\CQU}{43}
\author{
 F.~P.~An$^{\ECUST}$ \and
A.~B.~Balantekin$^{\Wisconsin}$ \and
H.~R.~Band$^{\Yale}$ \and
M.~Bishai$^{\BNL}$ \and
S.~Blyth$^{\NTU,\NUU}$ \and
D.~Cao$^{\NJU}$ \and
G.~F.~Cao$^{\IHEP}$ \and
J.~Cao$^{\IHEP}$ \and
W.~R.~Cen$^{\IHEP}$ \and
Y.~L.~Chan$^{\CUHK}$
% ...
\and
}
\maketitle 
\address{
\vspace{0.3cm}
{\normalsize (Daya Bay Collaboration)} \\ 
\vspace{0.3cm}
$^{\ECUST}$Institute of Modern Physics, East China University of Science and Technology, Shanghai \\ 
$^{\Wisconsin}$University~of~Wisconsin, Madison, Wisconsin 53706, USA \\ 
$^{\Yale}$Department~of~Physics, Yale~University, New~Haven, Connecticut 06520, USA \\ 
$^{\BNL}$Brookhaven~National~Laboratory, Upton, New York 11973, USA \\ 
$^{\NTU}$Department of Physics, National~Taiwan~University, Taipei \\ 
$^{\NUU}$National~United~University, Miao-Li \\ 
$^{\NJU}$Nanjing~University, Nanjing \\ 
$^{\IHEP}$Institute~of~High~Energy~Physics, Beijing \\ 
$^{\CUHK}$Chinese~University~of~Hong~Kong, Hong~Kong \\ 
$^{\NCTU}$Institute~of~Physics, National~Chiao-Tung~University, Hsinchu \\ 
$^{\SDU}$Shandong~University, Jinan \\ 
$^{\TsingHua}$Department~of~Engineering~Physics, Tsinghua~University, Beijing \\ 
$^{\NCEPU}$North~China~Electric~Power~University, Beijing \\ 
$^{\SZU}$Shenzhen~University, Shenzhen \\ 
$^{\ZSU}$Sun Yat-Sen (Zhongshan) University, Guangzhou \\ 
$^{\Dubna}$Joint~Institute~for~Nuclear~Research, Dubna, Moscow~Region \\ 
$^{\Siena}$Siena~College, Loudonville, New York  12211, USA \\ 
$^{\IIT}$Department of Physics, Illinois~Institute~of~Technology, Chicago, Illinois  60616, USA \\ 
$^{\UIUC}$Department of Physics, University~of~Illinois~at~Urbana-Champaign, Urbana, Illinois 61801, USA \\ 
$^{\LBNL}$Lawrence~Berkeley~National~Laboratory, Berkeley, California 94720, USA \\ 
$^{\SJTU}$Department of Physics and Astronomy, Shanghai Jiao Tong University, Shanghai Laboratory for Particle Physics and Cosmology, Shanghai \\ 
$^{\BNU}$Beijing~Normal~University, Beijing \\ 
$^{\UH}$Department of Physics, University~of~Houston, Houston, Texas  77204, USA \\ 
$^{\VirginiaTech}$Center for Neutrino Physics, Virginia~Tech, Blacksburg, Virginia  24061, USA \\ 
$^{\CIAE}$China~Institute~of~Atomic~Energy, Beijing \\ 
$^{\USTC}$University~of~Science~and~Technology~of~China, Hefei \\ 
$^{\NanKai}$School of Physics, Nankai~University, Tianjin \\ 
$^{\UC}$Department of Physics, University~of~Cincinnati, Cincinnati, Ohio 45221, USA \\ 
$^{\TempleUniversity}$Department~of~Physics, College~of~Science~and~Technology, Temple~University, Philadelphia, Pennsylvania  19122, USA \\ 
$^{\DGUT}$Dongguan~University~of~Technology, Dongguan \\ 
$^{\UCB}$Department of Physics, University~of~California, Berkeley, California  94720, USA \\ 
$^{\HKU}$Department of Physics, The~University~of~Hong~Kong, Pokfulam, Hong~Kong \\ 
$^{\Charles}$Charles~University, Faculty~of~Mathematics~and~Physics, Prague, Czech~Republic \\ 
$^{\Princeton}$Joseph Henry Laboratories, Princeton~University, Princeton, New~Jersey 08544, USA \\ 
$^{\XJTU}$Xi'an Jiaotong University, Xi'an \\ 
$^{\CUC}$Instituto de F\'isica, Pontificia Universidad Cat\'olica de Chile, Santiago, Chile \\ 
$^{\CalTech}$California~Institute~of~Technology, Pasadena, California 91125, USA \\ 
$^{\WM}$College~of~William~and~Mary, Williamsburg, Virginia  23187, USA \\ 
$^{\RPI}$Department~of~Physics, Applied~Physics, and~Astronomy, Rensselaer~Polytechnic~Institute, Troy, New~York  12180, USA \\ 
$^{\CGNPG}$China General Nuclear Power Group \\ 
$^{\NUDT}$College of Electronic Science and Engineering, National University of Defense Technology, Changsha \\ 
$^{\IowaState}$Iowa~State~University, Ames, Iowa  50011, USA \\ 
$^{\CQU}$Chongqing University, Chongqing \\ 
}


\begin{abstract}
A new measurement of the reactor antineutrino flux and energy spectrum by the Daya Bay reactor neutrino experiment is reported.
The antineutrinos were generated by six 2.9~GW$_{\mathrm{th}}$ nuclear reactors and detected by eight antineutrino detectors deployed in two near (560~m and 600~m flux-weighted baselines) and one far (1640~m flux-weighted baseline) underground experimental halls.
With 621 days of data, more than 1.2 million inverse beta decay (IBD) candidates were detected.
The IBD yield in the eight detectors was measured, and the ratio of measured to predicted flux was found to be $0.946\pm0.020$ ($0.992\pm0.021$) for the Huber+Mueller (ILL+Vogel) model.
A 2.9~$\sigma$ deviation was found in the measured IBD positron energy spectrum compared to the predictions.
In particular, an excess of events in the region of 4-6~MeV was found in the measured spectrum, with a local significance of 4.4~$\sigma$.
A reactor antineutrino spectrum weighted by the IBD cross section is extracted for model-independent predictions.
\end{abstract}



\maketitle

\begin{multicols}{2}
\section{Introduction}

Since the discovery of the neutrino in 1956 at the Savannah River reactor power plant by Cowan, Reines and collaborators~\cite{cowan1956}, reactor antineutrinos have played a crucial role in the development of the standard model of particle physics~\cite{bib:PDG2014}, and in the exploration of neutrino oscillation.
Near the beginning of this century, the CHOOZ and Palo Verde experiments attempted to measure the neutrino mixing angle $\theta_{13}$ using reactor antineutrinos at $\sim$1~km~baselines and obtained upper limits~\cite{bib:chooz1, bib:chooz2, bib:paloverda}.
In 2003, the KamLAND experiment observed terrestrial neutrino oscillations with a flux-average baseline of 180 km~\cite{KamLAND}, confirming the large mixing angle (LMA) solution to the solar neutrino problem.
In 2012, the Daya Bay experiment reported the first observation of a non-zero $\theta_{13}$~\cite{bib:prl_rate} with more than 5~$\sigma$ significance, consistent with the results from T2K~\cite{bib:t2k}, MINOS~\cite{bib:minos}, Double CHOOZ~\cite{bib:doublechooz} and RENO~\cite{bib:reno} experiments.
The discovery of a non-zero $\theta_{13}$ opened the way to determining the neutrino mass hierarchy and searching for CP violation in neutrino oscillation experiments.
In the future, reactor neutrino experiments at $\sim$km~baselines will continue to improve the precision of $\theta_{13}$ measurements, while reactor neutrino experiments at baselines of $\sim$50~km~\cite{bib:JUNOYB, bib:reno50} are aiming to determine the neutrino mass hierarchy and precisely measure the neutrino mixing angle $\theta_{12}$ and the mass-squared splittings $\Delta m^2_{21}$ and $\Delta m^2_{32}$.
In addition, reactor neutrino experiments at baselines of $\sim$10~m will probe physics beyond the three-neutrino framework through the search for short-baseline neutrino oscillation~\cite{Ashenfelter:2013oaa, Lane:2015alq, solid, stereo}.
A recent review of reactor neutrino oscillation experiments can be found in Ref.~\cite{Vogel:2015wua}.

Reactors are a pure source of electron antineutrinos, $\bar{\nu}_e$.
Inside a reactor core, fission processes are maintained by neutrons produced through the fission of $^{235}$U nuclei.
A portion of the neutrons are captured by $^{238}$U nuclei and subsequent beta decays and neutron captures lead to the production of fissile isotopes $^{239}$Pu and $^{241}$Pu.
The beta-decay chains of the fission products of these four isotopes are the main source of $\bar{\nu}_e$.
On average, about six antineutrinos are released per fission.
Before 2011, the prediction of antineutrino flux and spectrum was based on the beta spectra measured at ILL Grenoble for the thermal-neutron induced fission of $^{235}$U, $^{239}$Pu, and $^{241}$Pu~\cite{bib:ILL_1,bib:ILL_2,bib:ILL_3} and the theoretical calculation of Vogel for $^{238}$U~\cite{bib:vogel}, which was shown to be in good agreement with available data~\cite{reactor_review}.
In 2011, re-evaluation of the reactor antineutrino flux and spectrum~\cite{bib:mueller2011,bib:huber} with improved theoretical treatments was carried out, and the new predicted reactor antineutrino flux was shown to be higher than the experimental data.
This discrepancy is commonly referred to as the ``Reactor Antineutrino Anomaly''~\cite{bib:mention2011}.
One possible explanation of the reactor antineutrino anomaly is through neutrino oscillation with a frequency corresponding to a mass squared difference at the eV-scale, by introducing at least one additional sterile neutrino.
Meanwhile, it was pointed out in Ref.~\cite{bib:hayes} that the uncertainty due to the spectral shape of numerous first forbidden beta decays may be larger, which could largely reduce the significance of the anomaly.
In addition to the anomaly of the integrated reactor antineutrino flux, recent results from the current generation of $\theta_{13}$ experiments have also highlighted the presence of a spectral anomaly consisting of an excess of detected events with respect to predictions in the region of 4-6~MeV of the reconstructed prompt energy~\cite{reno_bump, prl_abs, dc_bump}.
This feature is unlikely to be the result of active-sterile neutrino oscillations, and raises further questions on the accuracy of some existing reactor antineutrino flux and spectrum predictions.

To shed light on these issues and probe the nuclear physics underlying current reactor antineutrino flux models, it is crucial to compare model predictions with precision measurements of reactor antineutrino flux and spectrum.
While the modeling of the reactor antineutrino spectrum is less critical for oscillation experiments employing relative measurements between multiple detectors, an accurate determination of the reactor antineutrino spectrum is critical to realize the full potential of the next-generation single-detector medium-baseline reactor antineutrino oscillation experiments~\cite{lisi1}.

This article will present Daya Bay's reactor antineutrino flux and spectral analyses utilizing the dataset from its most recent spectral oscillation analysis~\cite{bib:prl_shape2}.
The dataset is comprised of more than 1.2 million antineutrino candidates collected in eight antineutrino detectors (ADs) in two near experimental halls (with flux-weighted baselines of 560~m and 600~m) and one far hall (flux-weighted baseline 1640~m), providing a factor of 3.6 times more statistics over the results presented in Ref.~\cite{prl_abs}.
This paper also aims to provide detailed description of key inputs to these analyses not described in previous Daya Bay publications, such as the method of predicting the flux and spectrum from each Daya Bay core, as well as the method of determining the IBD detection efficiencies of the Daya Bay ADs.
Finally, a more detailed description will be provided regarding how the flux and spectrum analyses were carried out, and how the observed prompt spectra are unfolded into a reactor antineutrino spectrum, which is a useful input for future reactor antineutrino experiments.

This paper is organized as follows: Sec.~\ref{sec:flux} summarizes in detail the treatment of the reactor antineutrino flux and spectrum prediction in Daya Bay's neutrino oscillation analysis with the full eight-detector configuration~\cite{bib:prl_shape2}.
Sec.~\ref{sec:ibdselection} overviews the standard IBD selections used by Daya Bay, while Sec.~\ref{sec:Eff} provides an in-depth explanation of the analysis performed to determine the detection efficiency of the Daya Bay detectors.
The updated measurements of the reactor antineutrino flux and the positron prompt energy spectrum are presented in detail in Sec.~\textbf{\ref{sec:abs_flux}} and Sec.~\ref{sec:abs_spectrum}.
Based on the measured prompt energy spectrum, an extracted reactor antineutrino spectrum weighted by the IBD cross section is presented in Sec.~\textbf{\ref{sec:generic_spectrum}}.
Finally, a summary is given in Sec.~\textbf{\ref{sec:summary}}.


\section{Flux Prediction}\label{sec:flux}

\subsection{Reactor Description}
The Daya Bay nuclear power complex is situated at Daya Bay in southern China, approximately 55 kilometers northeast of Hong Kong.
As shown in Fig.~\ref{fig:layout}, the nuclear power complex consists of three nuclear power plants (NPPs): the Daya Bay NPP, the Ling Ao NPP, and the Ling Ao II NPP.
Each of them has a pair of reactor cores generating 2.9 GW thermal power each, during normal operation.
The distance between the two cores in each NPP is about 88 m.
The Ling Ao II cores reached full power in July 2011 while the other cores were running in commercial operation.
The uncertainty of the baseline measurement is estimated to be 18~mm.
Details of the baseline measurement are described in \cite{bib:detector}.

\begin{center}
\includegraphics*[trim=0.0cm 0.0cm 0.0cm 0.0cm, clip=true, width=\columnwidth]{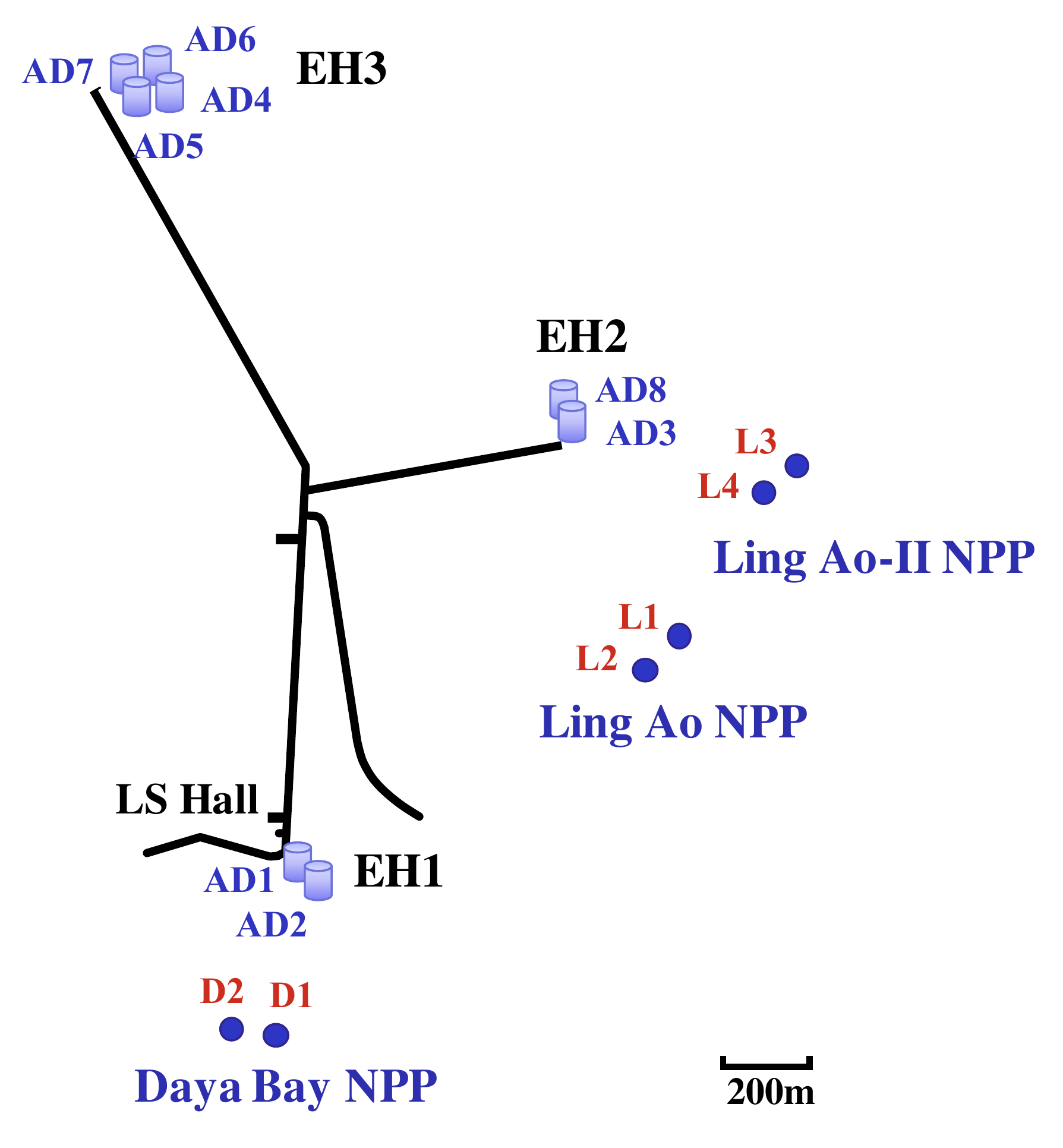}
\figcaption{Layout of the full configuration of the Daya Bay experiment with eight antineutrino detectors (ADs) installed in three underground experimental halls (EHs). The dots represent reactor cores, labeled as D1, D2, L1, L2, L3 and L4.
}
\label{fig:layout}
\end{center}

The Daya Bay and Ling Ao NPPs use the French Framatome Advanced Nuclear Power 990 MW$_\mathrm{e}$ (electric power) three cooling loop design, and Ling Ao II NPP uses an updated Chinese version (CPR 1000) of 1080-MW$_\mathrm{e}$.
Each cooling system consists of a primary loop and a secondary loop connected with a steam generator.
Figure~\ref{fig:ReactorLoop} shows a schematic diagram of one cooling system.
Inside each reactor core, 157 fuel elements are bonded to socket plates in the water-filled reactor pressure vessel.
The water absorbs the heat generated by fissions in the fuel and then circulates through inverted U-shape tubes of the steam generators, which are immersed in water of the secondary loops.
The heat is then transferred to the water in the secondary loop and the water is vaporized into saturated steam, which flows to the turbine-alternator unit.
The cooled water in the primary loop is then pumped back to the vessel and goes to the next cycle.
The water is slightly doped with boric acid, which acts as the thermal neutron absorber.
Boron concentration, controlled by the NPPs, decreases during the refueling cycle to compensate for the power loss caused by the depletion of fuel, helping to keep the total power of the reactor stable at a nominal level.
\begin{center}
\includegraphics*[trim=0.0cm 0.0cm 0.0cm 0.0cm, clip=true, width=\columnwidth]{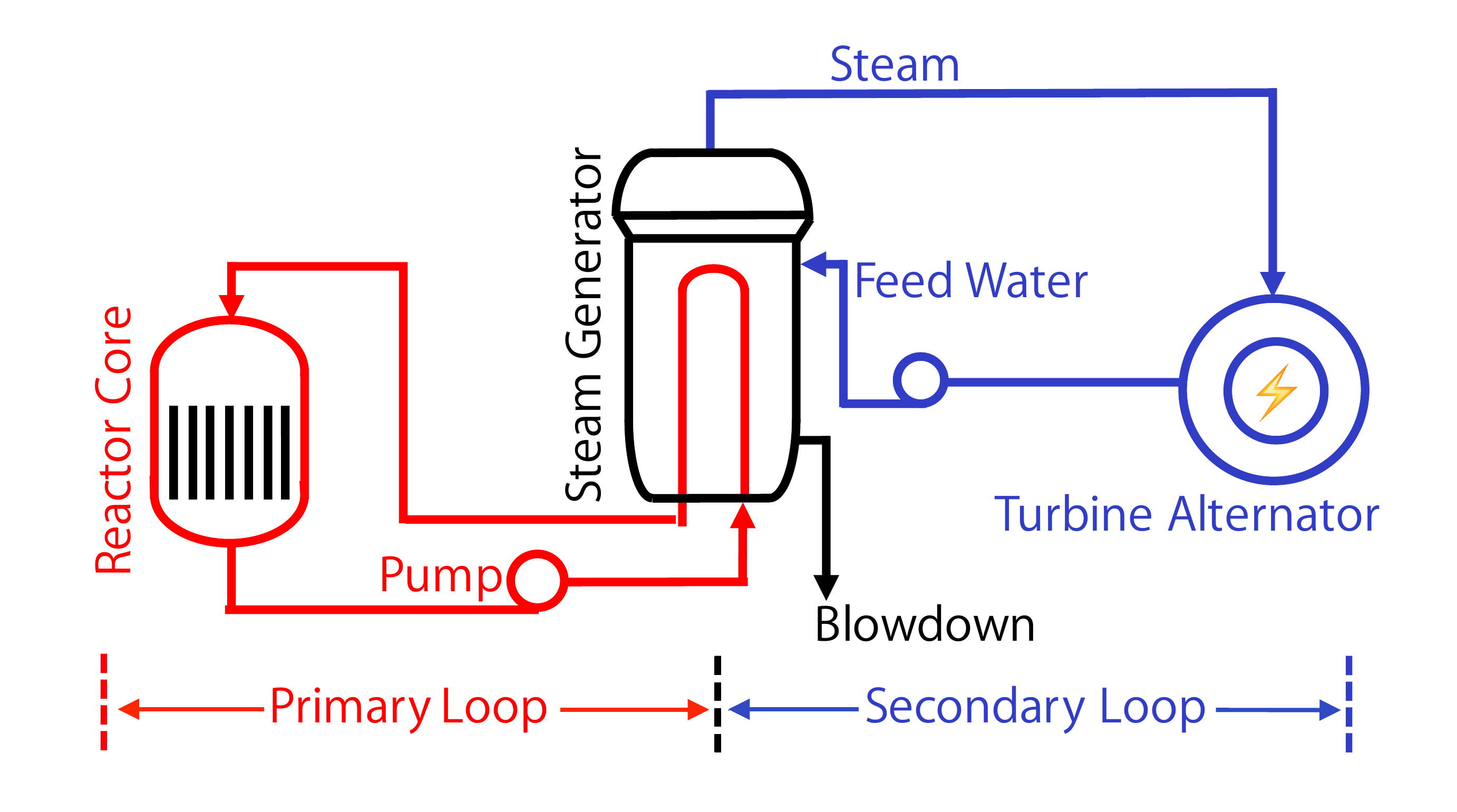}
\figcaption{ Schematic diagram of the reactor cooling
system.  At Daya Bay, each reactor core is connected with 3 cooling systems in parallel.}
\label{fig:ReactorLoop}
\end{center}

\subsection{Reactor Power Measurements and Monitoring Systems}
Three different systems, RPN (Nuclear Instrumentation System)~\cite{bib:RPN_KIT_KDO}, KME (Test Instrumentation System)~\cite{bib:KME}~\cite{bib:CTEC}, and KIT/KDO (Centralized Data Processing System/Test Data Acquisition System)~\cite{bib:RPN_KIT_KDO,bib:CTEC}, were deployed to monitor the power of the reactor cores in Daya Bay. Table~\ref{tab:Power_Systems} is a summary of the three power monitoring systems.
\begin{center}
 	 \tabcaption{Power monitoring systems in Daya Bay. `FP' stands for `Full Power'}
 	 \footnotesize
	  \begin{tabular}{c | c | c  }
	  \toprule
	    System& Frequency & Uncertainty  \\\hline
	    KME & Weekly/Monthly & $<0.5\%$\\
	    KIT/KDO & Online & $|P_{KIT}-P_{KME}|<0.1\% FP$ \\
	    RPN & Online & $|P_{RPN}-P_{KME}|<1.5\%FP$\\
	    \bottomrule
	   \end{tabular}
	   \label{tab:Power_Systems}
\end{center}

The RPN system is used for reactor monitoring and protection by measuring the neutron flux with four neutron detectors placed around the reactor core.
The reactor power is supposed to be proportional to the neutron flux.
However, as the nuclear fuel burns, the power as measured by RPN gradually differs by an increasing amount from the actual power due to the change of the isotope content in the core.
To guarantee accuracy, the RPN system's measured powers are compared with the more accurate KIT/KDO system every day.
Once the difference exceeds 1.5$\%$ of full power, the RPN system is re-calibrated.

The KME and KIT/KDO systems are based on the heat balance method.
The KME is the secondary loop power measurement system, and has the best accuracy among all three systems.
This system measures the parameters such as water flow rate, temperature and pressure in the secondary loop, and calculates the enthalpy increase when the water passes through the steam generator.
Other heat sources such as pumps in the secondary loop are also considered.
By considering the power of all three steam generators and heat from other sources, the reactor core thermal power can be calculated as
\begin{equation}\label{equ:thermal_power}
W_{R}=\sum_{i=1}^{3} W_{SG_{i}}-W_{\Delta Pr} \, ,
\end{equation}
where $W_{R}$ is the reactor core thermal power, $W_{SG_{i}}$ is the thermal power of the $i$-th steam generator, and $W_{\Delta Pr}$ is the heat input and power loss from the pump systems and other heat sources.

Daya Bay and Ling Ao reactors are all based on French Pressurized Water Reactors (PWRs).
For French PWRs, the measurement of nominal thermal power follows a procedure known as BIL100, which is performed on the secondary loop~\cite{bib:EFRI_report}.
The predominant term in the calculation of uncertainty for BIL100 is the uncertainty related to mass flow rate of the feed water, which accounts for up to 80\% of the uncertainty related to thermal power~\cite{bib:EFRI_report}.
To minimize this source of uncertainty, orifice plates were installed in the secondary loop to precisely measure water flow.
The uncertainty of the orifice water flow measurement is typically 0.72\% (90\% C.L.), and could be improved to 0.4\% (90\% C.L.) according to lab tests~\cite{bib:caojPower}.
For Daya Bay's KME system, four benchmark tests were made to compare the core power result between the KME system and an EDF (Electricite de France)-developed high precision SAPEC system (EDF's standardized system for enhanced safety and performance periodic tests on the PWR fleet), which has its own sensors, databases and data processing systems~\cite{bib:KME_BenchMark}.
The tests showed the relative difference between the two systems was 0.031\% to 0.065\%.
The power measurement uncertainty of the KME system is estimated to be less than 0.25\%.
This is comparable to the uncertainty estimated for the SAPEC system, which is $<$0.26\%~\cite{bib:KME_BenchMark}.

Although the KME system has the best precision, it is an offline system. The power plant usually does the KME measurements weekly or monthly, but this frequency does not meet the experimental requirement.
The KIT/KDO system is an online system for monitoring the core power, based on a primary loop heat balance method.
The system measures the temperature, pressure, and mass flow rate of the feed water in the primary loop to calculate the thermal power.
Installation of orifice plates in the primary loop is not allowed, thus the KIT/KDO system uses another flow meter to measure the water flow, which is less precise than the KME system.
However, the KIT/KDO system is calibrated monthly to the KME system by adjusting the feed water flow rate in the primary loop in the KIT/KDO system once the difference between the powers measured by the two systems exceeds 0.1\% of full power.
Conservatively, considering that the uncertainty between the steam generators is fully correlated in the KME system, and accounting for the difference between the KIT/KDO system and the KME system, the uncertainty of the KIT/KDO system is estimated to be 0.5\%.
In the Daya Bay Experiment the KIT/KDO measured thermal power is provided hourly to calculate the reactor antineutrino flux.

\subsection{Reactor Core and Refueling}
The reactor core consists of 157 fuel elements, and each element contains 264 fuel assemblies of uranium dioxide with a $^{235}$U enrichment of 4\%.
The height of the elements is 3.7~m and the diameter of the core is 3~m.
The six reactors shut down alternately for refueling and overhaul.
The refueling cycle period for the Daya Bay NPP is about 18 months, with 1/3 of all fuel elements replaced with fresh fuel during the refueling period.
For the Ling Ao NPPs, the refueling cycle period is 12 months, and 1/4 of the fuel elements are replaced. Refueling usually takes one month.
At the beginning of each burning cycle, the positions of the fuel elements in the core are rearranged.
Fresh fuel is placed in the core center, while the old fuel is moved outward.
This scheme has the advantages of reducing neutron leakage, enhancing activity, and increasing fuel burn-up.
On the other hand, the scheme results in a non-uniform power distribution in the core and increases the power peaking factor.
To reduce this effect, burnable gadolinium fuel ``poison'' rods are installed in some elements to absorb neutrons.
Figure~\ref{fig:coremap} shows an example of the reactor core map of the fuel elements with different burn-up at the end of a refueling cycle.


When refueling for a new cycle, the fuel elements are configured in the reactor core around the center as symmetrically as possible.
Because of this, the reactor core can be considered as a point source of antineutrinos with a center of gravity stable at the core center, as will be discussed later.

\subsection{Fuel Evolution and Core Simulation}
Burn-up describes the energy extracted from the fuel element per ton of initial uranium mass since its placement into the reactor core, defined as
\begin{equation}\label{equ:burnup}
\textrm{burn-up} \equiv \frac{W \cdot D}{M_{U^{\textrm{in}}}} \, ,
\end{equation}
where $W$ is the average power of the fuel element, $D$ is the days since the fuel element begins to burn in the core, and $M_{U^{\textrm{in}}}$ is the initial uranium mass of the fuel element.
The unit of burn-up is $\mathrm{MW} \cdot \mathrm{day} \cdot \mathrm{ton_U}^{-1}$.
A similar quantity, cycle burn-up, is used to describe the aging of the whole reactor core in a refueling cycle.
Cycle burn-up can also be calculated using Eq.~\ref{equ:burnup}, where $W$, $D$, and $M_{U^{\textrm{in}}}$ in this case represent the total nuclear power of the reactor core, the days since the beginning of the refueling cycle, and the initial uranium mass of all the fuel elements in the reactor core.

\begin{center}
\includegraphics*[trim=0.0cm 0.0cm 0.0cm 0.0cm, clip=true, width=\columnwidth]{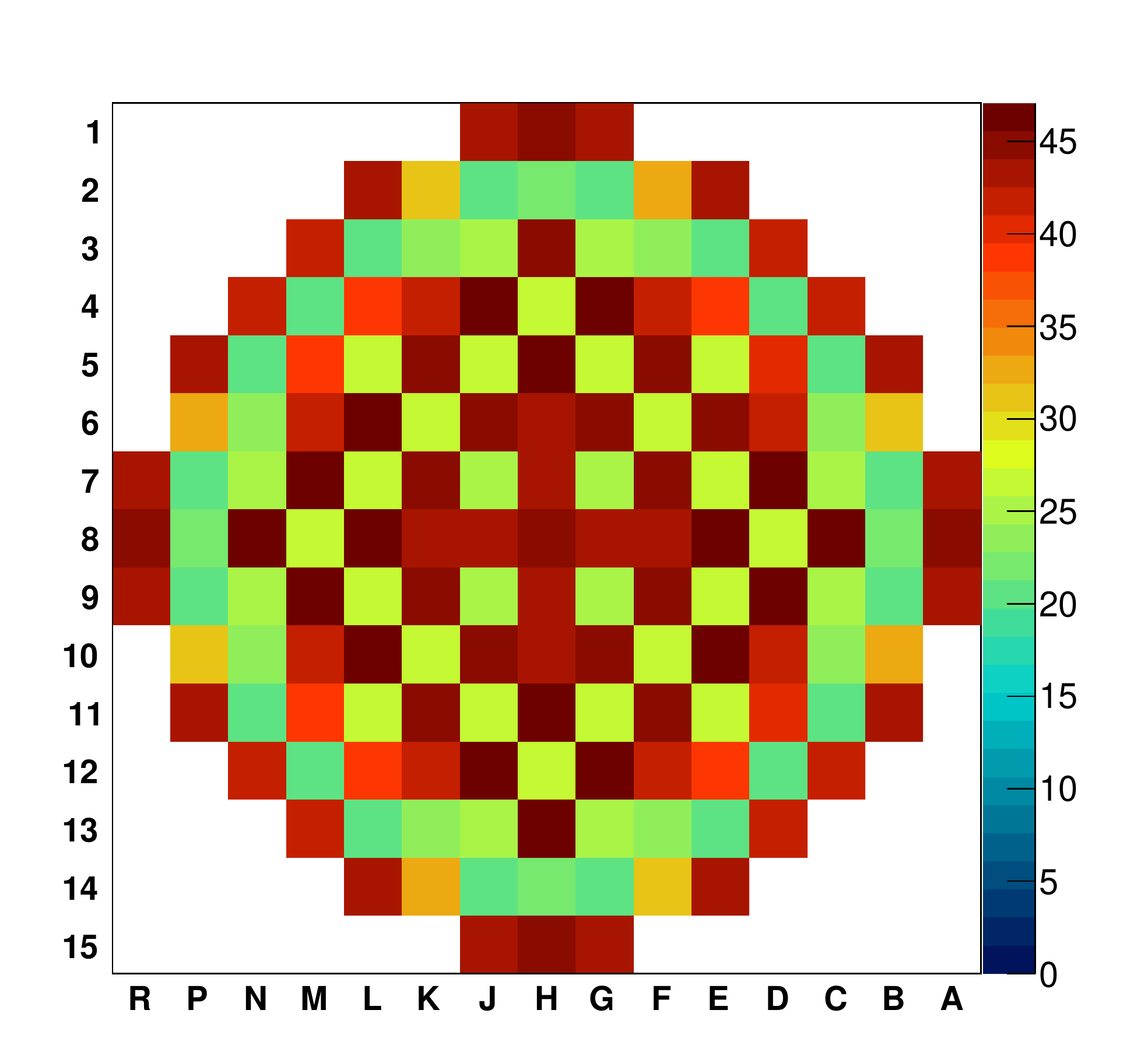}
\figcaption{An example of the reactor core map of fuel elements with different burn-up (unit: $\textrm{GW} \cdot \textrm{day} \cdot \textrm{ton}^{-1}$) shown in color scale at the end of a refueling cycle.
}
\label{fig:coremap}
\end{center}

In reactors, electron antineutrinos are emitted primarily from the fissions of four isotopes: $^{235}$U, $^{238}$U, $ ^{239}$Pu, and $^{241}$Pu.
Fissions of other isotopes contribute less than 0.3\%.
Fissions of $^{238}$U are only induced by fast neutrons, while fissions of the other three isotopes are mainly induced by thermal neutrons.
Fresh fuel elements contain only uranium isotopes.
The plutonium isotopes are gradually generated through neutron captures on $^{238}$U and subsequent neutron captures and beta decays of its successor isotopes.

Fuel evolution is a dynamic process related to many factors such as power, neutron flux, fuel composition, type and position of fuel elements, and boron concentration.
For safe operation of the reactors, NPPs do calculations and simulations of the fuel evolution in every refueling cycle by considering all of the factors above.
These detailed simulations are performed by validated and licensed commercial software.
The simulation package used by the Daya Bay NPP is SCIENCE, which was developed by CEA, France.
It uses the APOLLO2 code~\cite{bib:Appollo_Validation} as the core component.
The simulation results are provided to the Daya Bay collaboration in a table which uses cycle burn-up as the index.
The fission fractions are provided by the simulation in the form of $f_i(\beta)$,
where $f_{i}$ is the fission fraction of isotope $i$, and $\beta$ is the cycle burn-up.
Figure~\ref{fig:core_evolution} shows an example of the fission fraction evolution as a function of cycle burn-up within a refueling cycle~\cite{bib:cpc_rate}.

\begin{center}
\includegraphics*[trim=0.0cm 0.0cm 0.0cm 0.0cm, clip=true, width=\columnwidth]{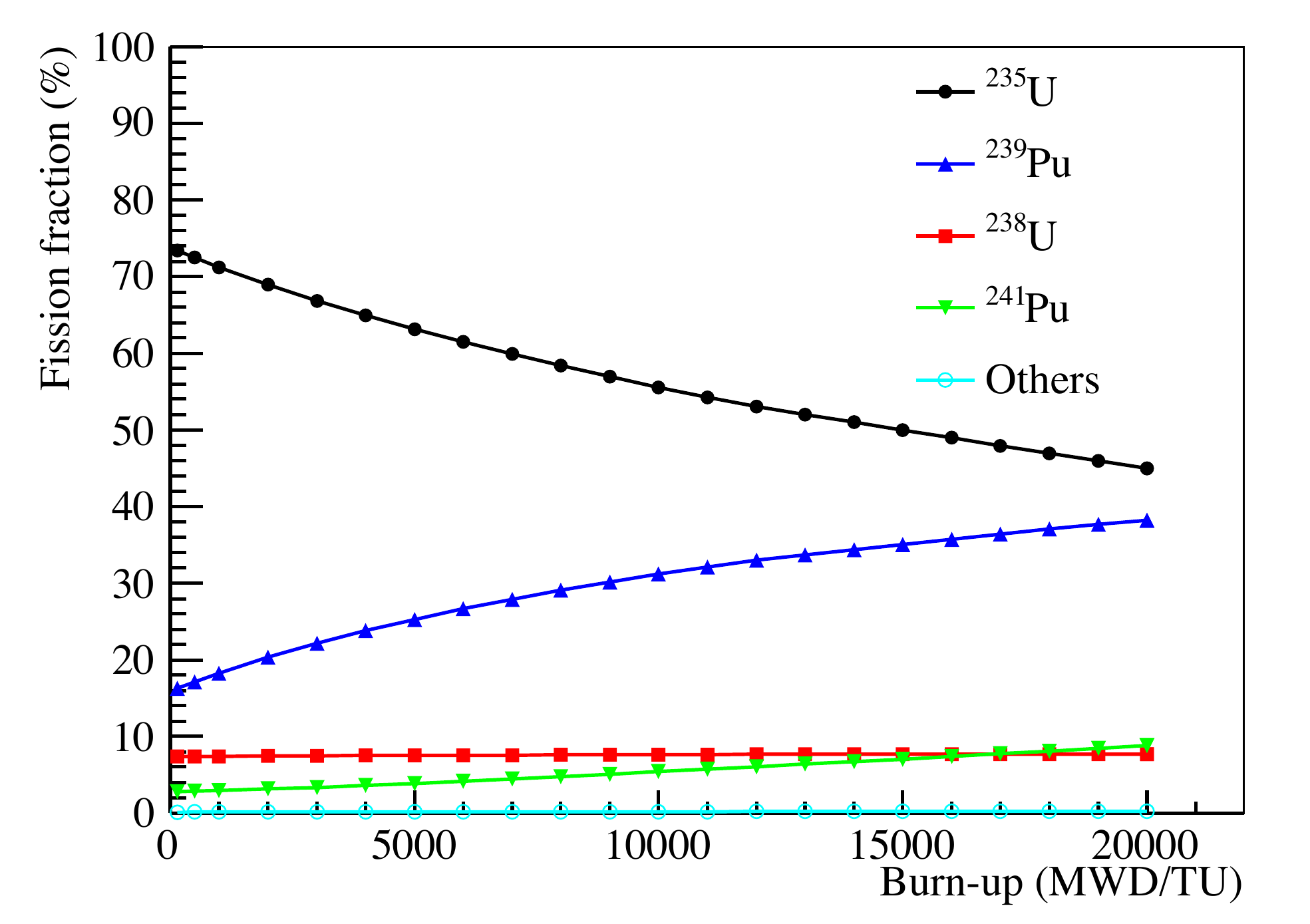}
\figcaption{Fission fractions of isotopes in reactor core D1 as a function of cycle burn-up from a simulation of a complete refueling cycle. Other isotopes contribute less than 0.3\%.}
\label{fig:core_evolution}
\end{center}


The APOLLO2 code is widely used for cross section generation and neutron transport calculations in commercial reactor cores.
It adopts rigorous methodology for its validation, including comparison with the reference calculation using the same nuclear data libraries, and with the experimental measurements~\cite{bib:Appollo_Validation}.
Measurements of spent fuel isotopic content were made and compared with the results calculated using the APOLLO2 code~\cite{bib:Appollo_2010}.
The comparison shows that the measurement-model deviations are less than 5\%.
Therefore, the uncertainty of the calculated fission fraction is conservatively estimated to be 5\% for each isotope.

The NPPs also provide 3D core simulation results for different burn-up stages, which enable an investigation of the spatial distribution of the antineutrino production inside the core.
The reactor can be considered as a point source of $\bar\nu_{e}$ for the Daya Bay experiment because the fuel elements are symmetrically arranged in the reactor core as shown in Fig~\ref{fig:coremap}.
The relative difference between treating the reactor as a point source and as a finite source is negligible and the variation of the effective fission center in the reactor is estimated to be 2 cm horizontally. 
The impact on the baselines of the vertical variation of the fission center is negligible.
Combined with the 18~mm uncertainty in the baseline measurements, the total uncertainty of the baselines is conservatively estimated to be 27~mm.

The open source simulation code DRAGON~\cite{bib:dragon}was also used to calculate the fission fractions, and to estimate their uncertainty.
The impact of many reactor parameters was taken into account, including power, neutron flux, fuel composition, type and position of fuel elements, and boron content.
DRAGON was originally developed for CANDU (CANada Deuterium Uranium) reactors, but also yields reliable predictions for PWRs~\cite{bib:MIT_DRAGON,bib:MaXubo_DRAGON}.
The fission fraction uncertainty of each isotope was found to be less than 5\%, consistent with the results of APOLLO2 validation.
The fission fractions of four isotopes are correlated with each other because $ ^{239}$Pu and $^{241}$Pu are gradually produced while $^{235}$U is continuously consumed and the sum of the fission fractions is normalized to be 100\%.
DRAGON was used to calculate correlations among fission fractions using the fission fraction data from several cycles of the NPPs. The results are given in Table~\ref{tab:fisFrac_cor}.
The correlations were used as an input when propagating the fission fraction uncertainties to the reactor antineutrino flux uncertainty.

\begin{center}
   \tabcaption{Correlation coefficients of fission fractions for the four isotopes.}
  \begin{tabular}{c | c | c | c | c}
  \toprule
    Isotope &  $^{235}$U & $^{238}$U &$^{239}$Pu &  $^{241}$Pu  \\\hline
    $^{235}$U &  1.00 & -0.22 & -0.53 & -0.18\\
    $^{238}$U & -0.22 & 1.00 & 0.18 & 0.26\\
    $^{239}$Pu & -0.53 & 0.18 & 1.00 & 0.49\\
    $^{241}$Pu & -0.18 & 0.26 & 0.49 & 1.00\\
    \bottomrule
   \end{tabular}
   \label{tab:fisFrac_cor}
\end{center}

\subsection{Expected Unoscillated Spectrum}
Electron antineutrinos are generated in the reactors from the beta decays of the fission fragments produced by the four isotopes.
Each fission isotope produces a unique $\bar\nu_e$ spectrum through its fission and subsequent decay chains.
In principle, using cumulative fission yields and beta decay information for each fission production, it is possible to compute the antineutrino spectrum ab initio.
However, this requires reliable beta decay information on more than 1000 
isotopes~\cite{bib:dan}, many of which have never been observed. The lack of decay information combined with nuclear structure-related uncertainties and the uncertainties of the fission yields, results in an overall ~10--20\% energy dependent uncertainty in the predicted antineutrino spectrum.

To improve on the purely ab initio method described above, several direct measurements were done at ILL~\cite{bib:ILL_1, bib:ILL_2, bib:ILL_3} in the 1980s to determine the electron energy spectra from the individual fission isotopes $^{235}$U,  $^{239}$Pu, and $^{241}$Pu.
In these measurements, foils of isotope samples were placed inside the reactor and exposed to thermal neutron fluxes for 1--2 days.
A high-precision electron spectrometer measured the electrons emitted by the samples.
The observed electron spectrum was then converted into an antineutrino spectrum by fitting with a set of hypothetical $\beta$-decay branches and adding up the antineutrino spectrum from each fitted branch.
The uncertainty of the antineutrino spectrum by this conversion process was estimated to be 2.7\%.
These experiments did not perform similar measurements for $^{238}$U, which only fissions with fast neutrons.
Theoretical antineutrino flux calculations for $^{238}$U were carried out by Vogel~\cite{bib:vogel}, with overall uncertainties $<10\%$.
Since $^{238}$U only contributes to $\sim$8\% of the total reactor antineutrino flux, the error introduced to the total flux is less than 1\%.
These calculations of antineutrino spectra are referred to as the ILL+Vogel model.

The prediction of antineutrino spectra from $^{235}$U, $^{239}$Pu, and $^{241}$Pu was recently improved~\cite{bib:mueller2011, bib:huber}, where the ILL electron spectra were reanalyzed by taking into account several higher-order corrections to the $\beta$-decay spectra.
The ab initio calculation of the $^{238}$U antineutrino spectrum was updated by Mueller {\it et al.}~\cite{bib:mueller2011}.
These new calculations are referred to as the Huber+Mueller model. The claimed uncertainty of the predicted total flux from the Huber+Mueller model is 2.4\%.
Both the ILL+Vogel model and the Huber+Mueller model are used to calculate the expected antineutrino spectrum from a single reactor core.
A measurement of the $^{238}$U beta spectrum was performed and the corresponding antineutrino spectrum was determined in Ref.~\cite{bib:munich}.
Replacing the Mueller $^{238}$U antineutrino spectrum with this measurement only changes the total integrated flux by 0.2\% since $^{238}$U only contributes 8\% of the total integrated flux.

The total antineutrino spectrum is calculated once the time evolution of reactor power and fission fractions are provided by the Daya Bay NPP,
 \begin{equation}\label{equ:total_reactor_spectrum}
\frac{d\phi(E_{\nu})}{dE_{\nu}}=\sum_{i} F_{i} \cdot \frac{d\phi_{i}(E_{\nu})}{dE_{\nu}},
\end{equation}
where $i$ is the index of individual fission isotope in the reactor fuel, that is $^{235}$U, $^{238}$U, $^{239}$Pu, or $^{241}$Pu. $d\phi_{i}(E_{\nu})/dE_{\nu}$ is the antineutrino spectrum of the $i$-th isotope per fission, and $F_{i}$ is the total fission rate of the $i$-th isotope.
The total fission rate is directly related to the total thermal power of the reactor core, and can be calculated as follows:
 \begin{equation}\label{equ:total_fission_rate}
F_{i}=\frac{W_{th}}{\sum_{j} f_{j} \cdot e_{j}} \cdot f_{i} \, ,
\end{equation}
where $W_{th}$ is the total thermal power of the reactor core, $e_{i}$ is the energy released per fission of the $i$-th isotope, and $f_{i}$ is the fission fraction of the $i$-th isotope.
The term $\sum_{i} f_{i} \cdot e_{i}$ represents the average energy released per fission from the four isotopes.

The energy released per fission ($e_{i}$) is defined as the amount of energy from a fission event that transforms into heat over a finite time interval~\cite{bib:kopeikin}, which has a slight dependence on the reactor burning history.
They were calculated by considering the neutron captures in the reactor and decays of long-lived fission daughters, using typical PWR reactor parameters~\cite{bib:kopeikin}.
The improved calculation of the energy released per fission~\cite{bib:fr_ma} used in this analysis includes using updated nuclear databases, considering the production yields of fission fragments from both thermal and fast incident neutrons, and an updated calculation of the average energy taken away by antineutrinos.
This new calculation gives slightly larger values of $e_{i}$ with smaller uncertainties than in~\cite{bib:kopeikin}, resulting in a 0.32\% decrease of the calculated antineutrino flux.
The values of $e_{i}$ and their uncertainties are listed in Table~\ref{tab:Xubo_Energy}.

\begin{center}
\tabcaption{Energy released per fission for the four main isotopes and their uncertainties.~\cite{bib:fr_ma}}
    \begin{tabular}{c | c}
    \toprule
      Isotope & Energy per Fission (MeV)  \\\hline
      $^{235}$U & 202.36 $\pm$ 0.26\\
      $^{238}$U & 205.99 $\pm$ 0.52\\
      $^{239}$Pu & 211.12 $\pm$ 0.34\\
      $^{241}$Pu & 214.26 $\pm$ 0.33\\
    \bottomrule
     \end{tabular}
  \label{tab:Xubo_Energy}
\end{center}

In the Daya Bay experiment, the electron antineutrinos are detected via the inverse beta decay (IBD) reaction: $\bar{\nu}_{e} + p \to e^{+} + n $.
The expected antineutrino spectrum weighted by the IBD cross section in the detector $d$ from reactor $r$ is calculated by
\begin{equation}\label{equ:singleReactor_prediction}
S_{dr}(E_{\nu})=\frac{1}{4 \pi L_{dr}^{2}} \frac{d\phi(E_{\nu})}{dE_{\nu}} \epsilon^d N^d_{p} \sigma (E_{\nu}),
\end{equation}
where $L_{dr}$ is the distance from reactor $r$ to detector $d$,  $\epsilon^d$ is the IBD selection efficiency, $N^d_{p}$ is the number of target protons, and $\sigma (E_{\nu})$ is the inverse beta decay cross section calculated using the formalism in~\cite{Vogel:1999zy}, with the updated neutron lifetime of 880.3$\pm$1.1~s taken from PDG 2014~\cite{bib:PDG2014}.
The uncertainty of the cross section is dominated by the uncertainty of neutron lifetime.
The total reactor antineutrino spectra for a detector $d$ is the sum of antineutrino spectra from all reactors:
\begin{equation}\label{equ:single_Reactor_prediction}
S_{d}(E_{\nu})=\sum_{r} S_{dr}(E_{\nu}) \, .
\end{equation}
As an example, the expected total antineutrino spectrum at the near site ADs is shown in Fig.~\ref{fig:SpectrumPrediction}.

%
\begin{center}
\includegraphics*[trim=0.0cm 0.0cm 0.0cm 0.0cm, clip=true, width=\columnwidth]{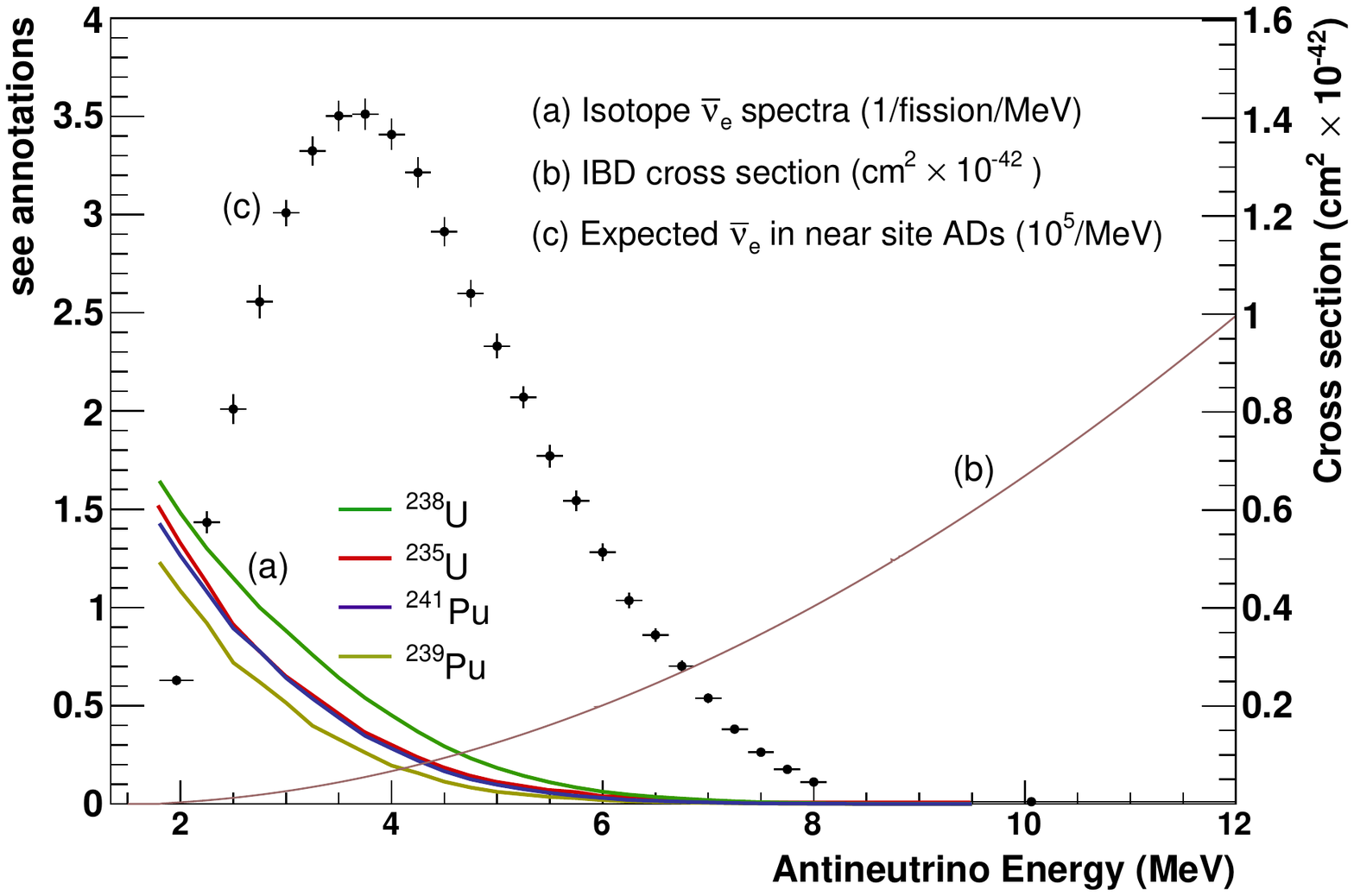}
\figcaption{ (a) The antineutrino spectra for four isotopes in Huber+Mueller model.(b) The inverse beta decay (IBD) cross section. (c) The expected antineutrino spectrum weighted by the IBD cross section without oscillation in the near site ADs. The error bars are systematic only (see text for details). }
\label{fig:SpectrumPrediction}
\end{center}

\subsection{Non-equilibrium Effect and Spent Nuclear Fuel Correction}

In the ILL measurements, fissile samples were exposed to the thermal neutron flux for only 1--2 days.
The rate of beta decays from some long-lived fission fragments did not reach equilibrium with their production rates.
When using converted antineutrino spectra from the ILL measurements, this non-equilibrium effect needs to be corrected, since the long-lived fission fragments accumulate in the reactor core and their beta decays contribute to the total antineutrino flux.

After burning in the core, the nuclear fuel is removed from the reactor and stored as spent nuclear fuel (SNF) in a cooling pool near the reactor core.
The long-lived isotopes in the SNF will decay and act as another source of antineutrinos.

The total neutrino spectrum is then modified:
\begin{equation}\label{equ:reactor_snf_long}
S_{\nu}=S_{ILL} + S_{neq} + S_{SNF}
\end{equation}
where $S_{ILL}$ is the expected antineutrino spectrum with ILL measurement-based models, $S_{neq}$ is the contribution from the non-equilibrium effect and $S_{SNF}$ is the contribution from the spent fuel.

The non-equilibrium correction is a function of antineutrino energy, the burn-up and irradiation history of nuclear fuel~\cite{bib:mueller2011}.
Taking into account the information of the refueling history of reactors provided by the China General Nuclear Power Corporation, the cumulative contribution of the non-equilibrium effect at Daya Bay and Ling Ao reactors was calculated.
On average, the effect contributed $\sim$0.6\% additional IBD events, which is illustrated in Fig.~\ref{fig:SNF}.
The uncertainty of the non-equilibrium effect is taken to be 30\% from the estimation in Ref.~\cite{bib:mueller2011}.

The contribution of SNF can be evaluated by using the cumulative yields and spectra of the known long-lived fission fragments.
The candidate isotopes were selected from the fission products with the condition that they have a half-life longer than 10 hours and either the isotope or its daughter nuclei undergoes beta decay with end point energy larger than the IBD reaction threshold ($1.8$~MeV).
The antineutrino spectra of these candidate isotopes were calculated based on their beta decay process.
The cumulative yields of the SNF were calculated with the input from the refueling history and SNF inventory information provided by the China General Nuclear Power Corporation.
The calculated SNF antineutrino spectrum is illustrated in Fig.~\ref{fig:SNF}. The contribution to the total number of IBD events is $\sim 0.3\%$, which is consistent with previous calculations~\cite{bib:fengpeng, bib:zhoubin}.
The uncertainty is conservatively estimated to be 100\% after the investigation on the uncertainty of the SNF inventory history information.
We neglect an additional low energy correction~\cite{bib:huber_ncap} which has a smaller effect than SNF.
\begin{center}
\includegraphics*[trim=0.0cm 0.0cm 0.0cm 0.0cm, clip=true, width=\columnwidth]{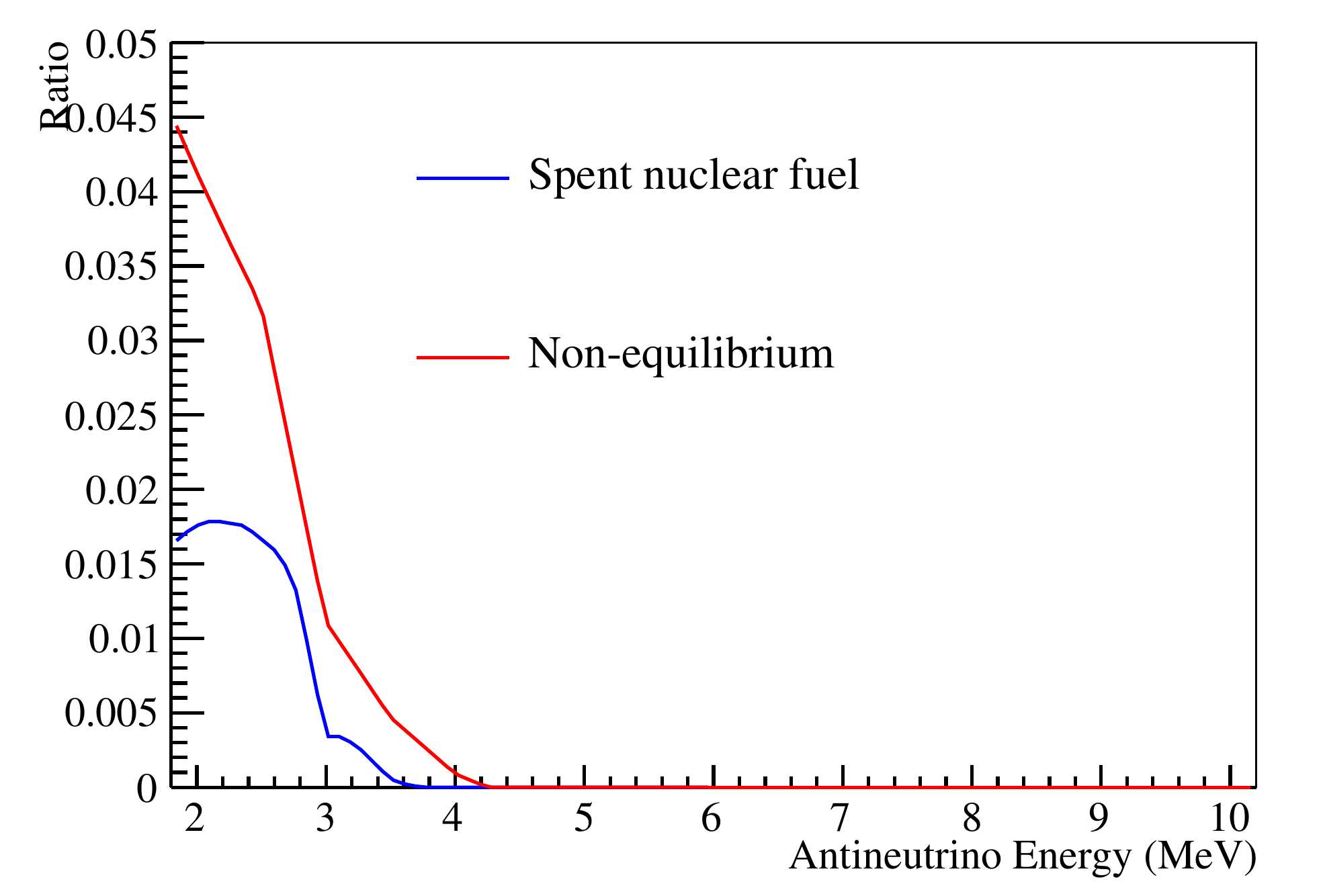}
\figcaption{The ratio of calculated antineutrino spectrum the non-equilibrium effect (red) and spent nuclear fuel (blue) to that from the four fissile isotopes in reactor core. The drop at 3 MeV is due to the end point energy of $^{144}$Pr beta decay, which contributes the most with its mother nuclide $^{144}$Ce to SNF antineutrinos.}
\label{fig:SNF}
\end{center}

\subsection{Systematic Uncertainties of the Predicted Reactor Antineutrino Spectrum}
The systematic uncertainties of the predicted reactor antineutrino spectrum can be categorized as either correlated or uncorrelated among different reactor cores.
The list of systematic uncertainties, and their values for the integrated reactor antineutrino flux, are shown in Table~\ref{tab:ReactorUncertainty}.
The combined correlated uncertainty is taken to be 2.7\% from the ILL+Vogel model (or  2.4\% from the Huber+Mueller model).
The correlated uncertainties are common for all reactor cores, therefore they are irrelevant in the neutrino oscillation analysis where only the relative rate and spectrum between the near and the far detectors are compared.
The combined uncorrelated uncertainty is 0.9\%, as a square root of the quadratic sum of the uncorrelated items, including power, energy/fission, fission fraction, spent fuel, and non-equilibrium in Table~\ref{tab:ReactorUncertainty}.

\begin{center}
\footnotesize
\tabcaption{Summary of the systematic uncertainties of the predicted integrated reactor antineutrino flux associated with a single reactor core.
}
\begin{tabular}{c|c}
\toprule
& uncertainty   \\\hline
power & 0.5\%   \\\hline
energy/fission & 0.2\%  \\\hline
isotope spectrum & 2.7\%  \\\hline
IBD cross section & 0.12\%  \\ \hline
fission fraction & 0.6\%  \\\hline
baseline & negligible \\\hline
spent fuel & 0.3\%  \\\hline
non-equilibrium & 0.2\% \\
\bottomrule
\end{tabular}
\label{tab:ReactorUncertainty}
\end{center}

\begin{center}
\includegraphics*[trim=0.0cm 0.0cm 0.0cm 0.0cm,  width=\columnwidth]{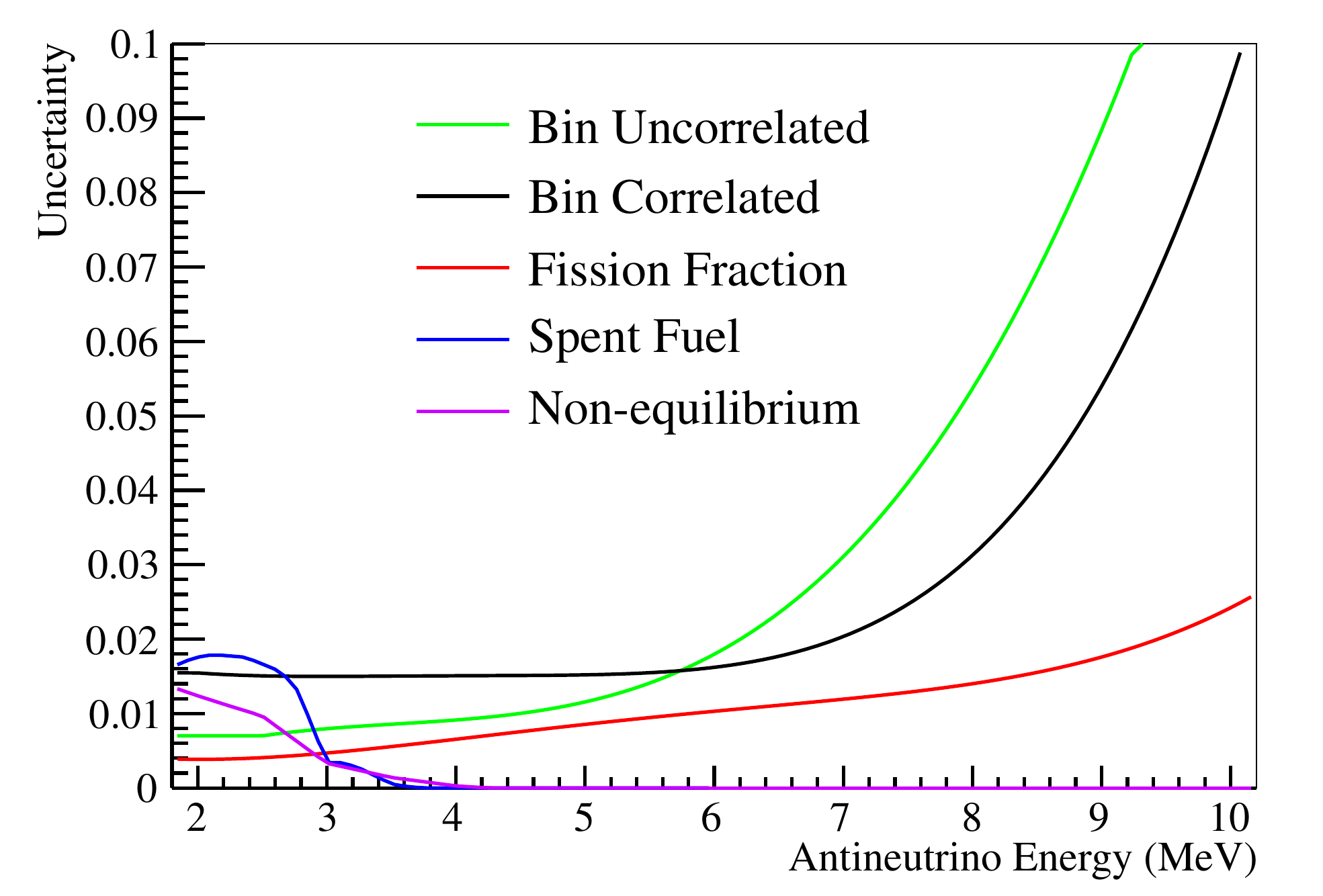}
\figcaption{The systematic uncertainties of the predicted reactor antineutrino spectrum from each energy-dependent component. The bin uncorrelated/correlated uncertainty (see text for details) is the average value of the four primary isotopes, weighted by their fission fractions.}
\label{fig:reactor_uncertainty}
\end{center}

Some uncertainties are dependent on antineutrino energy, and can induce fluctuations in the energy spectrum, while the others only impact the integrated antineutrino flux.
The contribution from each energy-dependent component is broken down and shown in Fig.~\ref{fig:reactor_uncertainty}.
The energy-dependent uncertainties can be further categorized as correlated or uncorrelated between energy bins.
The isotope antineutrino spectra of $^{235}$U,  $^{239}$Pu, and $^{241}$Pu are converted from the respectively measured beta decay spectra.
The uncertainties of these spectra have both bin-to-bin correlated and uncorrelated components.
The bin-to-bin correlated uncertainty is induced by the electron to antineutrino spectrum conversion models.
The bin-to-bin uncorrelated uncertainty is induced by the statistical uncertainty of the measured beta decay spectra.
The antineutrino spectrum of $^{238}$U is based on theoretical calculation, and its uncertainty is bin-to-bin correlated.

The size of the total uncertainty is shown as the error bars on the predicted antineutrino spectrum in Fig.~\ref{fig:SpectrumPrediction}.

\section{Inverse Beta Decay Event Selection} \label{sec:ibdselection}

After production in the six Daya Bay reactor cores as described above, $\bar{\nu}_e$ are detected in identically designed Daya Bay antineutrino detectors (ADs).
Each AD consists of three nested cylindrical vessels.
The inner acrylic vessel (IAV) with a thickness of 11~mm is filled with 0.1\% gadolinium-doped liquid scintillator (GdLS), which constitutes the primary antineutrino target.
The outer acrylic vessel surrounding the target is filled with undoped LS, increasing the efficiency of detecting gamma rays produced in the target.
The outermost stainless steel tank is filled with mineral oil.
A total of 192 8-inch photomultiplier tubes (PMTs) are radially positioned in the mineral-oil region of each AD.
Specular reflectors are deployed directly above and below the outer acrylic vessel.
Three automated calibration units (ACUs) capable of deploying radioactive sources into the AD along three vertical z-axes are located on the top of each AD's outer tank~\cite{bib:acu}.
At each site, ADs are submerged in two-zone water Cherenkov muon detection systems, composed of inner and outer water shields (IWS and OWS), in three experimental halls, as shown in Fig.~\ref{fig:layout}.
A more detailed description of all detector systems can be found in ~\cite{bib:detector,bib:nim_rate}.

For the first seven months of Daya Bay data-taking from December 2011 until July 2012, six ADs were deployed and utilized for data analysis, two at the Daya Bay near site, one at the Ling Ao near site, and three at the Far Site.
For the additional 13 months of the data to be used in this publication, from October 2012 to November 2013, the full eight-AD detector deployment was utilized, with two ADs at each near site and four ADs at the Far Site.
During a special calibration period in Summer 2012, one ACU was temporarily removed to facilitate deployment of a Manual Calibration System, which was capable of deploying an articulating acrylic arm down the AD's center axis, allowing for full-volume calibration of the GdLS volume at a variety of vertical Z-positions and radial R-positions with a PuC neutron/gamma source.

A series of cuts are applied to the data to select high purity time-coincident trigger pairs in the AD that match the characteristics of IBD signals: a prompt energy deposition from ionization and annihilation of the IBD positron, followed by an energy deposition from Gd-capture of the IBD neutron 30~$\mu$s later on average.
The selection process and various cuts have been described in detail in a previous Daya Bay publication~\cite{bib:cpc_rate}, and have remained unchanged for this analysis.
We briefly list the sequence of IBD selection cuts below.

\begin{itemize}
\item{\textit{Flasher Cut:} Spurious single triggers caused by PMT light emission are efficiently removed using light collection topology cuts described in~\cite{bib:cpc_rate}.}
\item{\textit{Capture Time Cut:} Candidate trigger pairs are selected by requiring time-coincident triggers be separated by 1--200 $\mu$s.}
\item{\textit{Prompt Energy Cut:} The prompt trigger in the time-coincident pair must have an energy of 0.7--12~MeV.}
\item{\textit{Delayed Energy Cut:} The delayed trigger in the time-coincident pair must have an energy of 6--12~MeV.}
\item{\textit{Muon Veto Cut:} Candidate pairs are rejected if their delayed signals occur (i) within a (-2~$\mu$s, 600~$\mu$s) time window with respect to a water shield muon trigger with a PMT multiplicity $>$12 either in the inner or outer water shield, or (ii) within a (0, 1000~$\mu$s) time window with respect to triggers in the same AD with an energy ranging from 20~MeV to 2.5~GeV, or (iii) within a (0, 1~s) time window with respect to triggers in the same AD with an energy above 2.5~GeV.}
\item{\textit{Multiplicity Cut:} To remove ambiguities in the IBD pair selection when multiple triggers are in time-coincidence, candidate pairs are removed if there is an additional candidate with $E>0.7$~MeV in the interval 200~$\mu$s before the prompt-like signal, 200~$\mu$s after the delay-like signal, or between the prompt-like and delayed-like signals.}
\end{itemize}

Total IBD candidate event rates after applying these cuts are listed in Table~\ref{tab:ibd}.
Due to the near-identical response of the Daya Bay ADs, the efficiencies of most IBD selection cuts are the same for all detectors.
Muon veto efficiency ($\epsilon_{\mu}$) and multiplicity cut efficiencies ($\epsilon_m$) are dependent on muon fluxes and intrinsic background levels, which vary among different sites and ADs.

Backgrounds from accidental coincidences, fast neutrons, cosmogenic $^8$He/$^9$Li production, AD-intrinsic alpha radioactivity, and AmC neutron calibration sources remain in the sample of IBD candidates and have been estimated using a variety of techniques described in detail in previous publications~\cite{bib:cpc_rate,bib:prl_shape2}.
Background rate estimates remain unchanged for this analysis.

~\\

\end{multicols}
\begin{center}
  \tabcaption{Summary of signal and backgrounds. Rates are corrected for the muon veto and multiplicity cut efficiencies $\varepsilon_{\mu}\cdot\varepsilon_{m}$. Rate differences between detectors at the same site result from differences in fluxes between detector locations.
\label{tab:ibd}}
  \begin{minipage}[c]{\textwidth}
  \resizebox{\textwidth}{!}{
\begin{tabular}{c|cc|cc|cccc} \hline
\toprule
  & \multicolumn{2}{c|}{EH1}&\multicolumn{2}{c|}{EH2}&\multicolumn{4}{c}{EH3} \\
  & EH1-AD1  & EH1-AD2  & EH2-AD1 & EH1-AD2 & EH3-AD1 & EH3-AD2 & EH3-AD3 & EH3-AD4 \\
\hline
IBD candidates & 304459 &309354 & 287098 & 190046 & 40956 & 41203 & 40677 & 27419 \\
DAQ live time(days) & 565.436 & 565.436 & 568.03 & 378.407 & 562.451 & 562.451 & 562.451 & 372.685 \\
$\varepsilon_{\mu}$ & 0.8248 & 0.8218   &0.8575  &0.8577         &0.9811        &0.9811 &0.9808 &0.9811 \\
$\varepsilon_{m}$ & 0.9744      &0.9748 &0.9758 &0.9756  &0.9756         &0.9754        &0.9751 &0.9758 \\
Accidentals(per day) & $8.92\pm0.09$ & $8.94\pm0.09$ & $6.76\pm0.07$ & $6.86\pm0.07$ & $1.70\pm0.02$ & $1.59\pm0.02$ & $1.57\pm0.02$ & $1.26\pm0.01$ \\
Fast neutron(per AD per day) & \multicolumn{2}{c|}{$0.78\pm0.12$} & \multicolumn{2}{c|}{$0.54\pm0.19$} & \multicolumn{4}{c}{$0.05\pm0.01$} \\
$^9$Li/$^8$He(per AD per day) & \multicolumn{2}{c|}{$2.8\pm1.5$} & \multicolumn{2}{c|}{$1.7\pm0.9$} & \multicolumn{4}{c}{$0.27\pm0.14$} \\
Am-C correlated 6-AD(per day) & $0.27\pm0.12$ & $0.25\pm0.11$ & $0.27\pm0.12$ & & $0.22\pm0.10$ & $0.21\pm0.10$ & $0.21\pm0.09$  \\
Am-C correlated 8-AD(per day) & $0.20\pm0.09$ & $0.21\pm0.10$ & $0.18\pm0.08$ & $0.22\pm0.10$ & $0.06\pm0.03$ & $0.04\pm0.02$ & $0.04\pm0.02$ & $0.07\pm0.03$ \\
$^{13}$C($\alpha$, n)$^{16}$O(per day) & $0.08\pm0.04$ & $0.07\pm0.04$ & $0.05\pm0.03$ & $0.07\pm0.04$ & $0.05\pm0.03$ & $0.05\pm0.03$ & $0.05\pm0.03$ & $0.05\pm0.03$ \\ \hline
IBD rate(per day) & $657.18\pm1.94$ & $670.14\pm1.95$ & $594.78\pm1.46$ & $590.81\pm1.66$ & $73.90\pm0.41$ & $74.49\pm0.41$ & $73.58\pm0.40$    & $75.15\pm0.49$ \\ 
\bottomrule
\end{tabular}}
  \end{minipage}
\end{center}
\begin{multicols}{2}

\section{Event Selection Efficiencies}
\label{sec:Eff}

In order to estimate the total number of inverse beta decay interactions in each AD, the efficiencies of all signal selection cuts must be estimated.
All cut efficiencies have been estimated in previous Daya Bay publications~\cite{bib:cpc_rate, bib:prl_shape2}.
Many of these efficiencies remain unchanged in this analysis, and are only briefly described here.
A few key efficiencies common to all detectors have been recalculated with respect to those reported in~\cite{bib:cpc_rate} utilizing new comparisons between data and Monte Carlo (MC) simulation.
The improved data-constrained detection efficiencies and systematics will be described below in detail.
The recalculation and application of these key efficiencies and systematics result in a robust measurement of the overall reactor $\overline{\nu}_e$ flux from Daya Bay.
Since these key systematics for detector efficiencies are largely correlated among all Daya Bay detectors, this reanalysis does not affect the previous measurement of oscillation parameters reported by Daya Bay.

To produce improved efficiency determinations, a variety of new MC samples were generated utilizing an updated version of Daya Bay's simulation framework NuWa, which is based on the Geant4 simulation package~\cite{bib:Geant4} and the Gaudi framework~\cite{bib:Gaudi}.
A few key MC improvements with respect to the version utilized to produce previous efficiency estimates in~\cite{bib:cpc_rate} are briefly highlighted.
Models used to generate the spectrum of gammas released by neutron capture on Gd were altered based on new Daya Bay and bench-top datasets. These alterations, which affect the efficiency in detecting neutron captures on Gd, will be described in further detail below.
Adjustment was also made to the model describing the thermalization and scattering of neutrons at all energies.  This adjustment will also be described in further detail below, as it has a small impact on the capture time cut and on the position distribution of IBD events.

\subsection{Flasher Cut Efficiency}
Spontaneous light emission from the Daya Bay PMT bases can mimic particle interactions of various energies.
Flasher triggers can be rejected using charge topology cuts, as described in detail in~\cite{bib:cpc_rate}.
The IBD signal efficiency of these cuts is estimated to be 99.98\%.

\subsection{Capture Time Cut Efficiency}
To be selected as an IBD signal the time separation between the trigger pair must be within a (1$\mu$s, 200$\mu$s) range.
As described in~\cite{bib:cpc_rate}, the vast majority of signal events meet this criterion, with 98.70\% passing this cut in the most recent Daya Bay MC simulations.
An uncertainty of 0.12\% is assigned to this cut by noting small differences in trigger coincidence time distributions between AmC, AmBe, and PuC fast neutron source deployments and MC.

\subsection{Muon Veto Cut Efficiency}
Cuts are applied to reject coincident triggers correlated in time with muons traversing the water pools or ADs.
The characteristics and performance of these cuts are described in~\cite{bib:cpc_rate}.
Total signal efficiencies for these cuts depend on the muon flux at each site and are around 82\%, 86\%, and 98\% at EH1, EH2, and EH3, respectively, as shown in Table~\ref{tab:ibd}.
Muon veto cut efficiencies are calculated based on the actual number of muon vetos enforced in the dataset, and thus have negligible uncertainties.

\subsection{Multiplicity Cut Efficiency}
Some trigger coincidences containing more than two triggers are also rejected to avoid ambiguities in identifying the true IBD prompt-delayed pair.
The definitions of these multiplicity cuts are described in~\cite{bib:cpc_rate}, and have an efficiency of 97.5\% for all ADs, within 0.1\%.
Multiplicity cut efficiencies are calculated on-the-fly and have negligible associated uncertainty.

\subsection{Prompt Energy Cut Efficiency}
While the 0.7~MeV prompt energy cut is significantly below the 1~MeV annihilation gamma energy, a small proportion of events (0.2\%) deposit most of their energy in the non-scintillating inner acrylic vessel and fall below the threshold.
The efficiency is $99.81\%\pm 0.10\%$ determined by the most recent Daya Bay MC data which is consistent with that cited in~\cite{bib:cpc_rate}.

\subsection{Delayed Energy Cut: Gd Capture Fraction}

Inefficiency in detection of neutrons from IBD interactions in the target GdLS region is the result of three primary physical processes:
\begin{itemize}
\item{Capture on hydrogen in the target, producing a single 2.2 MeV gamma well below the applied 6~MeV threshold.}
\item{Capture on hydrogen outside the target where no Gd is present, producing the same 2.2 MeV gamma (spill-out effect).}
\item{Deposition of significant neutron-Gd (nGd) capture gamma energy outside the scintillating detector region, producing a detected delayed energy below the applied 6~MeV threshold.}
\end{itemize}
We choose to describe and quantify each of these contributions to the delayed energy cut efficiency separately in this analysis to produce robust and transparent efficiency and uncertainty estimates fully constrained by data.
We begin by describing our estimates of inefficiency from the first two of these processes, collectively described as the Gd capture fraction.

\subsubsection{Gd Concentration and AD-Center Gd Capture Fraction}
The keV-range kinetic energy neutrons created in IBD interactions in the GdLS thermalize in the detector and capture principally on either H or Gd nuclei.
Because of their low capture energy (2.2~MeV), neutron-hydrogen (nH) captures are completely excluded from the IBD signal by the 6~MeV cut used in this analysis.
Determining the Gd capture fraction is vital in determining the predicted reactor antineutrino flux.
This Gd capture fraction is physically determined by the Gd concentration in the GdLS, which is $\sim$0.1\% by weight.
The Gd capture fraction resulting from the Gd concentration can be measured largely independently of spill-out effects by looking at AD-center Gd capture events from various non-IBD (i.e., from calibration)datasets.

The AD-center Gd capture fraction was first measured utilizing muon spallation neutrons.
This dataset was obtained by selecting all AD non-flasher triggers within a time window of 20-300~$\mu$s after traversal of the AD by a muon, which is identified by an AD trigger with more than 3000 photoelectrons ($\sim$20~MeV).
Triggers from events other than neutron captures are then removed from the sample by subtracting a similar dataset occurring 520-800~$\mu$s after a muon traversal.
AD center events are then selected by removing all events having reconstructed positions $R>0.8$~m or $|Z|> 0.8$~m, where the position reconstruction follows the second method described in~\cite{bib:nim_rate}.
The background-subtracted spallation neutron capture spectrum for all four near ADs is shown in Fig.~\ref{fig:Spall_HGd}, along with the background spectrum.
\begin{center}
\includegraphics*[trim=0.25cm 0.0cm 0.5cm 0.75cm, clip=true, width=\columnwidth]{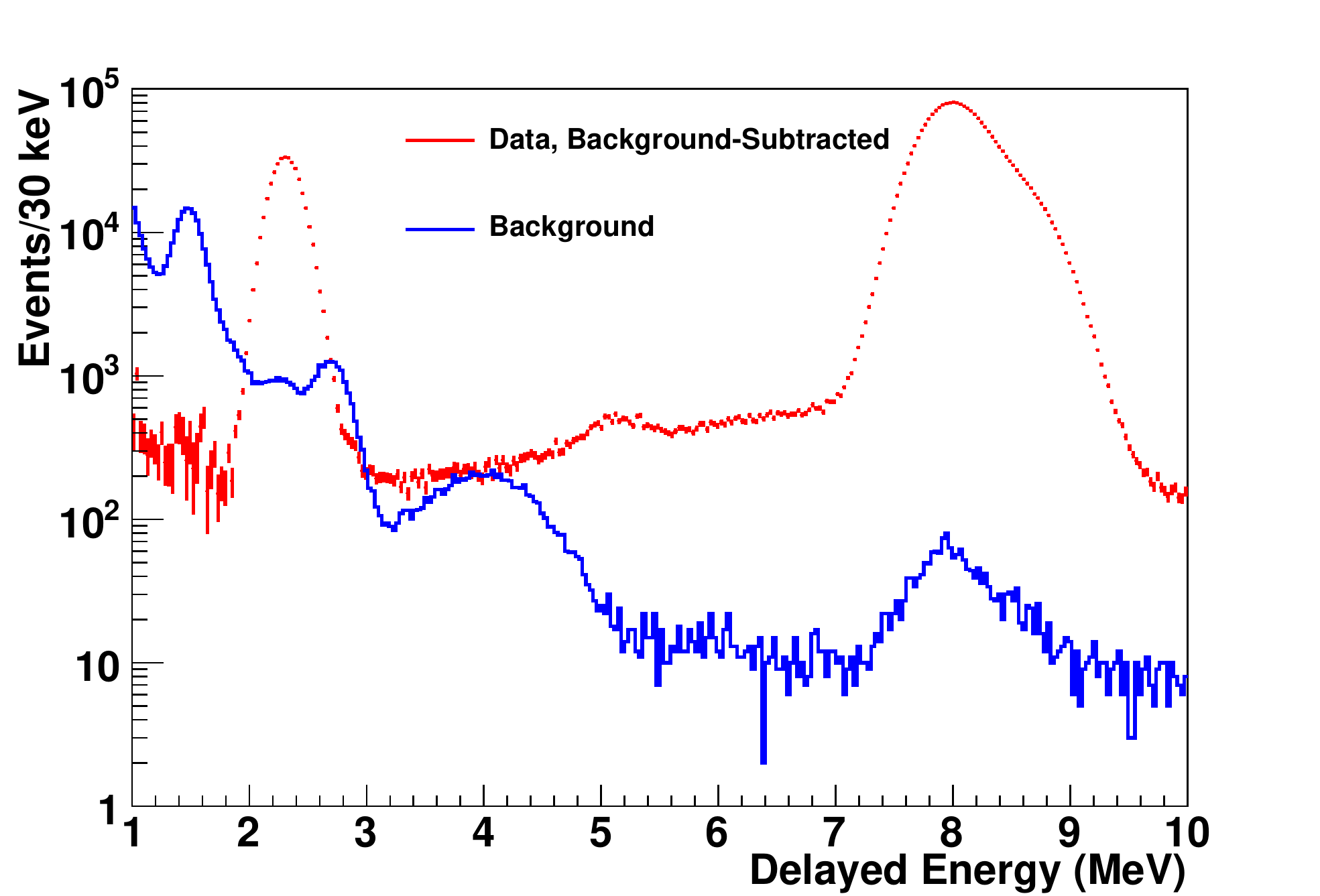}
\figcaption{The background-subtracted spallation neutron capture spectrum and associated background spectrum in all near-site ADs combined for the full Daya Bay dataset.
The main contribution to the background at low energy is natural radioactivity.}
\label{fig:Spall_HGd}
\end{center}

The Gd capture fraction for this dataset can be calculated using the following definition:
\begin{eqnarray}
  F_{Gd} & = & \frac{N_{Gd}}{N_{Gd}+N_H} \nonumber \\
 & = & \frac{N(6-12~\mathrm{MeV})}{N(6-12~\mathrm{MeV})+N(1.7-2.7~\mathrm{MeV})}.
\label{eq:GdCapFrac}
\end{eqnarray}
Low-energy cuts $\sim$30\% below the nH and nGd peak values were chosen to exclude roughly similar proportions of the nH and nGd low-energy tails from the calculation of this metric.
By this definition, the Gd capture fraction for a specific dataset is determined independently of any MC inputs.

For this dataset, we obtain a Gd capture fraction of 85.4\%, as seen in Table~\ref{tab:Neutron_Sources}, with a statistical uncertainty of $<$0.1\%.
We estimate the systematic uncertainty in this ratio by looking for the variation in $F_{Gd}$ with variation in the selection parameters.
To probe possible uncertainties arising from unequal inclusion of nGd and nH low-energy tails, $F_{Gd}$ low-energy cut values were independently adjusted to values between 10\% and 50\% below each peak's energy.
For all variations, $F_{Gd}$ was found to be consistent within 0.2\%.
When signal and background subtraction time windows are altered in absolute length (180 to 280 $\mu$s), relative length (few-$\mu$s difference in signal and background window length), or in start time (from 20$\mu$s to 40$\mu$s for signal, for example), $F_{Gd}$ is altered by $<0.1\%$.
As AD-center position cuts are varied from the nominal 0.8~m to either 0.5~m or 1.0~m, $F_{Gd}$ is altered by 0.3\%.
We also note that fractional contributions of target spallation neutron capture on other isotopes, such as carbon, are below 0.1\%, negligible in the scope of the efficiency analysis.
Adding the uncertainties quadratically, we obtain a Gd capture fraction of 85.4\% $\pm$ 0.4\% from spallation neutrons.

The Gd capture fraction has also been measured by deploying AmC, AmBe, and PuC neutron calibration sources at the centers of the two ADs at the Daya Bay Near Site (EH1) during a period of special calibration runs coincident with installation of the final two Daya Bay detectors at the other experimental halls~\cite{bib:acu, bib:mcs}.
These neutron sources produce time-correlated triggers, with proton recoils and excitation gammas forming the prompt signal, and the subsequent neutron capture forming the delayed signal.
The neutron kinetic energy ranges and excitation gamma energies for various prompt energy ranges for these sources are listed in Table~\ref{tab:Neutron_Sources}.
Some excited states with low neutron energies closer to that of $\sim$keV-scale IBD neutrons, such as the second excited state of $^{16}$O produced by the PuC ($\alpha$,n) reaction, are easily separable from other calibration source decays exhibiting higher neutron kinetic energies.
This is because minimally quenched de-excitation gammas from these excited states produce a much higher prompt energy than the highly-quenched prompt proton recoils generated by energetic neutrons produced in the ground state.
Meanwhile, other excited states produce either a variety of neutron kinetic energies (AmBe), or have prompt energies indistinguishable from the ground state.
Daya Bay's standard gamma-less AmC sources produce no transitions to excited states, since alphas in these sources are moderated with thin gold foils~\cite{AmC-paper}.
For all sources, removal of uncorrelated triggers was accomplished by subtracting a set of accidental coincidences formed by randomly ordering in time that calibration run's single triggers according to the calculated singles rate for that run.
As the sources were deployed at the detector center, cuts on reconstructed position were not utilized.  


\end{multicols}
\begin{center}
  \tabcaption{Characteristics and AD-center nGd capture fractions for neutron
calibration sources with varying prompt $E_{rec}$ categories. Prompt signals are
provided by muons (spallation neutrons), proton recoils (calibration source
decaying to ground states), excitation gammas (calibration source decaying
to excited states), or IBD positrons (IBD MC). Results for spallation
neutrons are the average of all ADs, while results for calibration sources
are the average of the two EH1 ADs, where all of the differing neutron
sources were deployed.  Measured Gd capture fractions are consistent within
the associated systematic uncertainty range of 0.4\%.}
    \begin{tabular}{l | l | l | l |l|l}
    \toprule
      Data Set & $E_{rec,prompt}$ (MeV) & $KE_{n}$ (MeV) & $E_{\gamma}$ (MeV) & $F_{Gd}$ (\%)& $\sigma_{stat}$ (\%)\\\hline
      Spallation Neutron & - & 0-100+ & - & 85.4 & $<$0.1\\ \hline
      AmC & 0-4 & 3-5.5 & - & 85.2 & 0.2 \\ \hline
      AmBe, Ground State & 0-4 & 4-10 & - & 85.3 & 0.1 \\
      AmBe, First Excited & 4-7 & 0-5 & 4.4 & 85.4 & 0.1 \\ \hline
      PuC, Ground State & 0-4 & 3-7.5 & - & \multirow{2}{*}{85.5} & \multirow{2}{*}{$<$0.1}\\
      PuC, 1$^{st}$ Excited & 0.5-1 & $<$0.6 & - & & \\
      PuC, 2$^{nd}$ Excited & 5.5-7 & $<$0.6 & 6.13 & 85.5 & $<$0.1 \\ \hline
      IBD MC & 0.7-12 & $<$0.1 & - & 85.5 & $<$0.1 \\
      \bottomrule
    \end{tabular}
  \label{tab:Neutron_Sources}
\end{center}
\begin{multicols}{2}

For the calibration source data, an alternate procedure utilizing MC input for determining the Gd capture fraction was used to cross-check the spallation results utilizing the $F_{Gd}$ metric.
First, the total fraction of background-subtracted time-coincident triggers passing the 6~MeV delayed energy cut was determined:
\begin{eqnarray}
  F_{Gd,all} & = & \frac{N_{Gd}}{N_{All}} = \frac{N(6-12~\mathrm{MeV})}{N(1.7-12.0~\mathrm{{MeV}})}.
\label{eq:GdCapTot}
\end{eqnarray}
Delayed energies below 1.7~MeV are excluded from the dataset as the statistical uncertainties from the accidental background subtraction for some datasets in this region are too high.
Next, MC simulations including sources and the radioactive sources, source enclosures, deployment weights and suspension lines analogous to the actual AD-center source deployments were used to calculate the total number of coincidences in the $<$1.7~MeV delayed energy region ($\sim$0.3\%), as well as the number of nGd captures below 6~MeV reconstructed energy ($\sim$1.5\%).
These numbers were used to correct $F_{Gd,all}$ to provide a semi-independent measure of the total ratio of nGd to other capture types, similar to $F_{Gd}$.
This method of estimating the Gd capture fraction avoids the uncertainty from defining nH and nGd energy windows, but has added uncertainty because of 0.1\%-level disagreements in low-energy contributions between the source deployment in data and MC.

Resultant $F_{Gd}$ values from extended runs of these three sources are shown in Table~\ref{tab:Neutron_Sources}, with delayed spectra from each source shown in Figure~\ref{fig:Source_HGd}.
While neutron capture tail shapes from the different sources deviate slightly from one another, likely due to differing source packaging material and optical properties, values of $F_{Gd}$ from all sources agree to within 0.3\%.
These source $F_{Gd}$ values are also consistent within 0.4\% between data and MC for all source types and neutron energy ranges.
Similar variations of energy and timing cuts applied to the spallation neutron dataset above produce $<$0.1\% changes in $F_{Gd}$ values.
These differences provide a conservative estimate of systematic uncertainty on the Gd capture fraction similar to that reported from spallation neutrons.

\begin{center}
\includegraphics*[trim=0.25cm 0.0cm 0.5cm 0.75cm, clip=true, width=\columnwidth]{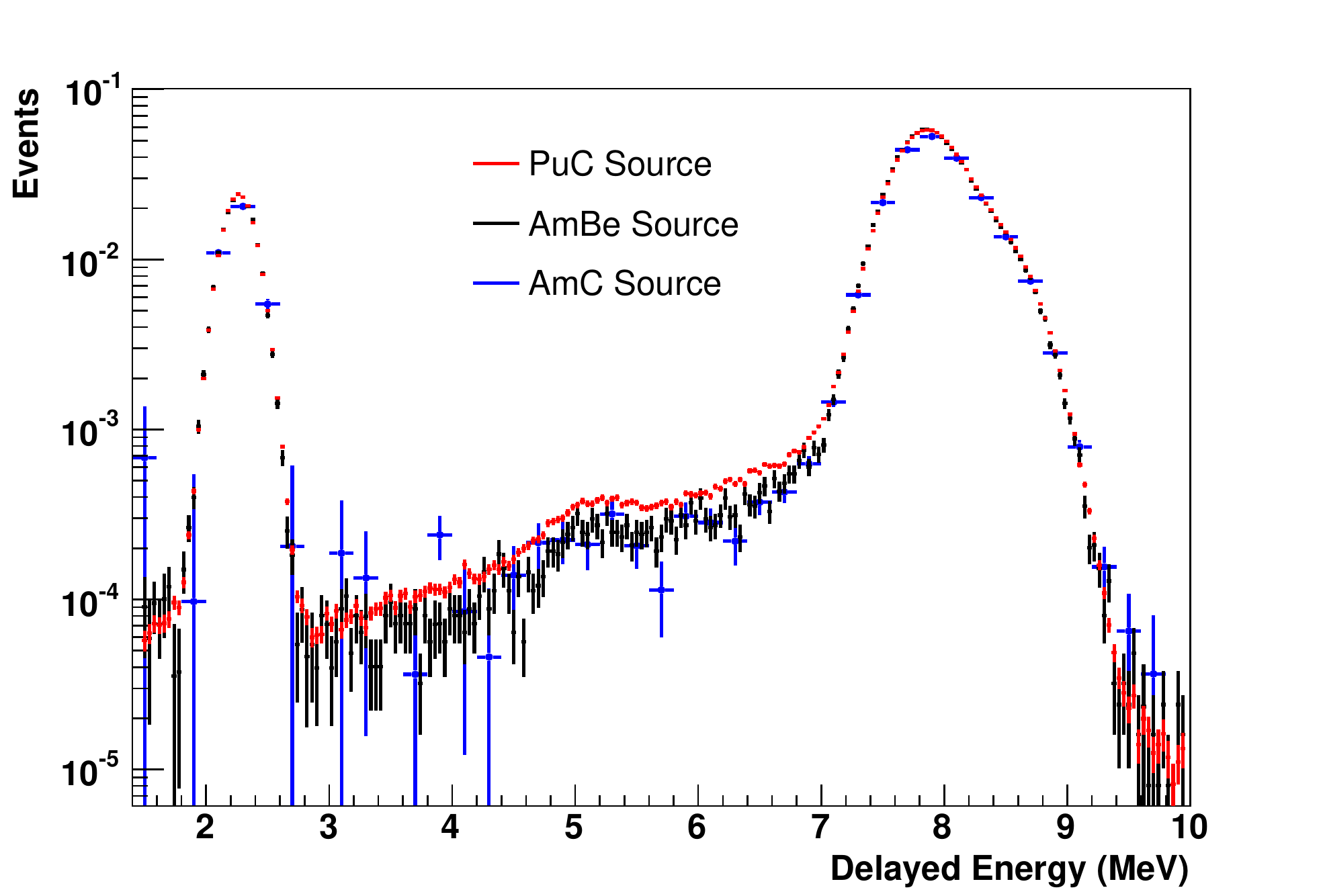}
\figcaption{Background-subtracted calibration neutron capture spectra from three different neutron sources deployed in the EH1 detector centers.  The AmC data is binned more coarsely to reduce error bar sizes in the low-statistics tails.}
\label{fig:Source_HGd}
\end{center}

After the completion of these studies, the Gd capture fraction for AD-center inverse beta decay interactions was studied in the MC simulation and were determined to be 85.5\%.
This agreement with a wide variety of studies indicates the initial values of Gd concentrations of the Daya Bay scintillator were properly measured and implemented in simulation.

\subsubsection{Spill-out Effects and Full-Volume Gd Capture Fraction}

The previous section concerned itself with finding the Gd capture fraction at the detector center and matching this value between data and MC.
In order to determine the Gd capture fraction for the entire target volume, which is the relevant number for the total detection efficiency, one must take into account the proportion of IBD neutrons created in the GdLS that escape the target and capture outside the GdLS, where all captures are non-Gd.
This process, termed as the ``spill-out'' effect, is naturally dependent on the proximity of the IBD interaction point to the boundary of the GdLS volume.

The MC is used to provide the full-volume Gd-capture fraction for the detection efficiency analysis.
This value is calculated with MC to be 84.17\%.
The accuracy and systematic uncertainty of this total Gd capture fraction were then estimated by comparing total H/Gd ratios for existing non-IBD datasets between data and MC.
We note that since the sizes of the nH and nGd low-energy tails in the neutron energy spectra increase with increasing R and $|$Z$|$ due to increased gamma energy leakage, it is difficult to fully disentangle Gd detection inefficiencies from spill-out effects.
As the position dependences of the nH and nGd tails are correlated, the previously-defined metric $F_{Gd}$ in Eq.~\ref{eq:GdCapFrac} is relatively insensitive to these gamma energy leakage effects.
For this reason, we utilize the $F_{Gd}$ metric described above for a data-MC comparison of full-volume Gd capture fractions.
An comparison of PuC calibration source data with MC data using an alternate metric is described in Sec.~\ref{subsec:combined} as a cross-check.

The Daya Bay MCS~\cite{bib:mcs} deployed a PuC neutron source on an articulating arm at a wide variety of positions throughout the GdLS volume with an accuracy of 2.5~cm in r and 1.2~cm in z; this dataset can be used to calculate a full-volume Gd capture fraction.
Due to the geometry of the source and articulating arm, deployments were limited to source positions with -1.45~m $< Z < $ 1.25~m and $R < $ 1.35~m.
PuC deployments at similar positions were then simulated, including the attendant MCS articulating arm infrastructure.
For each source placement position, $F_{Gd}$ was calculated in data and MC utilizing a process identical to that described in the previous section, except that backgrounds were subtracted utilizing an off-window method as was done in the previously described spallation neutron study.
This was necessary to remove coincidences formed by closely-spaced neutrons from the intense ($\sim$~1kHz) PuC source.

Figure~\ref{fig:SO_RZ} demonstrates the change in PuC neutron $F_{Gd}$ separately as a function of R and Z for data and MC.
One can see good agreement at most positions.
By fitting the distributions in R along the detector's Z-center and in Z along the detector axis, one can integrate over the full target volume to obtain a full-volume Gd capture fraction.
A variety of fit methods are utilized to account for the lack of data near the GdLS top and bottom ($|$Z$|$$\sim$1.5~m).
This process yields a full-volume Gd capture fraction of 84.1\% for data and 83.5\% for MC.
The Gd capture fraction for the PuC source in the MCS differs from that of IBDs due to  the higher kinetic energy of neutrons and the MCS deployment arm; hence, we use the MCS data to benchmark
the full-volume Gd captures fractions between data and MC.
Therefore, these results should be utilized not as an indicator of the true IBD Gd capture fraction, but as a benchmark of the agreement between full-volume Gd capture fractions between MC and data.
An additional check on this analysis using interpolated values between all available PuC MCS deployment positions yields differences of up to 0.7\% between data and MC for all PuC neutron kinetic energies.

After performing these benchmark comparisons between MC and data, the precision of the MC-reported full-volume Gd capture fraction is estimated as the maximum difference between these reported MC and data values above, 0.7\%.
Adding quadratically the approximate 0.4\% uncertainty in the AD-center Gd capture fraction, we obtain a predicted Gd capture fraction of 84.17 $\pm$ 0.80\%.
The difference from early Daya Bay publications (83.8 $\pm$ 0.8\%) is caused by the improved Geant4 neutron thermalization models on which this analysis is based, which produce a lower rate of IBD neutron spill-out.

\begin{center}
\includegraphics*[trim=0.0cm 0.0cm 0.75cm 0cm, clip=true, width=0.9\columnwidth]{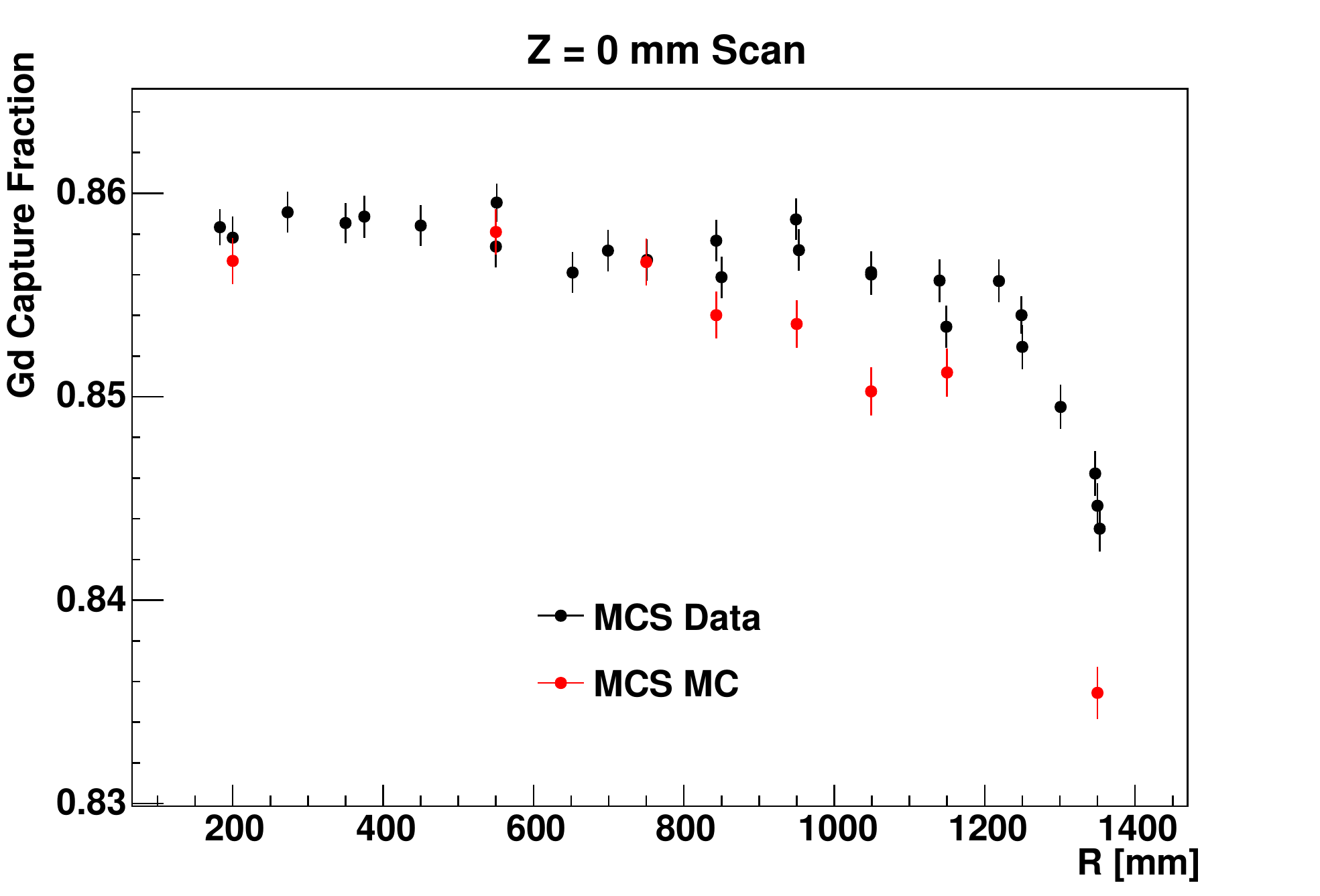}
\includegraphics*[trim=0.0cm 0.0cm 0.75cm 0cm, clip=true, width=0.9\columnwidth]{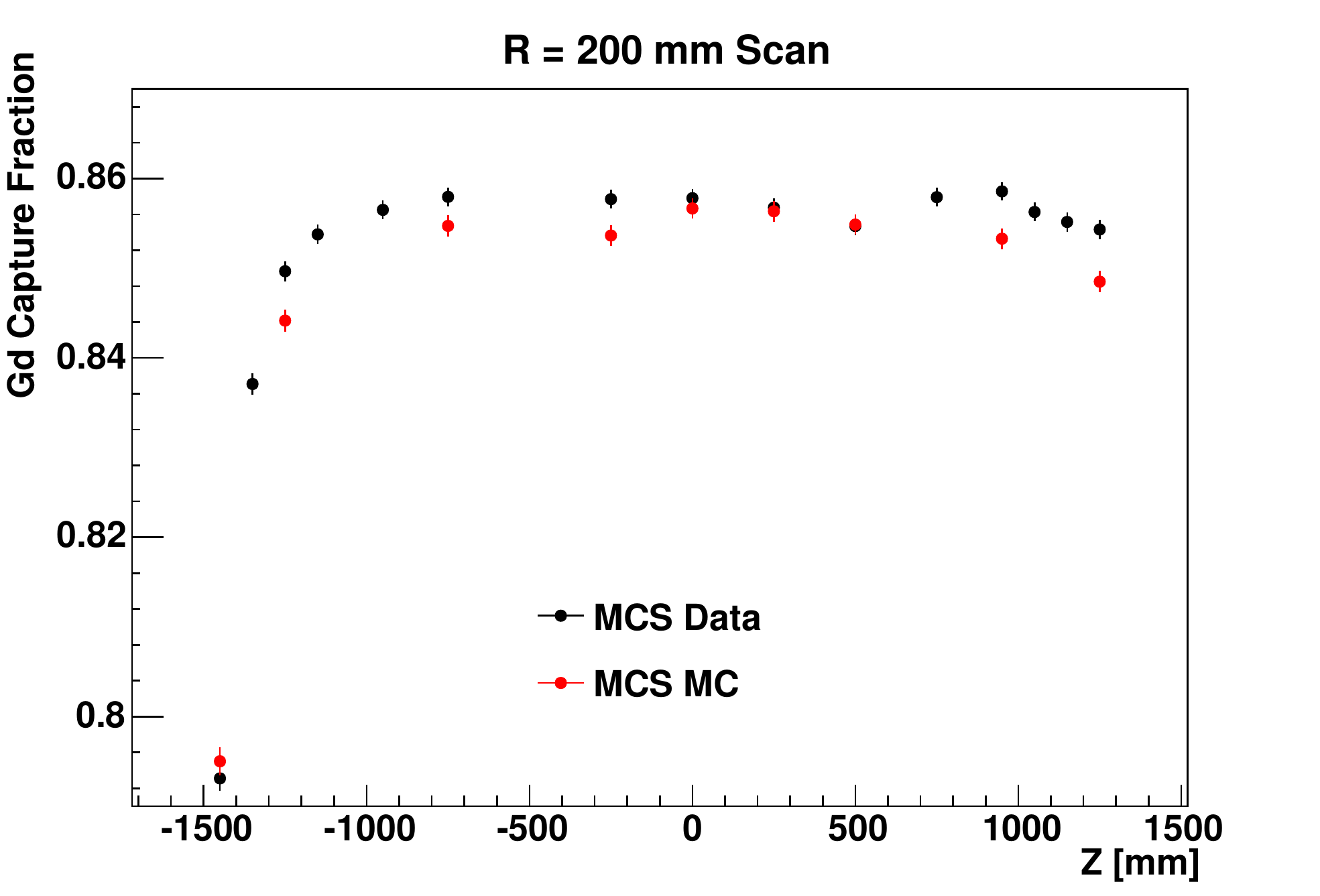}
\figcaption{Variation in the Gd capture fraction $F_{Gd}$ reported by a MCS-deployed PuC source as a function of R (top) and Z (bottom) position in the detector.  For the Z scan, MCS runs at R=200~mm were utilized.  For the R scan, MCS runs at Z=0~cm (AD middle) were utilized.  A drop in the Gd capture fraction is clearly visible near the target boundary.  Despite small visible differences, data and MC yield similar integrated full-volume Gd capture fractions within 0.6\%.}
\label{fig:SO_RZ}
\end{center}

\subsection{Delayed Energy Cut: Gd Capture Detection Efficiency}

Of the 84.17\% of target IBD neutrons capturing on Gd, a small percentage will have delayed reconstructed energy below the 6~MeV delayed energy cut.
This inefficiency arises as a portion of gammas from some Gd captures exit the scintillating region of the detector before depositing their energy.  In order to properly estimate the predicted reactor antineutrino flux, this Gd capture detection efficiency must be properly estimated.
As with the full-volume Gd capture fraction, the Gd capture detection efficiency is determined using MC, since the full tail of the IBD delayed energy signal is obscured in data by nH captures and accidental backgrounds.

The shape of the Gd capture tail, and therefore the Gd capture detection efficiency, is  dependent on the model used to describe the gamma energies released by a nGd capture.
The excited states of $^{158}$Gd and $^{156}$Gd, the products of neutron capture on $^{157}$Gd and $^{155}$Gd, are numerous, making a first-principles determination and modelling of de-excitation pathways impractical.
Instead, the Daya Bay MC produces nGd capture gammas by performing an energy-conserving sampling of previously-measured Gd-capture gamma spectra.
The algorithm that performs this sampling is tuned to ensure that the energy conservation requirement does not bias aggregate sampled gamma spectra relative to the input spectrum.
In previous publications~\cite{bib:prl_rate, bib:cpc_rate}, Daya Bay utilized nGd gamma spectrum models based on early spectroscopic measurements~\cite{bib:gd_gammas}, shown in top panel of Fig.~\ref{fig:GammaModels}, which do not sufficiently reproduce the IBD extended nGd tail shapes now visible in Daya Bay's high-statistics datasets, pictured in middle panel of Fig.~\ref{fig:GammaModels}.  This gamma model is referred to in this paper as the ``M13A,Old'' model

We investigated additional nGd gamma models to obtain a better description of the data.
It was found that nGd gamma spectra included in Geant4 libraries~\cite{bib:Geant4}, shown in bottom panel of Fig.~\ref{fig:GammaModels}, produced reasonable agreement with observed data once energy conservation in gamma emission, not present in Geant4 by default, was implemented.
This model is referred to as the ``M14A,Geant'' model in this paper.
Another well-matching model, called ``M14A,Caltech'', was generated through direct measurement of nGd gamma production in a small cell of Daya Bay GdLS using a benchtop HPGe detector setup at Caltech.
In both new ``M14A,Geant'' and ``M14A,Caltech'' models, the total contribution of high-energy gammas is lower than in early spectroscopic measurements.

Figure~\ref{fig:GammaModels} shows the combined IBD nGd capture spectra from all Daya Bay detectors and from the various tested MC models.
The nGd tail is clearly visible with high statistical precision above 3.0~MeV, and provides a direct constraint on the delayed energy cut inefficiency above this energy.
The two MC models provide a bounding envelope around the observed spectrum when approaching the low-energy region where the nH peak obscures the true nGd tail shape.
The delayed energy cut inefficiency from this low energy region is estimated by the relative contribution from these new MC nGd capture models.
The data-constrained portion of the tail from 3--6~MeV provides a 6.6\% inefficiency, while the low-energy MC-constrained portion below 3.0~MeV contributes 0.4\% and 0.9\% for the different models.

\begin{center}
\includegraphics*[trim=0.0cm 0.0cm 1.4cm 0.0cm, clip=true, width=0.39\textwidth]{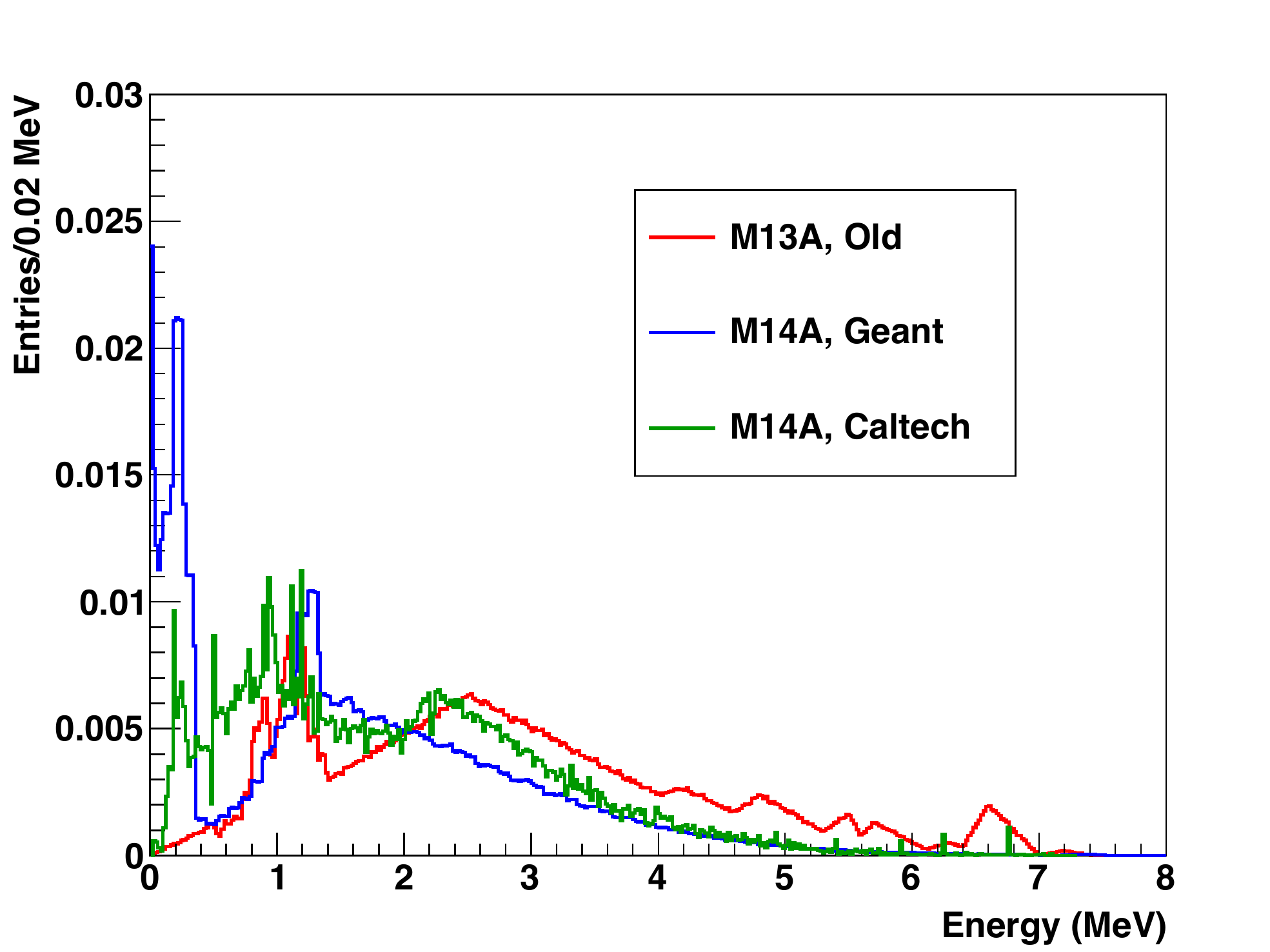}
\includegraphics*[trim=0.0cm 1.5cm 1.8cm 1.1cm, clip=true, width=0.39\textwidth]{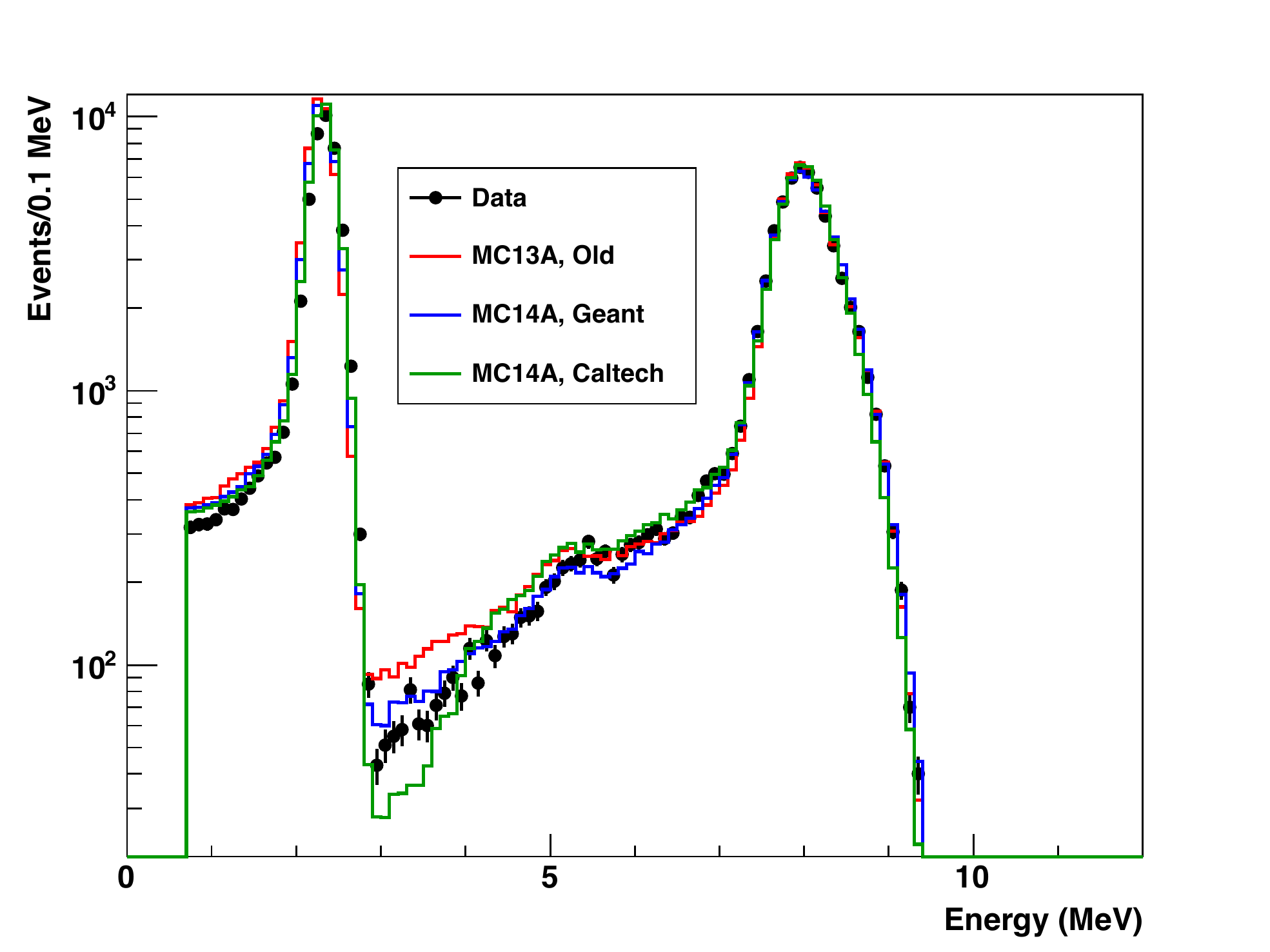}
\includegraphics*[trim=0.0cm 0.0cm 1.8cm 1.6cm, clip=true,width=0.39\textwidth]{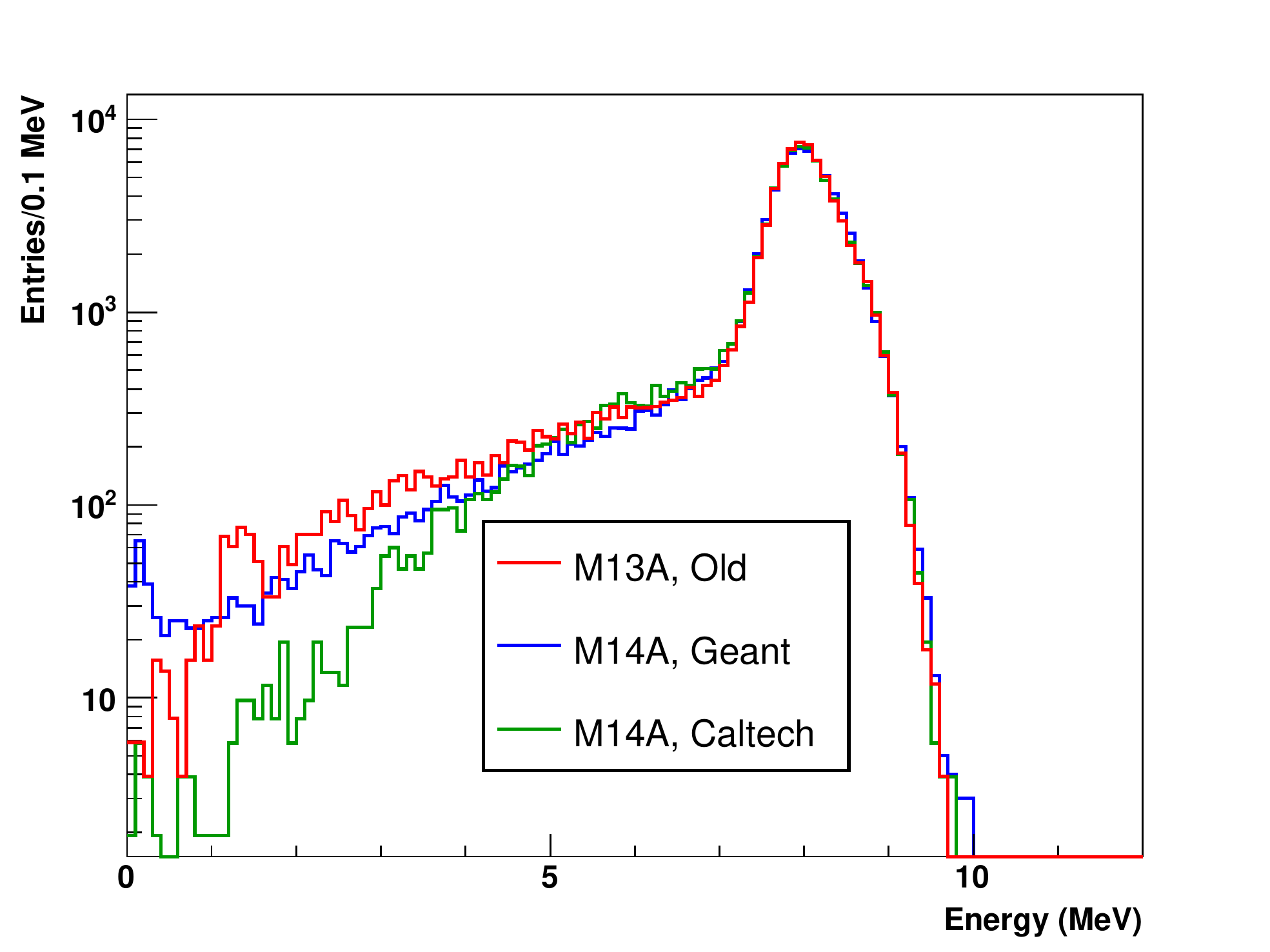}
\figcaption{Top: Models of produced nGd capture gamma spectra utilized in the previous (red) or current (blue and green) efficiency estimates.
Middle: $E_{rec}$ distribution of the IBD delayed signal for data and MC utilizing these nGd gamma models.
Bottom: The spectrum of IBD nGd delayed signals from MC for these models.  These distributions are used to compute the expected nGd capture detection efficiency for Daya Bay.}
\label{fig:GammaModels}
\end{center}

The total estimated nGd detection efficiency using the ``M14A,Geant'' model is 92.71\%.
A conservative 100\% uncertainty is assigned to the total contribution below 3~MeV due to the lack of direct data constraints.
The 0.5\% difference in the low-energy contribution from the data-enveloping MC models provides good motivation for this choice.
Further uncertainty contributions from statistical and other systematics, such as the MC-data difference in energy scale near the GdLS-LS boundaries, are negligible in comparison.


This Gd capture detection efficiency estimate, 92.71\%, differs from previous estimates, 90.9\%, in~\cite{bib:prl_rate} by 1.8\%, a $\sim$3~$\sigma$ change with respect to previous systematic uncertainty estimates.
As previously mentioned, this difference stems from improved modelling of the nGd gamma spectrum in the updated Daya Bay MC simulations.
Due to the limited available statistics, previous uncertainty estimates were made using comparisons between the previous MC model and data only in a narrow higher-energy window (6-7~MeV) bordering the nGd tail region.
In contrast, the updated efficiency estimate is directly constrained by data with $<$0.1\% statistical uncertainty for the bulk of the nGd tail, with 100\% uncertainty assumed in regions where no direct data constraint exists.
This results in a robust and conservative estimate of the delayed energy cut efficiency.
Nevertheless, the change of the Gd capture detection efficiency does not affect the measurement of the oscillation parameters reported by Daya Bay using relative comparison between Near and  Far detectors.

\subsection{Combined Delayed Energy Cut Efficiency Cross-Check}
\label{subsec:combined}

In addition to estimating them separately as we have done above, we can cross-check the accuracy of the MC in modelling combined effects of the Gd capture fraction and nGd detection efficiency by comparing the previously-defined $F_{Gd,all}$ metric between data and MC for a representative non-IBD dataset.
This $F_{Gd,all}$ metric effectively achieves both of these efficiencies above the applied 1.7~MeV analysis threshold.
The MCS data set was re-analyzed to determine $F_{Gd,all}$ for points spaced evenly throughout the target volume.
The position-weighted average value for all data points using the full-volume fits described above were 80.3\%, compared to a MC value of 79.5\%.
This difference is well within the uncertainties of 0.8\% and 0.9\% defined for the full-volume Gd capture fraction and nGd detection efficiency,  providing further confidence that MC modelling of these two sub-efficiencies is accurate within the estimated systematics.

%

\subsection{Spill-in Effects}
When calculating the total number of expected Gd capture detections, one must take into account IBD neutrons generated outside the GdLS that are captured in the GdLS.
This process, termed as the spill-in effect, effectively increases the size of the target volume.
As with the spill-out effect, the size of the spill-in effect and the net increase in effective target volume is calculated using MC simulation of IBD neutrons in the detector.
The calculated value of the effective target size in the default Daya Bay MC due to spill-in is 104.9\%.

The spill-in correction obtained by the MC is dependent on the choice of neutron scattering models.
Daya Bay's default MC for neutron scattering includes inelastic scattering of thermal neutrons below 4~eV where molecular effects due to hydrogen bonds and their energy transfer with neutrons must be considered.
This is in contrast to ``free-gas'' models of neutron scattering which forego this detailed modelling at low energies.
In its G4NDL3.13 physics library, Geant4 has inherited several neutron thermal scattering models and parameters from the Evaluated Nuclear Data Files (ENDF/B-VI) database~\cite{bib:endf} for a variety of moderators such as water and polyethylene.
As database entries are unavailable for the primary Daya Bay target materials GdLS, acrylic, and mineral oil, ENDF models for water and polyethylene were used to describe each of these target materials in the MC.
Variations between these different models individually for each Daya Bay target material produced $<$1.0\% changes in MC-reported total IBD rates.
A free-gas neutron scattering model produced effective target masses roughly 2\% larger than the default MC.
Due to a variety of mismatches between data and the free-gas model MC to be described below, the free-gas model was ruled out as a viable description of the physics in the detector and not considered when calculating systematics envelopes.


A wide variety of data-MC comparisons of calibration and IBD data have been implemented to determine the uncertainty in this MC-produced spill-in estimate.
Spill-in estimates from the default MC can be directly benchmarked to data by comparing extended deployments of a combined AmC/Ge source at a single position at the detector Z-center in the LS volume 22~cm radially outward from the GdLS edge.
Given the low spill-in rate from this position, a main background in this calibration dataset is IBDs from reactor antineutrinos, which were reduced by choosing a low prompt energy window (0.9-1.3~MeV) and applying a cut on the reconstructed distance from the source ($<$1~m), with the remaining IBDs statistically subtracted utilizing time-adjacent non-calibration runs.
The ratio of nGd to nH captures in this calibration dataset was found to be 4.1\%, in agreement with the default MC within 1.0\%, even after including wide variations in background subtraction methods, energy cut windows, and distance cuts.

While this calibration-based result appears quite robust, the difference in kinetic energy between AmC source neutrons and keV-scale IBD neutrons necessitates further studies to reliably estimate spill-in effects for IBD interactions.
In the absence of a low-energy neutron source, indirect determination of MC spill-in accuracy was also accomplished by comparing spill-in-correlated IBD time and position observables between data and MC.

The comparison between data and MC reconstructed prompt position distributions is shown in Fig.~\ref{fig:SpillIn_Pos}.
Care has been taken in this comparison to correct for cm-level relative differences in reconstructed IBD positions between MC and data for the MCS measurement, which can also cause event rate differences at large $R_{p,rec}$.
With these biases corrected, the magnitude of the MC-data differences near the detector boundary ( $R^2_{p,rec} > 1.5$~m$^2$) sum to $\sim$0.5\% of the total number of IBD events shown in bottom panel of Fig.~\ref{fig:SpillIn_Pos}.
With variation of neutron scattering models, the sums of MC-data differences for all cases are $<0.6\%$,  which translates to a 1.0\% difference of spill-in.  

\begin{center}
\includegraphics*[trim=0.45cm 0.1cm 1.5cm 0.1cm, clip=true, width=0.45\textwidth]{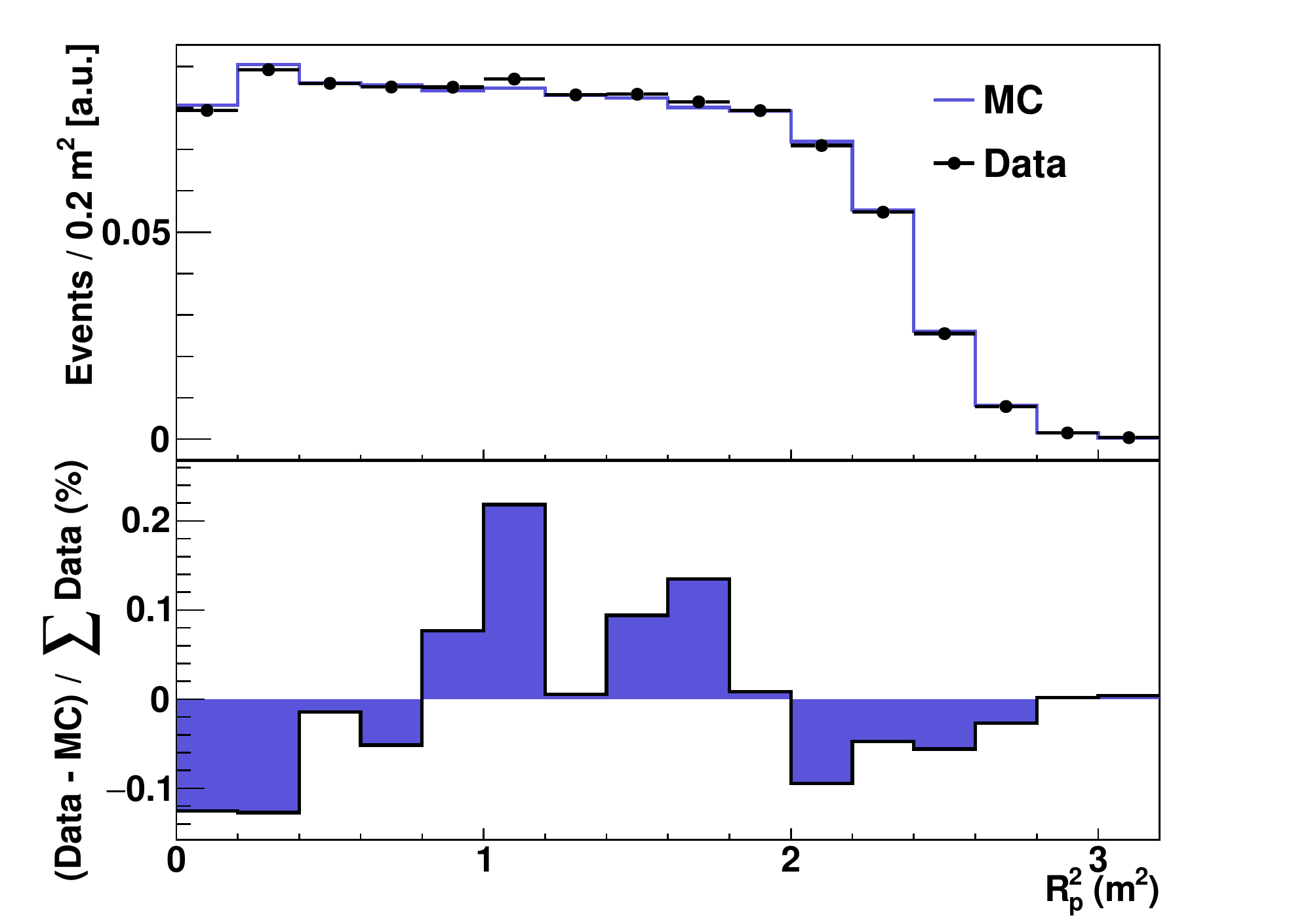}
\figcaption{Prompt event reconstructed $R_p^2$ positions for the full sample of data and MC IBD events with a requirement of $>$ 3.5~MeV on the prompt event energy to suppress the accidental background in data. }
\label{fig:SpillIn_Pos}
\end{center}

The IBD coincidence time distributions have been compared between data and MC.
Since spill-in events originate in the LS, they tend to have longer coincidence times and contribute heavily to the tail of the IBD coincidence time distribution.
This relation was determined with a MC IBD event dataset by calculating both the spill-in fraction and the fraction of signal events with greater than 50~$\mu$s coincidence time for subsets of events in common reconstructed position bins, and the result is shown in top of Fig.~\ref{fig:SpillIn_Time}.
The coincidence time peak-to-tail ratios in each $R_p$ bin were computed for both data and MC with a requirement of $>$ 3.5~MeV on the prompt event energy, shown in bottom of Fig.~\ref{fig:SpillIn_Time}.
The difference between data and MC at the boundary of GdLS region reflects the spill-in difference.
According to the relation between capture time distortion (peak-to-tail ratio) and the fraction of spill-in event, the spill-in fraction is evaluated for the data.
The relative contribution of spill-in events was found to agree between data and MC to within 0.5\% of the total event sample for a wide variety of systematic variations including coincidence time tail definitions, and assumed position reconstruction biases.

\begin{center}
\includegraphics*[trim=0.1cm 0.1cm 1.5cm 0.1cm, clip=true, width=0.45\textwidth]{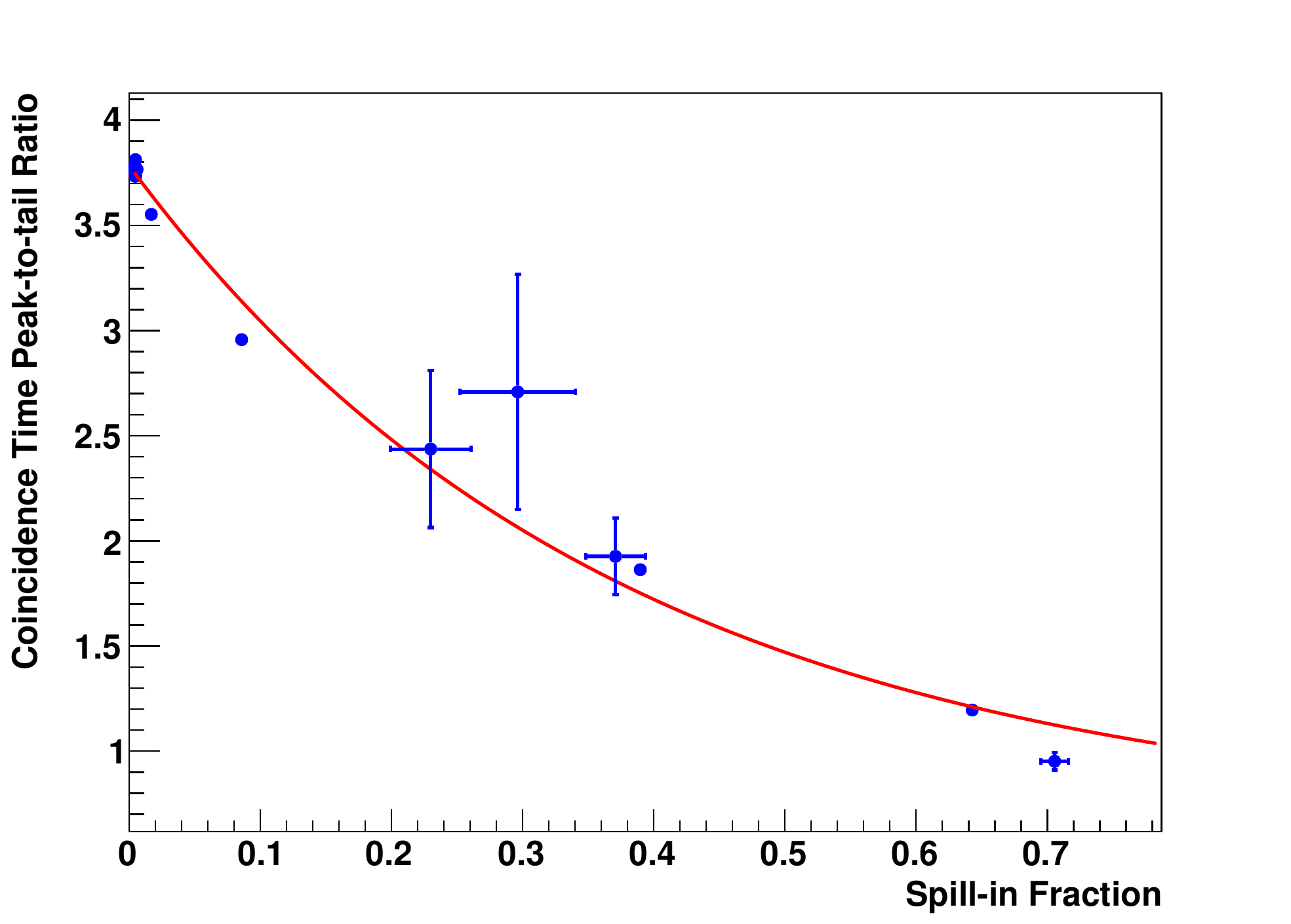}
\includegraphics*[trim=0.1cm 0.1cm 1.5cm 0.1cm, clip=true, width=0.45\textwidth]{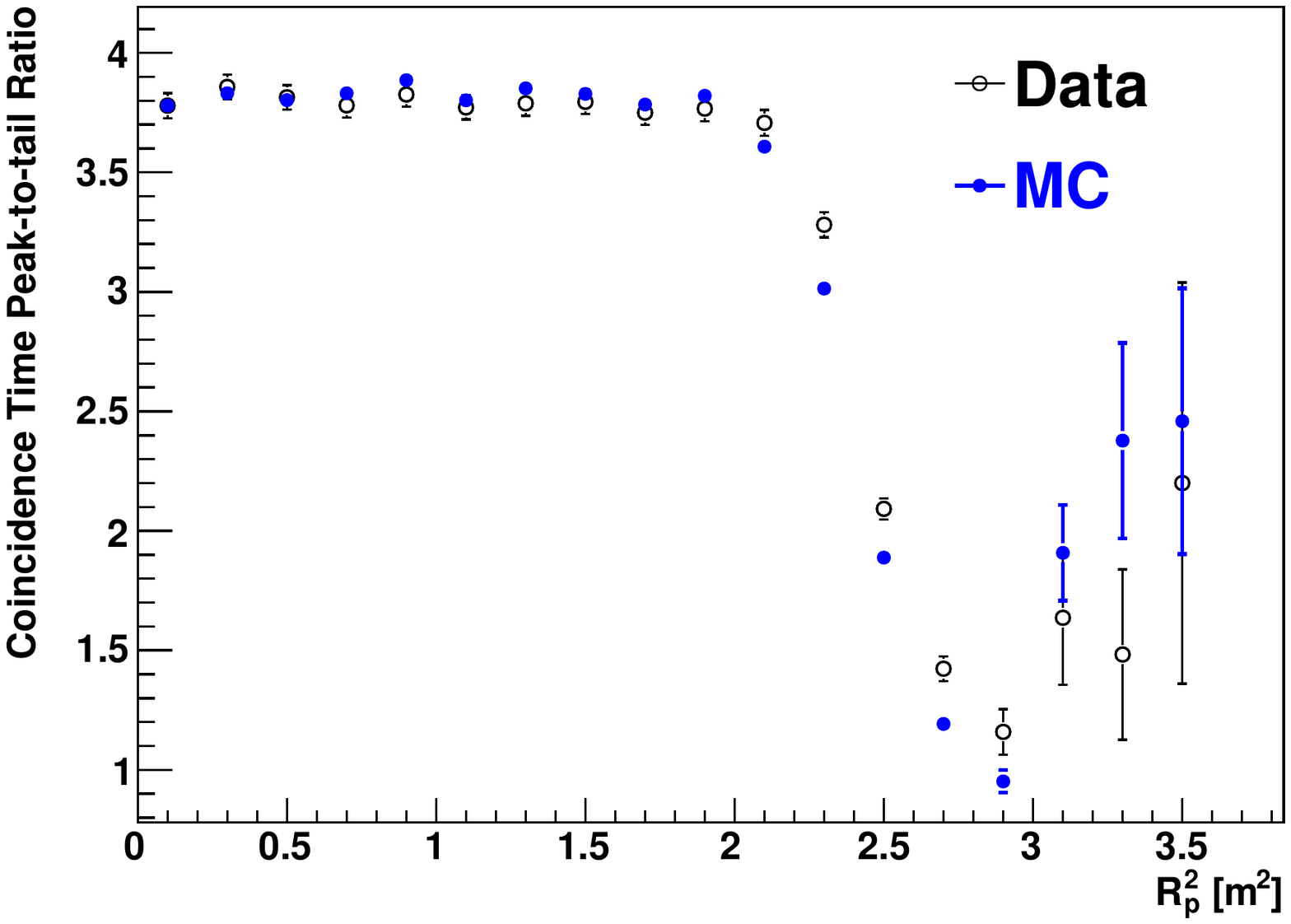}
\figcaption{
Top: Points show the relation between true spill-in percentage and coincidence time peak-to-tail ratio in Monte Carlo simulation for event groupings at common $R_{p,rec}$. The peak-to-tail ratio compares IBD events in the (1,50) and (50,200)~$\mu$s capture time regions. A curve is fitted to infer the spill-in fraction for the IBD candidate dataset.
Bottom: The coincidence time peak-to-tail ratio in different bins along the radius.
The spill-in percentage in each bin of data is predicted with the relation of spill-in and peak-to-tail ratio obtain from MC.}
\label{fig:SpillIn_Time}
\end{center}

To provide a conservative estimate of the inaccuracy of the IBD spill-in percentage reported by MC, the maximum MC-data difference observed in any of these studies, 1.0\%, is used as the uncertainty in the spill-in contribution to the efficiency estimate.


\subsection{Target protons}
The uncertainty in the number of target protons is included with the detection efficiency uncertainties as the IBD rate is proportional to the number of target protons.
The number of target protons in the GdLS is calculated as
\begin{equation}
    N_p = M\cdot F_H\cdot N_A\cdot I_{^{1}H}/m_H,
\end{equation}
where $M$ is the mass of GdLS in the target, $F_H$ is the mass fraction of hydrogen in GdLS, $N_A$ is the Avogadro constant, $I_{^1H}$ is the isotope abundance of $^1$H in natrual hydrogen, and $m_H$ is the atomic mass of hydrogen.

The target mass was precisely measured during the detector filling and monitored during the data taking.
The uncertainty of the target mass is 3.0~kg~\cite{bib:detector}, corresponding to 0.015\% of the 20~ton target mass.
The hydrogen mass fraction of $F_H=12.02\pm0.11\%$ was obtained from the combination of two sets of independent combustion measurements, one of which is tabulated in Ref.~\cite{bib:detector}.
The combined fractional uncertainty in $N_p$ is 0.92\%.
The previous reported uncertainty in $N_p$ of 0.47\%~\cite{prl_abs} was incorrect.

\subsection{Efficiency Summary}

Calculated detection efficiencies and their related uncertainties are listed in Table~\ref{tab:AllEffs}.
The detection efficiency common to all detectors is $\epsilon = $80.6\%.
Including the efficiencies that vary among detectors, as given in Table~\ref{tab:ibd}, total detection efficiencies range from 64.6\% to 77.2\%.
The total systematic uncertainty of detection efficiencies is $\delta\epsilon/\epsilon = $1.93\%.

\begin{center}
  \tabcaption{Summary of the detection efficiencies and systematic uncertainties.
Muon veto and multiplicity cut efficiencies vary between sites and have negligible uncertainty.}
    \begin{tabular}{l | c | c } \hline
      Source & $\epsilon$ & $\delta\epsilon/\epsilon$  \\\hline
      Target protons & - & 0.92\%  \\
      Flasher cut & 99.98\% & 0.01\%  \\
      Capture time cut & 98.70\% & 0.12\%  \\
      Prompt energy cut & 99.81\% & 0.10\%  \\
      Gd capture fraction & 84.17\% & 0.95\%  \\
      nGd detection efficiency & 92.71\% & 0.97\%  \\
      Spill-in correction & 104.86\% & 1.00\%  \\ \hline
      Combined & 80.60\% & 1.93\% \\\hline
    \end{tabular}
  \label{tab:AllEffs}
\end{center}

\section{Measurement of Reactor Antineutrino Flux}\label{sec:abs_flux}

Naively, the reactor antineutrino flux can be measured directly using the Daya Bay near-site data. However, due to the relatively large size of $\theta_{13}$, even at the near sites (360--500 m baselines) there is an approximately 1--2\% deficit of the antineutrino flux caused by neutrino oscillations. Therefore, far-site data are required in order to extract the value of $\theta_{13}$ independent of other experiments. In this section, we describe two methods to measure the reactor antineutrino flux from the Daya Bay experiment. In the first method, the data from all ADs are fit based on neutrino oscillation theory and a reference reactor antineutrino flux model. The value of $\sin^22\theta_{13}$ and the flux normalization $R$ are simultaneously obtained from the fit, the latter being the measured reactor antineutrino flux. In the second method, we use the measured value of $\sin^22\theta_{13}$ and the near-site data only. The measured reactor antineutrino flux is then expressed in a model-independent way in terms of $\sigma_f$ (cm$^2$/fission) and $Y$ (cm$^2$/day/GW$_{\textrm{th}}$). Finally, we combine our measurement with the past short-baseline experiments to obtain a global average value, and compare it with different model predictions.

\subsection{Measurement of $\sin^22\theta_{13}$ and Flux Normalization $R$}

The IBD event candidates are selected as described in Sec.~\ref{sec:ibdselection}. Figure~\ref{fig:absflux_ratevstime} shows the daily averaged rates of IBD candidate events per AD in the three experimental halls as a function of time. The expected backgrounds are subtracted and the detection efficiencies are corrected in the figure.
The measured IBD rates are highly correlated with the reactor operations.

\begin{center}
\includegraphics[width=\columnwidth]{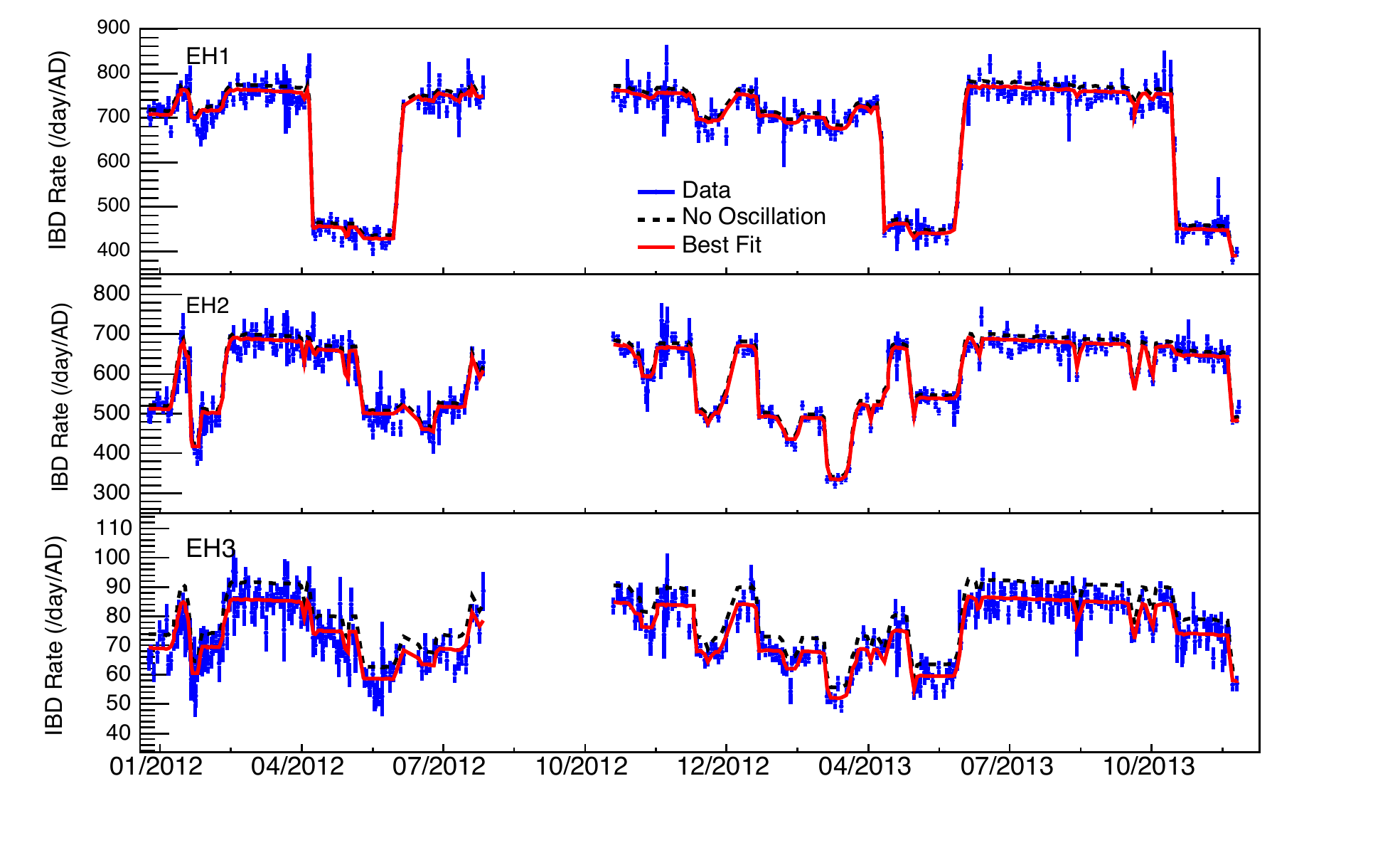}
\figcaption{Daily averaged rates of IBD candidate events per AD in the three experimental halls as a function of time. The discontinuity from July 2012 to Oct 2012 corresponds to the period when the last two ADs were installed. The dotted curves represent no-oscillation predictions based on reactor antineutrino flux analyses and detector simulation. The predictions incorporated the best-fit normalization parameter ($R$). The rates predicted with the best-fit $\sin^22\theta_{13}$ are shown as the red solid curves. }
\label{fig:absflux_ratevstime}
\end{center}

Figure~\ref{fig:absflux_ratedeficit} shows the integrated rate of the detected $\bar\nu_e$ signals at each AD, divided by the no-oscillation predictions. A signal deficit of about 6\% at the far hall relative to the near halls is observed, indicating the size of the oscillation driven by $\theta_{13}$. A normalization factor $R$ was defined to scale the signal predicted by a reactor model. The value of $R$, together with the value of $\sin^22\theta_{13}$, was simultaneously determined with a $\chi^2$ constructed similarly as in Ref.~\cite{bib:prl_rate} using only the integrated rate information,
\begin{multline}\label{eqn:norm}
\chi^{2} = \sum_{d=1}^{8} \frac{[M_{d} - R \cdot T_{d}(1 + \epsilon_{D} + \sum_{r}\omega_{r}^{d}\alpha_{r} + \epsilon_{d}) + \eta_{d}]^{2}}{M_{d} + B_{d}} \\
+ \sum_{r=1}^6\frac{\alpha_{r}^{2}}{\sigma_{r}^{2}} + \sum_{d=1}^{8}\biggl(\frac{\epsilon_{d}^{2}}{\sigma_{d}^{2}} + \frac{\eta_{d}^{2}}{\sigma_{B,d}^{2}}\biggr) + \frac{\epsilon_{D}^{2}}{\sigma_{D}^{2}},
\end{multline}
where $M_{d}$ is the number of measured IBD events in the $d$-th detector with backgrounds subtracted, $B_{d}$ is the corresponding number of background events, $T_{d}$ is the number of IBD events predicted by a reactor model with neutrino oscillations, and $\omega_{r}^{d}$ is the fractional IBD contribution from the $r$-th reactor to the $d$-th detector determined by baselines and reactor antineutrino fluxes. $\sigma_{r}$ (0.9\%) is the uncorrelated reactor uncertainty, $\sigma_{d}$ (0.2\%) is the uncorrelated detection uncertainty, $\sigma_{B,d}$ is the background uncertainty listed in Ref.~\cite{bib:prl_shape2}, and $\sigma_{D}$ (1.93\%) is the correlated detection uncertainty, i.e. the uncertainty of detection efficiency in Table~\ref{tab:AllEffs}. Their corresponding nuisance parameters are $\alpha_{r}$, $\epsilon_{d}$, $\eta_{d}$, and $\epsilon_{D}$, respectively.

We use the rate-only fit in this analysis in order to fix the reference reactor model to its nominal value. Thus the obtained normalization $R$ can be directly compared with other experiments. Fixing the reactor model does not affect the oscillation result due to the relative measurement between far and near detectors. If we add the spectral information, we would need to include and inflate the model uncertainty in the fit in order not to bias the oscillation result. Consequently, even though we would obtain a more precise value of $\sin^22\theta_{13}$, the best-fit flux of the reference reactor model would deviate from its nominal value, making the comparison with other experiments impractical.

The minimization of the rate-only $\chi^2$ defined in Eq.~\ref{eqn:norm} yields $\chi^2/\textrm{NDF} = 5.7/6$. The best-fit value of $\sin^{2}2\theta_{13} = 0.085 \pm 0.006$ is insensitive to the choice of reactor models. The uncertainty in $\sin^{2}2\theta_{13}$ is statistically dominated. The 0.9\% reactor related uncertainty, treated as uncorrelated in the oscillation analysis in order to avoid a bias of the $\sin^{2}2\theta_{13}$ fit, is conservatively added quadratically to the uncertainty of $R$, effectively treating it as correlated among reactors in the rate measurement. The best-fit result of $R$ is $0.946 \pm 0.020$ ($0.992 \pm 0.021$) when compared with the Huber+Mueller (ILL+Vogel) model. 
Replacing the Mueller $^{238}$U spectrum with the measured spectrum in Ref.~\cite{bib:munich} yields an $R$ increased slightly by 0.002.
The contributions to the uncertainty in $R$ are summarized in Table~\ref{tab:flux_err}. The uncertainty is dominated by the detection uncertainty $\sigma_{D}$.

\begin{center}
\includegraphics[width=\columnwidth]{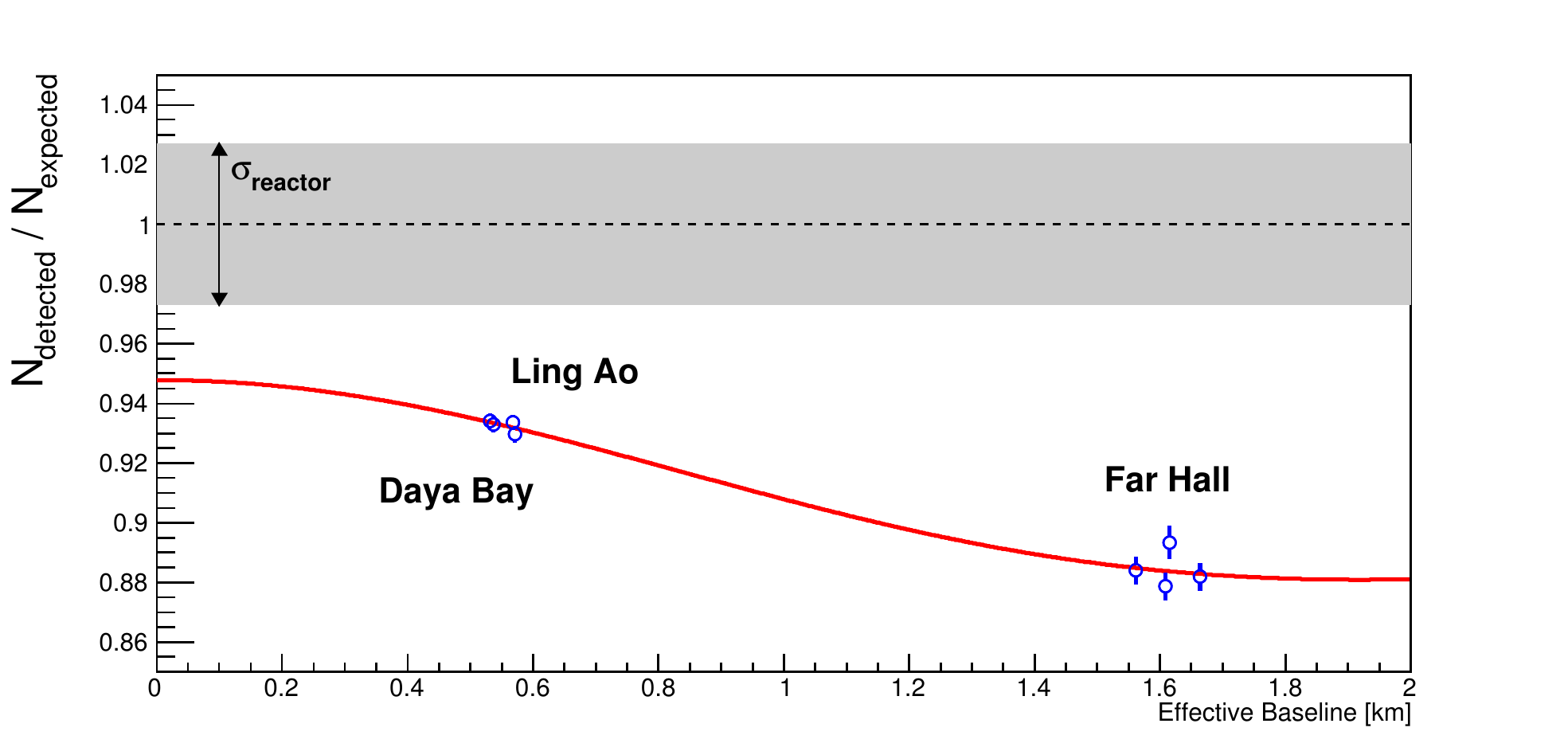}
\figcaption{Ratio of the detected to expected non-oscillation $\bar\nu_e$ signals at the 8 ADs located in three experimental halls as a function of the effective baseline, which is determined for each detector by equating the multicore oscillated flux to an effective oscillated flux from a single baseline. A 6\% signal deficit at the far hall relative to the near halls is observed, indicating the size of the $\theta_{13}$-driven oscillation. The oscillation survival probability at the best-fit value is given by the red curve. In addition, there is a 5\% normalization deficit when compared with the Huber+Mueller model prediction. The uncertainty of the model prediction is shown as the gray band around unity. Two far hall points are displaced by 50 m for visual clarity.}
\label{fig:absflux_ratedeficit}
\end{center}

\begin{center}
  \tabcaption{\label{tab:flux_err}Summary of contributions to the total uncertainty of the reactor antineutrino flux measurement.}
  \begin{tabular}{l | c }
    \toprule
    ~ & Uncertainty \\
    \hline
    statistics  ~&~ 0.1\%  \\
    oscillation ~&~ 0.1\%  \\
    reactor     ~&~ 0.9\%  \\
    detection efficiency ~&~ 1.93\% \\
    \hline
    Total ~&~ 2.1\% \\
    \bottomrule
  \end{tabular}
\end{center}


The best-fit oscillation curve is shown in Fig.~\ref{fig:absflux_ratedeficit}. Disregarding the normalization, the measurement is consistent within the three-neutrino paradigm. On the other hand, the normalization is inconsistent with the Huber+Mueller model prediction within the model uncertainties. We will further discuss the implication in Sec.~\ref{sec:absflux_global}.

\subsection{Measurement of IBD Yield}
In this subsection, we express the measurement in two model-independent ways: the IBD yield per nuclear fission ($\sigma_f$), and the IBD yield per GW$_{th}$ per day ($Y$).

$\sigma_f$ for each AD is determined by solving the following equation:
\begin{equation}\label{eqn:absflux_sigmaf}
  M_d = \sum_{r=1}^6 \frac{N^f_r}{4 \pi L_{dr}^2} \sigma_f^d N^{\textrm{T}}_d P_{sur}^{dr} \epsilon^D_d,
\end{equation}
where $N^f_r$ is the predicted number of fissions from the $r$th reactor core, which is calculated based on $W_r$ (average thermal power of $r$th core), $f^{iso}_r$ (average fission fraction of $r$th core for each isotope) and $E^{iso}$ (mean energy release per fission for each isotope), integrated over the live time of the detector:
\begin{equation}
    N^f_r = \int \frac{W_r}{\sum_{iso=1}^{4}f^{iso}_r E^{iso}} \textrm{d}t.
\end{equation}
 $L_{dr}$ is the distance between the $d$th detector and the $r$th reactor core. $N^{\textrm{T}}_d$ is the total number of target protons in the GdLS of each AD. The total detection efficiency, $\epsilon^D_d$, is different for each AD because of different effects of muon veto and multiplicity cuts on
each AD. $P_{sur}^{dr}$ is the survival probability given an AD-core pair, calculated using the best-fit value of $\sin^22\theta_{13}$ from the rate-only analysis described in the previous subsection. Due to the relatively large size of $\theta_{13}$, even at the near sites there are on average about 1.5\% rate deficits, as shown in Fig.~\ref{fig:absflux_ratedeficit}. The values of $\sigma_f^d$ for all ADs, from Eq.~\ref{eqn:absflux_sigmaf}, are summarized in Table~\ref{table:absflux_results}. Similar to the normalization $R$, the uncertainty in $\sigma_f^d$ (summarized in Table~\ref{table:absflux_results} as $\sigma_{exp}$) is dominated by the correlated detection uncertainty $\sigma_{D}$.

Theoretically, $\sigma_f$ represents the IBD cross section convolved with the reactor antineutrino spectra from all fission isotopes, and integrated over energy:
\begin{equation}
    \sigma_f = \sum_{iso=1}^4 f_{iso} \int S_{iso}(E_\nu) \sigma(E_\nu) dE_\nu
\end{equation}
Given a reactor model that predicts the antineutrino spectrum $S_{iso}(E_\nu)$ for each of the four main fission isotopes $^{235}$U, $^{238}$U, $^{239}$Pu and $^{241}$Pu, and the fission fractions $f_{iso}$ determined by NPP operations and simulations, $\sigma_f$ can be theoretically calculated and compared with the model-independent measurement. The ratios of the measurement versus the Huber+Mueller model prediction (R$_{\textrm{H+M}}$), and versus the ILL+Vogel model prediction (R$_{\textrm{I+V}}$) for each AD are summarized in Table~\ref{table:absflux_results}.

Alternatively, we can define $Y_d \equiv \sigma^d_f N_r^f / W_r$ as the IBD yield per GW thermal power per day. The above expression approaches a common value $Y$ after averaging multiple fuel burnup cycles, since all the reactor cores have the same average fuel composition. During the 6-AD data taking period, none of the reactor cores had completed a burnup cycle. The differences in fuel composition cause about 2\% variations in measured IBD yield (top panel of Fig.~\ref{fig:absflux_sigmafy}). These core-to-core variations can be corrected using known values of the fission fractions given by Table~\ref{table:absflux_results}.  On the other hand, all the reactor cores had roughly one full cycle during the 6-AD and 8-AD data taking period. Therefore measurements from eight detectors give the same value (within statistical fluctuation), and core-to-core variations are negligible (bottom panel of Fig.~\ref{fig:absflux_sigmafy}).

\end{multicols}
\begin{center}
\tabcaption{\label{table:absflux_results}Tabulated results of the flux measurement from each AD. $\sigma_f$ is the measured cross section in units of $10^{-43} $cm$^2$/fission. $Y$ is the IBD yield in units of $10^{-18}$cm$^2$/GW/day. $R_\textrm{H+M}$ and $R_\textrm{I+V}$ are the ratios of measured flux with respect to Huber-Mueller and ILL-Vogel model predictions, respectively. $\sigma_{exp}$ is the total fractional experimental uncertainty of the flux measurement. $^{235}$U, $^{238}$U, $^{239}$Pu, $^{241}$Pu are the flux-weighted fission fractions of each fission isotope. L is the flux-weighted baseline for each AD. $P_{sur}$ is the average $\bar\nu_e$ survival probability at each AD. See the text for more details.}
\footnotesize
\begin{tabular}{c|c|c|c|c|c|c|c|c|c|c|c}
\hline
\hline
  ~  & $\sigma^d_f \cdot 10^{43}$ & $Y \cdot 10^{18}$ & R$_\textrm{H+M}$   & R$_\textrm{I+V}$  & $\sigma_{exp}$ & $^{235}$U & $^{238}$U & $^{239}$Pu & $^{241}$Pu  &  $L_d$ & $P_{sur}$   \\
  ~  & (cm$^2$/fission)  & (cm$^2$/GW/day) & ~   & ~  & (\%) & ~ & ~ & ~ & ~  & (m) & ~   \\
\hline
EH1-AD1 & 5.907 & 1.531 & 0.945 & 0.991 & 2.1 & 0.564 & ~0.076~ & ~0.303~ & ~0.056~ & 566  & 0.985\\
EH1-AD2 & 5.912 & 1.536 & 0.946 & 0.992 & 2.1 & 0.564 & ~0.076~ & ~0.303~ & ~0.056~ & 561  & 0.986\\
EH2-AD1 & 5.925 & 1.538 & 0.948 & 0.994 & 2.1 & 0.557 & ~0.076~ & ~0.312~ & ~0.055~ & 594  & 0.983\\
EH2-AD2 & 5.894 & 1.529 & 0.944 & 0.990 & 2.1 & 0.552 & ~0.076~ & ~0.315~ & ~0.057~ & 598  & 0.983\\
EH3-AD1 & 5.819 & 1.521 & 0.940 & 0.986 & 2.2 & 0.559 & ~0.076~ & ~0.310~ & ~0.055~ & 1635 & 0.934\\
EH3-AD2 & 5.858 & 1.540 & 0.946 & 0.992 & 2.2 & 0.559 & ~0.076~ & ~0.310~ & ~0.055~ & 1636 & 0.934\\
EH3-AD3 & 5.842 & 1.536 & 0.944 & 0.990 & 2.2 & 0.559 & ~0.076~ & ~0.310~ & ~0.055~ & 1640 & 0.934\\
EH3-AD4 & 5.907 & 1.554 & 0.956 & 1.002 & 2.2 & 0.552 & ~0.076~ & ~0.315~ & ~0.057~ & 1641 & 0.934\\
\hline
\hline
\end{tabular}
\end{center}
\begin{multicols}{2}

Table~\ref{table:absflux_results} further summarizes a few characteristic parameters calculated for each AD, including the average fission fraction $f^{iso}_d$, flux-weighted baseline $L_d$ and average survival probability $P_{sur}^d$. These parameters can be trivially obtained in the case of a single reactor core, but require clear definitions in the multi-core case of Daya Bay. The average fission fraction $f^{iso}_d$ is defined as follows:
\begin{equation}\label{eqn:f_iso}
  f^{iso}_d = \frac{\sum_{r=1}^6 \beta_{dr} \cdot f^{iso}_r}{\sum_{r=1}^6 \beta_{dr}}, \qquad
  \beta_{dr} = \frac{N_r^f}{L_{dr}^2}
\end{equation}
where $\beta_{dr}$ is the flux-weighting factor calculated from $N_r^f$ and $L_{dr}$ (see Eq.~\ref{eqn:absflux_sigmaf} for definition). We note that the average fission fractions for the two newly installed ADs (EH2-AD2 and EH3-AD4) are slightly different from the ADs at the same site, because they are seeing different reactor core histories with respect to other detectors. The flux-weighted baseline $L_d$ is defined as
\begin{equation}\label{eqn:L_d}
    \frac{1}{L_d^2} = \frac{\sum_{r=1}^6 N^f_r \cdot 1/L_{dr}^2}{\sum_{r=1}^6 N^f_r}.
\end{equation}
Finally, the average survival probability $P_{sur}^d$ is calculated as follows:
\begin{equation}
    P^d_{sur} = \frac{\sum_{r=1}^6 N_{dr} P^{dr}_{sur}}{\sum_{r=1}^6 N_{dr}}
\end{equation}
where $N_{dr}$ is the predicted number of IBD events at the $d$th AD from the $r$th reactor core without oscillation, and $P_{sur}^{dr}$ is the average survival probability given an AD-core pair as defined in Eq.~\ref{eqn:absflux_sigmaf}.

\begin{center}
\includegraphics[width=\columnwidth]{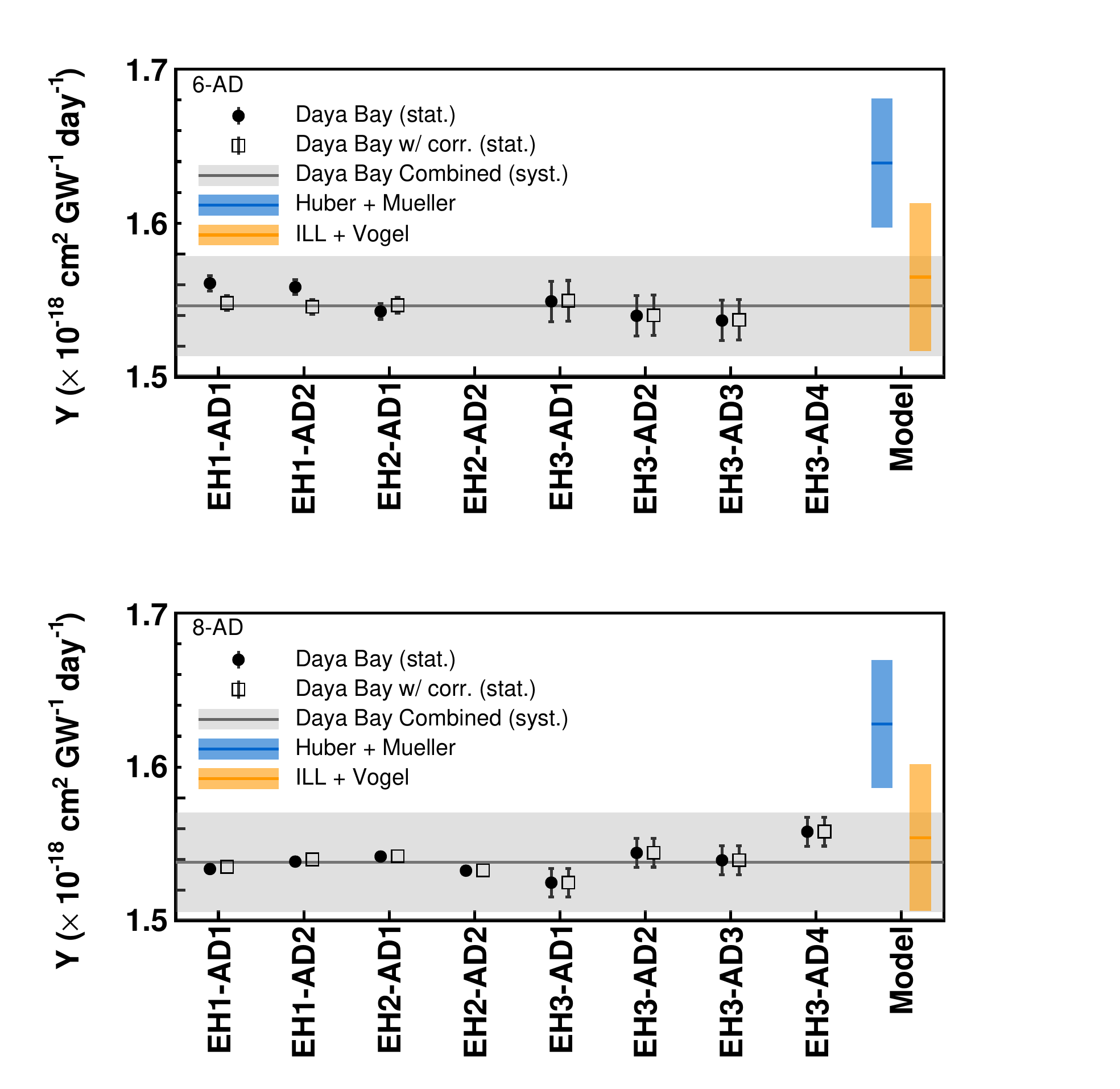}
\figcaption{Yield $Y$ for the IBD events in the 6-AD only (top) and 8-AD only (bottom) period with corrections of 3-flavor oscillations (closed circles), and additional corrections due to the variations of flux-weighted fission fractions at different sites (open squares). The horizontal line is the average yield of the near detectors, and the gray band is its $1\sigma$ systematic uncertainty. The rate predicted by the Huber+Mueller (ILL+Vogel) model and its uncertainty are shown in blue (orange) region.}
\label{fig:absflux_sigmafy}
\end{center}

The measured IBD yields for each AD are plotted in Fig.~\ref{fig:absflux_sigmafy}. The yields are consistent among all ADs after correcting for the small variations of fission fractions at the different sites.
The results are summarized in Table~\ref{tab:norm3m}.
\begin{center}
\tabcaption{The average IBD yields ($Y$ and $\sigma_f$) of the near halls, the flux normalization with respect to different reactor model predictions, and the flux-weighted average fission fractions of the near halls.}
\footnotesize
\begin{tabular}{c|c}
\toprule
\multicolumn{2}{c}{IBD Yield} \\
\hline
$Y$~(~cm$^2$/GW/day)       &  $(1.53 \pm 0.03) \times 10^{-18}$          \\
$\sigma_f$~(cm$^2$/fission)     &  $(5.91 \pm 0.12) \times 10^{-43}$          \\
\hline
\multicolumn{2}{c}{Data / Prediction} \\
\hline
$R$~(Huber+Mueller) & $0.946 \pm 0.020~ (\textrm{exp.})$ \\
$R$~(ILL+Vogel) & $0.992 \pm 0.021~(\textrm{exp.})$ \\
\hline
$^{235}$U : $^{238}$U : $^{239}$Pu : $^{241}$Pu & 0.561 : 0.076 : 0.307 : 0.056\\
\bottomrule
\end{tabular}
\label{tab:norm3m}
\end{center}

\subsection{Comparison with Past Reactor Experiments} \label{sec:absflux_global}

Recently, there was great interest in the so-called ``reactor antineutrino anomaly'', which arises from re-evaluations of the reactor $\bar\nu_e$ flux that resulted in an increase of the predicted $\bar\nu_e$ flux in the Huber+Mueller model~\cite{bib:mueller2011,bib:huber}. Combining the new predictions with the re-analysis of the past experimental data at baselines 10-100 m suggests a $\sim$4-6\% deficit between the measured and the predicted reactor $\bar\nu_e$ flux~\cite{bib:mention2011,bib:chao}. In this subsection, our measurement is compared with the past reactor neutrino experiments.

A global fit was performed for the past reactor neutrino experiments. Nineteen short-baseline ($<$100~m) measurements were included using the data from Ref.~\cite{bib:mention2011}. The measurements from CHOOZ~\cite{bib:chooz} and Palo Verde~\cite{bib:paloverde} were also included after correcting for the standard three neutrino oscillations using the best-fit value of $\sin^22\theta_{13}$ from the rate-only analysis described in a previous subsection.
The results of all 21 experiments are summarized in Table~\ref{table:absflux:past}.
In the ``ratio'' column, each measured flux was compared to the Huber+Mueller flux prediction (in Ref.~\cite{bib:mention2011}, the ``ratio'' column is calculated with respect to the Mueller~\cite{bib:mueller2011} model).
The $\sigma_\textrm{err}$ column summarizes the total uncertainty reported by each measurement, and the $\sigma_\textrm{cor}$ column summarizes the correlated uncertainty among experiments in the same group.
Both $\sigma_\textrm{err}$ and $\sigma_\textrm{cor}$ include the theoretical uncertainty of the model prediction $\sigma_{\textrm{model}}$. At the time of those measurements, all experiments reported results using a common $\sigma_{\textrm{model}} = 2.7\%$ from the ILL+Vogel model.
Since this $\sigma_{\textrm{model}}$ is correlated among all measurements, it is the minimum value of $\sigma_\textrm{cor}$.
Both $\sigma_\textrm{err}$ and $\sigma_\textrm{cor}$ were taken from Ref.~\cite{bib:mention2011} except for SRP-I, SRP-II, ROVNO88-1I, and ROVNO88-2I.
We adopted the uncertainty treatment of Ref.~\cite{bib:chao} for those four experiments.

\end{multicols}
\begin{center}
\tabcaption{\label{table:absflux:past}Tabulated results of 21 past reactor antineutrino flux measurements. Experiments are categorized into different groups
with horizontal lines. Within each group, the $\sigma_\textrm{cor}$ represent the correlated uncertainties among different experiments. This table is compiled from Ref.~\cite{bib:mention2011,bib:chao}. The ``ratio'' column shows the measured flux from each experiment with respect to the Huber+Mueller model prediction.}
\footnotesize
\begin{tabular}{@{} ccccccccccccc @{}}
\hline
\hline
\# & Exp.     & Det. type         & $^{235}$U & $^{239}$Pu & $^{238}$U & $^{241}$Pu & ratio & $\sigma_\text{err}$ &
$\sigma_\text{cor}$ & L & $P_{sur}$   & Year \\
& & & & & & & & (\%) & (\%) & (m) & &  \\\hline
 1~ & ~Bugey-4    ~ & ~$^3$He+H$_2$O~ & ~0.538     ~ & ~0.328~ & ~0.078~ & ~0.056~ & ~0.926~ & ~3.0 ~ & ~3.0~ & ~15  ~ & ~$\approx$1~ & ~1994~ \\
 2~ & ~ROVNO91    ~ & ~$^3$He+H$_2$O~ & ~0.614     ~ & ~0.274~ & ~0.074~ & ~0.038~ & ~0.924~ & ~3.9 ~ & ~3.0~ & ~18  ~ & ~$\approx$1~ & ~1991~ \\\hline
 3~ & ~Bugey-3-I  ~ & ~$^6$LiLS     ~ & ~0.538     ~ & ~0.328~ & ~0.078~ & ~0.056~ & ~0.930~ & ~4.8 ~ & ~4.8~ & ~15  ~ & ~$\approx$1~ & ~1995~ \\
 4~ & ~Bugey-3-II ~ & ~$^6$LiLS     ~ & ~0.538     ~ & ~0.328~ & ~0.078~ & ~0.056~ & ~0.936~ & ~4.9 ~ & ~4.8~ & ~40  ~ & ~$\approx$1~ & ~1995~ \\
 5~ & ~Bugey-3-III~ & ~$^6$LiLS     ~ & ~0.538     ~ & ~0.328~ & ~0.078~ & ~0.056~ & ~0.860~ & ~14.1~ & ~4.8~ & ~95  ~ & ~$\approx$1~ & ~1995~ \\\hline
 6~ & ~Goesgen-I  ~ & ~$^3$He+LS    ~ & ~0.620     ~ & ~0.274~ & ~0.074~ & ~0.042~ & ~0.950~ & ~6.5 ~ & ~6.0~ & ~38  ~ & ~$\approx$1~ & ~1986~ \\
 7~ & ~Goesgen-II ~ & ~$^3$He+LS    ~ & ~0.584     ~ & ~0.298~ & ~0.068~ & ~0.050~ & ~0.976~ & ~6.5 ~ & ~6.0~ & ~45  ~ & ~$\approx$1~ & ~1986~ \\
 8~ & ~Goesgen-III~ & ~$^3$He+LS    ~ & ~0.543     ~ & ~0.329~ & ~0.070~ & ~0.058~ & ~0.909~ & ~7.6 ~ & ~6.0~ & ~65  ~ & ~$\approx$1~ & ~1986~ \\
 9~ & ~ILL        ~ & ~$^3$He+LS    ~ & ~$\approx$1~ & ~-    ~ & ~-    ~ & ~-    ~ & ~0.786~ & ~9.5 ~ & ~6.0~ & ~9   ~ & ~$\approx$1~ & ~1981~ \\\hline
10~ & ~Krasn. I   ~ & ~$^3$He+PE    ~ & ~$\approx$1~ & ~-    ~ & ~-    ~ & ~-    ~ & ~0.920~ & ~5.8 ~ & ~4.9~ & ~33  ~ & ~$\approx$1~ & ~1987~ \\
11~ & ~Krasn. II  ~ & ~$^3$He+PE    ~ & ~$\approx$1~ & ~-    ~ & ~-    ~ & ~-    ~ & ~0.937~ & ~20.3~ & ~4.9~ & ~92  ~ & ~$\approx$1~ & ~1987~ \\
12~ & ~Krasn. III ~ & ~$^3$He+PE    ~ & ~$\approx$1~ & ~-    ~ & ~-    ~ & ~-    ~ & ~0.931~ & ~4.9 ~ & ~4.9~ & ~57  ~ & ~$\approx$1~ & ~1987~ \\\hline
13~ & ~SRP-I      ~ & ~GdLS         ~ & ~$\approx$1~ & ~-    ~ & ~-    ~ & ~-    ~ & ~0.936~ & ~3.7 ~ & ~2.7~ & ~18  ~ & ~$\approx$1~ & ~1996~ \\
14~ & ~SRP-II     ~ & ~GdLS         ~ & ~$\approx$1~ & ~-    ~ & ~-    ~ & ~-    ~ & ~1.002~ & ~3.8 ~ & ~2.7~ & ~24  ~ & ~$\approx$1~ & ~1996~ \\\hline
15~ & ~ROVNO88-1I ~ & ~$^3$He+PE    ~ & ~0.607     ~ & ~0.277~ & ~0.074~ & ~0.042~ & ~0.901~ & ~6.9 ~ & ~5.7~ & ~18  ~ & ~$\approx$1~ & ~1988~ \\
16~ & ~ROVNO88-2I ~ & ~$^3$He+PE    ~ & ~0.603     ~ & ~0.276~ & ~0.076~ & ~0.045~ & ~0.932~ & ~6.9 ~ & ~5.7~ & ~18  ~ & ~$\approx$1~ & ~1988~ \\\hline
17~ & ~ROVNO88-1S ~ & ~GdLS         ~ & ~0.606     ~ & ~0.277~ & ~0.074~ & ~0.043~ & ~0.956~ & ~7.8 ~ & ~7.2~ & ~18  ~ & ~$\approx$1~ & ~1988~ \\
18~ & ~ROVNO88-2S ~ & ~GdLS         ~ & ~0.557     ~ & ~0.313~ & ~0.076~ & ~0.054~ & ~0.943~ & ~7.8 ~ & ~7.2~ & ~25  ~ & ~$\approx$1~ & ~1988~ \\
19~ & ~ROVNO88-3S ~ & ~GdLS         ~ & ~0.606     ~ & ~0.274~ & ~0.074~ & ~0.046~ & ~0.922~ & ~7.2 ~ & ~7.2~ & ~18  ~ & ~$\approx$1~ & ~1988~ \\\hline
20~ & ~Palo Verde ~ & ~GdLS         ~ & ~0.600     ~ & ~0.270~ & ~0.070~ & ~0.060~ & ~0.959~ & ~6.0 ~ & ~2.7~ & ~835 ~ & ~0.967~      & ~2001~ \\\hline
21~ & ~CHOOZ      ~ & ~GdLS         ~ & ~0.496     ~ & ~0.351~ & ~0.087~ & ~0.066~ & ~0.945~ & ~4.2 ~ & ~2.7~ & ~1052~ & ~0.954~      & ~1999~ \\\hline
\hline
\end{tabular}
\end{center}
\begin{multicols}{2}

To calculate the global average independent of the model uncertainty used by the past measurements,  we follow the method described in Ref.~\cite{bib:chao} by first removing $\sigma_{\textrm{model}}$ from both uncertainties, and define:
\begin{eqnarray}\label{eq:exp}
\sigma^{\textrm{exp}}_\textrm{err}  &=& \sqrt{\sigma^2_\textrm{err} - \sigma_\textrm{model}^2} \nonumber \\
\sigma^{\textrm{exp}}_\textrm{cor} &=& \sqrt{\sigma^2_\textrm{cor} - \sigma_\textrm{model}^2}.
\end{eqnarray}
$\sigma^{\textrm{exp}}_\textrm{err}$ and $\sigma^{\textrm{exp}}_\textrm{cor}$ now represent experimental uncertainties only. We then build a covariance matrix $V^{\textrm{exp}}$ such that
\begin{eqnarray}\label{eq:cov_exp}
    V^{\textrm{exp}}_{ij} &=& R_i^{\textrm{obs}} \cdot \sigma_{i,cor}^{\textrm{exp}} \cdot R_j^{\textrm{obs}} \cdot \sigma_{j,cor}^{\textrm{exp}},
\end{eqnarray}
where $R_i^{\textrm{obs}}$ is the ``ratio'' column in Table~\ref{table:absflux:past} corrected by the ``$P_{sur}$'' column for the $\theta_{13}$-oscillation effect. $R_i^{\textrm{obs}}$ represents the observed rate from each measurement.

We then calculate the best-fit average ratio $R_g^{\textrm{past}}$ by minimizing the $\chi^2$ function defined as:
\begin{equation}\label{eq:chi2}
    \chi^2(R_g^{\textrm{past}}) = (R_g^{\textrm{past}} - R_i) \cdot (V^{\textrm{exp}}_{ij})^{-1} (R_g^{\textrm{past}} - R_j),
\end{equation}
where $V^{-1}$ is the inverse of the covariance matrix $V$. This procedure yields the best-fit result $R_g^{\textrm{past}} = 0.942 \pm 0.009$, where the error is experimental only.

Since we now use the Huber+Mueller model as the reference model, we re-evaluate the model uncertainty using the correlated and uncorrelated uncertainty components given by Ref.~\cite{bib:huber,bib:mueller2011}.
Using the weighted average fission fraction from all experiments ($^{235}$U : $^{238}$U : $^{239}$Pu : $^{241}$Pu  =  0.642 : 0.063 : 0.252 : 0.0425), the model uncertainty is calculated to be 2.4\%, and the final result becomes:
\begin{eqnarray}
    R_g^{\textrm{past}} &=& 0.942 \pm 0.009~(\textrm{exp.}) \pm 0.023~(\textrm{model})
\end{eqnarray}


Finally, we compare the Daya Bay result with the past global average. In the previous subsection, we obtained the Daya Bay measured reactor antineutrino flux with respect to the Huber+Mueller model prediction: $R_{\textrm{DYB}}= 0.946 \pm 0.020 \, \textrm{(exp.)}$.
This result is consistent with the past global average $R_g^{\textrm{past}} = 0.942 \pm 0.009 \, \textrm{(exp.)}$. If we include the Daya Bay result in the global fit, the new average is $R_g = 0.943 \pm 0.008 \, \textrm{(exp.)} \pm 0.023 \, \textrm{(model)}$. The results of the global fit and the Daya Bay measurement are shown in Fig.~\ref{fig:absflux_globalfit}.

The consistency between Daya Bay's measurement and past experiments suggests that the origin of the ``reactor antineutrino anomaly'' is from the theoretical side. Either the uncertainties of the theoretical models that predict the reactor antineutrino flux are underestimated or more intriguingly, there exists an additional neutrino oscillation that suppresses the reactor antineutrino flux within a few meters from the reactor. Such an oscillation would imply the existence of one or more eV-mass-scale sterile neutrinos. To investigate this tantalizing possibility, future short baseline (10~m) experiments are required to observe the $L/E$ dependence of such an oscillation.

\begin{center}
\includegraphics[width=\columnwidth]{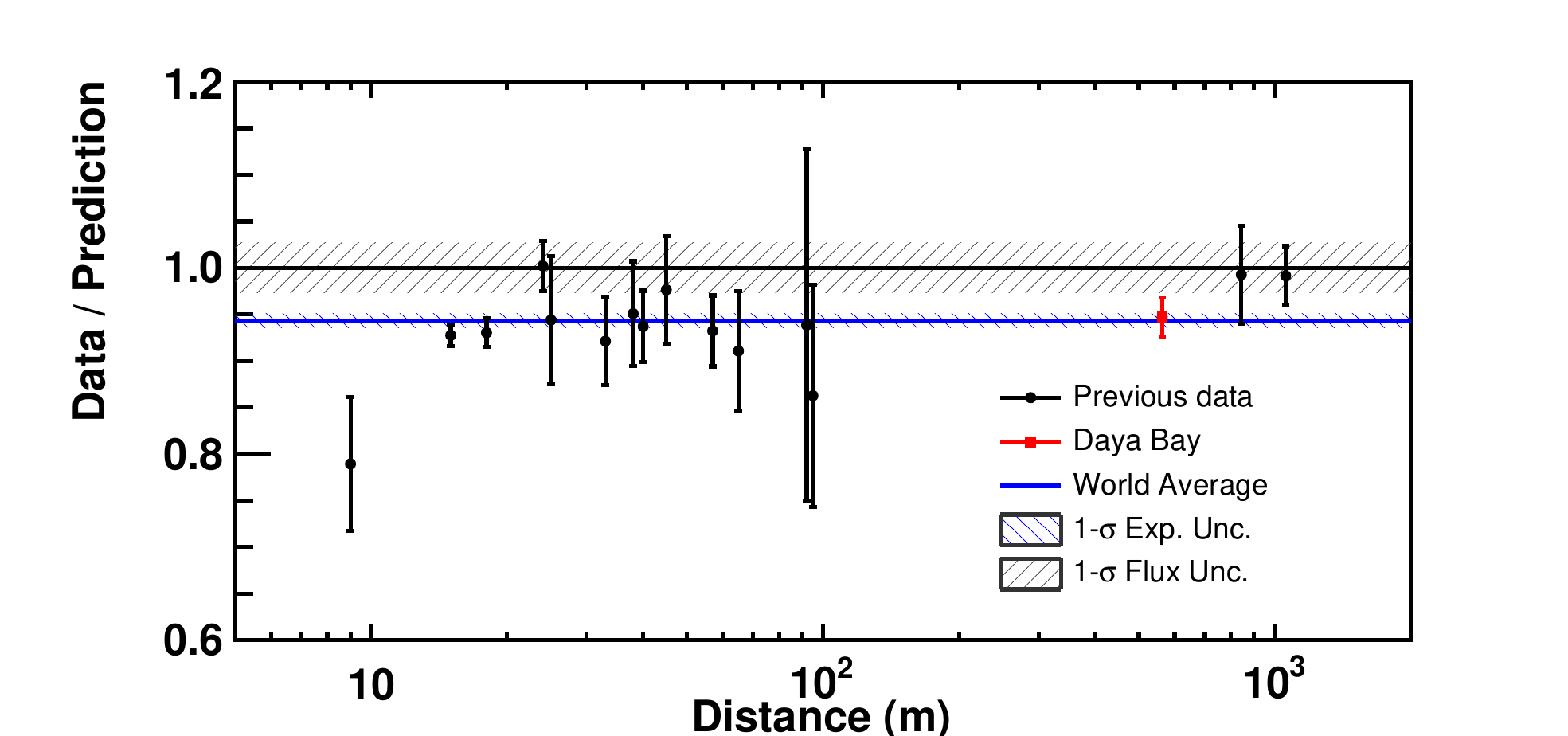}
\figcaption{The measured reactor $\bar{\nu}_e$ rate as a function of the distance from the reactor, normalized to the theoretical prediction of Huber+Mueller model. The rate is corrected by 3-flavor neutrino oscillations at the distance of each experiment. The purple shaded region represents the global average and its $1\sigma$ uncertainty. The 2.4\% model uncertainty is shown as a band around unity. The measurements at the same baseline are combined together for clarity. The Daya Bay measurement is shown at the flux-weighted baseline (573 m) of the two near halls.}
\label{fig:absflux_globalfit}
\end{center}

\section{Measurement of Reactor Antineutrino Spectrum}\label{sec:abs_spectrum}
In this section, we extend the study from reactor antineutrino flux to its energy spectrum.
The measured prompt energy spectra from the four near-site ADs were summed and compared with the predictions.
The detector response of the Daya Bay ADs was studied and used to convert the predicted antineutrino spectrum to the prompt energy spectrum for comparison.
A discrepancy was found in the energy range between 4 and 6 MeV with a maximum local significance of 4.4~$\sigma$.
The discrepancy and possible reasons for it were investigated.

\subsection{Detector Response}
The predicted antineutrino flux and spectrum were calculated via the procedure described in Sec.~\ref{sec:flux}.
At each AD, the reactor antineutrino survival probability was taken into account with the best fit oscillation parameters, $\sin^22\theta_{13} = 0.084$ and $|\Delta m^2_{ee}| = 2.42\times 10^{-3}~\mathrm{eV}^2$, based on the oscillation analysis of the same dataset~\cite{bib:prl_shape2}.
The relation of the antineutrino spectrum $S(E_{\bar{\nu}_{e}})$ and the reconstructed prompt energy spectrum $S(E_p)$ can be expressed as,
\begin{equation}
  \label{eq:energy_response}
  S(E_p) = \int S(E_{\bar{\nu}_e}) R(E_{\bar{\nu}_e}, E_p) dE_{\bar{\nu}_e}
\end{equation}
where $R(E_{\bar{\nu}_e},E_p)$ is the detector energy response and can be thought of as a response matrix, which maps each antineutrino energy to a spectrum of reconstructed prompt energies.
The energy response includes four main effects: the IBD prompt energy shift, IAV effect, non-linearity, and energy resolution, which are studied in the following.

\subsubsection{IBD Prompt Energy Shift}
The antineutrino energy is transferred to a positron and a neutron via the IBD reaction, $\bar{\nu}_{e^+} + p \rightarrow e^{+} + n$.
The positron kinetic energy is
\begin{equation}
  \label{eq:kinetic_positron}
  T_{e^+} = E_{\bar{\nu_e}} - (M_n + M_e - M_p) - T_n,
\end{equation}
where $E_{\bar{\nu}_e}$ is energy of the antineutrino, $M_n$, $M_p$ and $M_e$ are the neutron, proton and electron masses, and $T_n$ is the kinetic energy of the neutron.
The visible prompt energy is related to the antineutrino energy as
\begin{equation}
 E_p = T_{e^+} + 1.022~\mathrm{MeV} = E_{\bar{\nu}_e} - 0.78~\mathrm{MeV} - T_n.
 \end{equation}
The positron annihilation produces two gammas with total energy 1.022~MeV.
The shift is approximately 0.78~MeV, with a small correction from $T_n$, which has an average value of $\sim$10~keV for reactor antineutrinos.
The kinetic energies of the positron and the neutron are calculated based on the formula in~\cite{Vogel:1999zy} at the first order in $1/M$, where $M$ is the nucleon mass.

\subsubsection{IAV Effect}
In the Daya Bay ADs, the inner and outer acrylic vessels, as well as the supporting acrylic ribs are non-scintillating material.
In particular, when IBD reactions occur around or in the acrylic, the generated positrons and the annihilation $\gamma$-rays are likely to lose energy in the acrylic without producing scintillation light.
This effect will reduce the visible energy and distort the prompt energy spectrum.
This effect is called the IAV effect as most of the events that lose energy in acrylic cluster around the inner acrylic vessel.

\begin{center}
\includegraphics[width=\columnwidth]{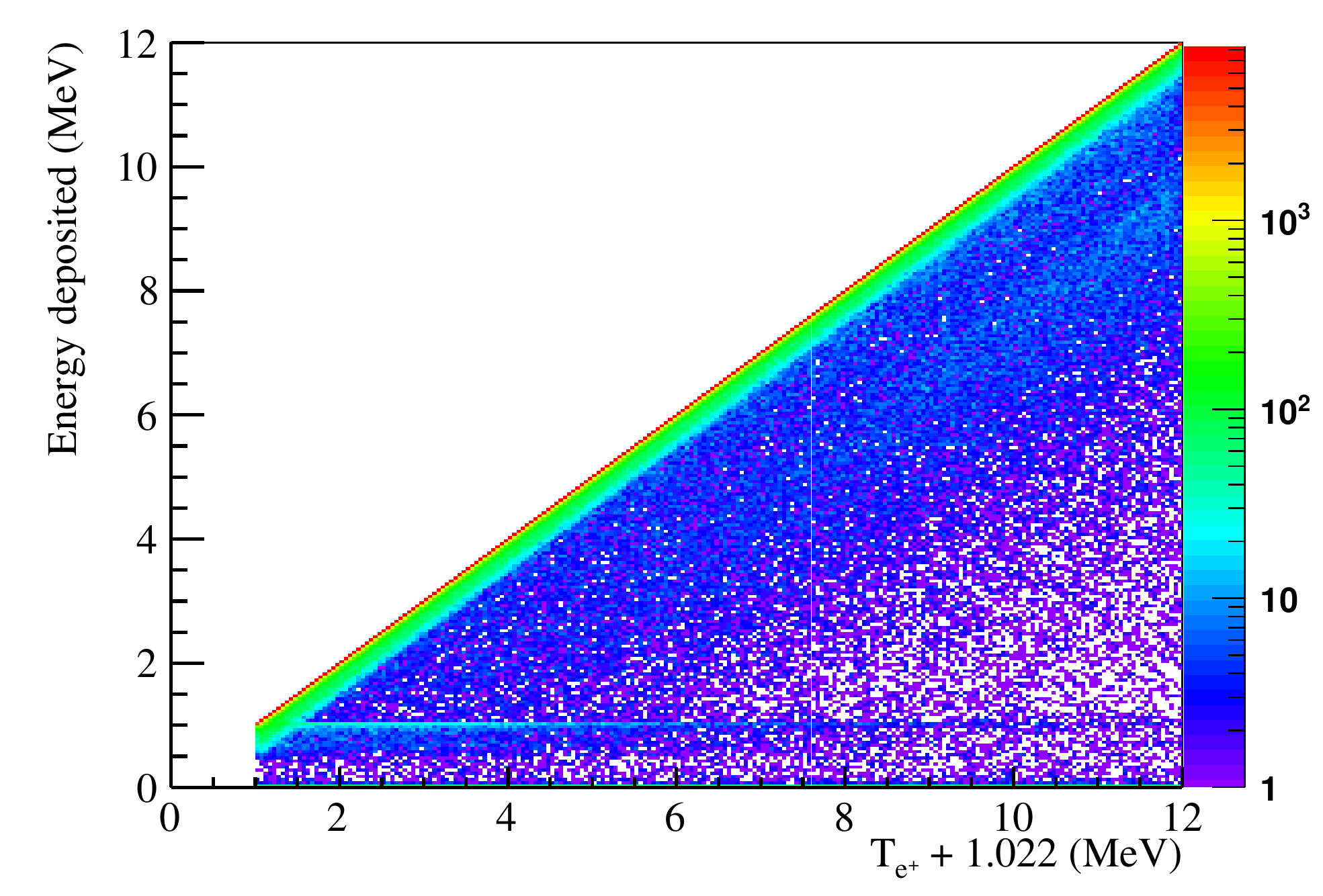}
\figcaption{The IAV effect from the Geant4 simulation.
For a small number of events, energy loss in the IAV reduces the energy deposited in the LS and GdLS so that it is less than the positron kinetic energy plus the 1.022~MeV from annihilation.}
\label{fig:iaveffect}
\end{center}

To study the IAV effect, simulated IBD reactions were uniformly generated based on the density of the target protons in the detector materials and determined the corresponding deposited energy.
From the MC truth information, 13\% of IBD events lose more than 50~keV in the acrylic vessel, which yields $E_{vis}<T_{e^+} + 1.022$~MeV.
Figure~\ref{fig:iaveffect} shows the deposited energy versus ($T_{e^+} + 1.022$~MeV) to
illustrate the effect of the IAV on the IBD prompt events.
Some positrons lose all of their kinetic energy in the acrylic vessel but the two annihilation $\gamma$-rays escape to the scintillator.
In this case, a deposited energy of about 1.022~MeV will be detected, which enhances the deposited energy spectrum at around 1~MeV as shown in Fig.~\ref{fig:iaveffect}.
The uncertainty of the IAV effect is studied by comparing the simulation results with the IAV thickness varying within a range of 0.4~mm.
The induced uncertainty on the prompt energy spectrum was estimated to be 4\% below 1.25~MeV, dropping rapidly to 0.1\% at higher energies.

\subsubsection{Non-linearity}
The energy response of the antineutrino detector is not linear due to the effects originating from the scintillator and the electronics. These two effects, both at a level of 10\%, are parameterized with two functions, $f_{scint}$ and $f_{elec}$.
The scintillator non-linearity is related to the ionization quenching, which is modeled by Birks' formula, and Cherenkov light emission.
The electronics non-linearity is introduced by the loss of the slow scintillation light in a limited charge collection time-window.
It is modeled using an exponential as a function of total visible energy based on the scintillation light timing profile and a charge collection study~\cite{bib:prl_shape}.

\begin{center}
\includegraphics[width=0.95\columnwidth]{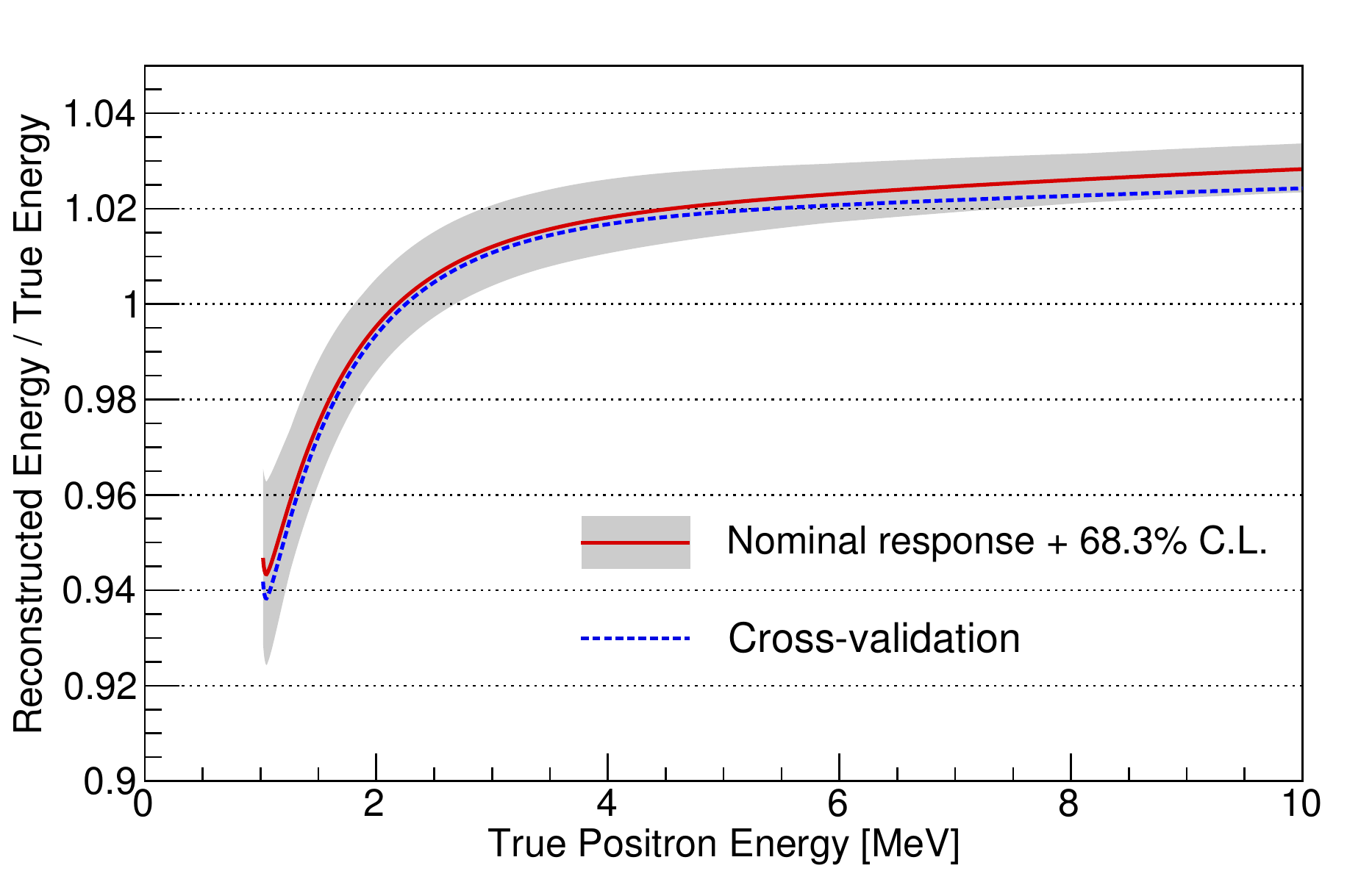}
\figcaption{
Estimated energy response of the detectors to positrons, including both kinetic and annihilation gamma energy (red solid curve).
Gamma rays from both deployed and intrinsic sources as well as spallation $^{12}$B $\beta$ decay determined the model, and provided an envelope of curves consistent with the data within a 68.3\% C.L. (grey band).
An independent estimate using the beta+gamma energy spectra from $^{212}$Bi, $^{214}$Bi, $^{208}$Tl, as well as the Michel electron spectrum produced a consistent result (blue dashed line).}
\label{fig:non-linearity}
\end{center}

The non-linearity model includes five parameters: detector energy scale, Birks' constant, relative contribution from Cherenkov light, and the amplitude and decay constant of the electronics model.
The parameters are determined by a combined $\chi^2$ fit to the mono-energetic $\gamma$ lines of calibration sources and continuous $\beta$ spectrum of $^{12}$B produced by the muon spallation inside the AD.
The Geant4 simulation is used to build the relation of non-linearity response of different particle species, such as gamma, $e^{+}$ and $e^{-}$.
The IBD positron non-linearity response derived from the best fit parameters is shown in Fig~\ref{fig:non-linearity}.
The uncertainty band is constructed by considering calibration and model uncertainties.
The positron non-linearity response was validated using the Michel electron spectrum from muon decay at rest and the continuous $\beta + \gamma$ spectra from internal radioactive $\beta$ decays of $^{212}$Bi, $^{214}$Bi and $^{208}$Tl (see Ref.~\cite{bib:prl_shape2} for detailed non-linearity treatment).
The non-linearity uncertainty has a negligible effect on the measured oscillation parameters because it is treated as correlated for all ADs.

\subsubsection{Energy Resolution}
The detector energy resolution was studied by a variety of calibration sources deployed at the detector center, IBD and spallation neutrons, and alpha sources from radioactivity.
For each source, the reconstructed energy is measured and the width and the energy of the peak are obtained from fits with Gaussian function to the peak of the energy distribution.
The results from both MC and experimental data are shown in Fig.~\ref{fig:Eresolution}.

\begin{center}
\includegraphics[width=0.95\columnwidth]{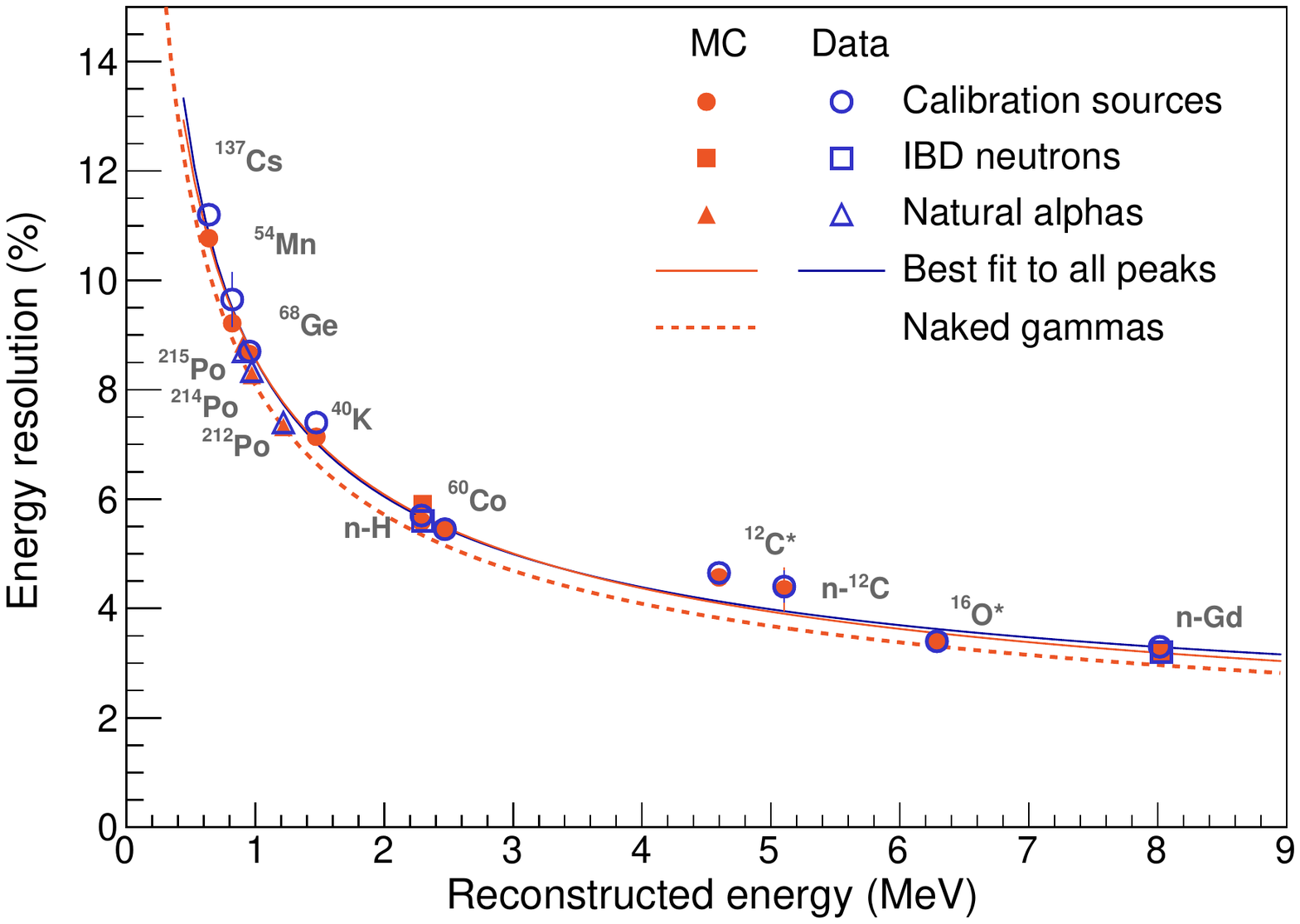}
\figcaption{
  Energy resolution for a variety of calibration sources as well as the IBD neutron capture gamma peaks for both MC and data.
  The parameters in the energy resolution function were extracted by fitting the calibration energy peaks and widths.
  Alpha source data were used to cross-check the result.
  The naked gamma sources are also simulated.}
\label{fig:Eresolution}
\end{center}

The relative energy resolution of an antineutrino detector as a function of energy is parameterized by
\begin{equation}
  \label{eq:resolution}
  \frac{\sigma_E}{E} = \sqrt{a^2 + \frac{b^2}{E} + \frac{c^2}{E^2}},
\end{equation}
where $\sigma_E$ is the uncertainty of the reconstructed energy distribution, $E$ is the peak of the distribution,and $a$, $b$ and $c$ are three parameters that quantify the contribution from spatial resolution of reconstructed energy, photon statistics, and PMT dark noise, respectively~\cite{bib:Eres_par}.
The parameters in Eq.~\ref{eq:resolution} were studied by fitting the energy resolution of the calibration sources as well as IBD and spallation neutrons, uniformly distributed in GdLS.
The internal radioactive alpha sources were used to cross-check the result.
Naked gamma sources are also simulated for comparison, and they have better energy resolution than the calibration data because they do not include the source shielding and calibration source deployment apparatus.
The best fit parameters are $a=0.016$, $b=0.081$~MeV$^{\frac{1}{2}}$ and $c=0.026$~MeV when the energy is given in the units of MeV.
A variation of the parameters within the uncertainties has negligible effects on the prompt spectrum when it is smeared, therefore the uncertainty of energy resolution is neglected in the analysis.

\subsubsection{Energy Response Matrix}
After taking into account the above effects, the detector response matrix (Eq.~\ref{eq:energy_response}) can be constructed to map the reconstructed energy to the antineutrino energy.
\begin{center}
\includegraphics[width=\columnwidth]{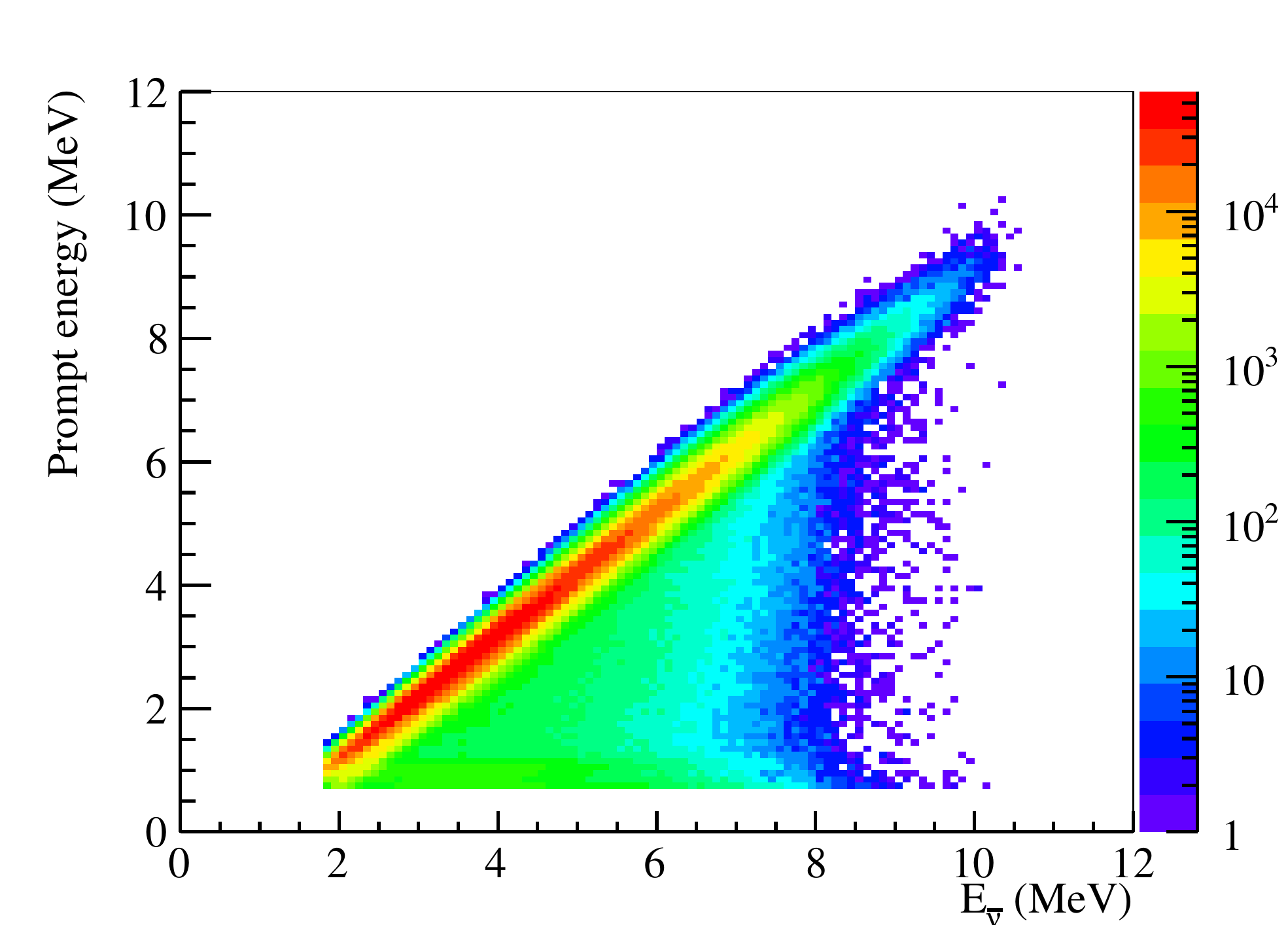}
\figcaption{
  The detector response matrix used to map antineutrino energy to the reconstructed energy.
  The IBD energy shift, IAV effect, non-linearity, and energy resolution are included.}
\label{fig:matrix}
\end{center}

Two methods were used to evaluate the energy response matrix. The first method estimates the IAV effect, non-linearity, and energy resolution step-by-step using analytical methods as described above.
The second method constructs the response matrix using a full-detector simulation based on Geant4~\cite{bib:Geant4}.
The detector geometry and material properties used in simulation are precisely determined by the various surveys and standalone measurements.
As an example, the thickness of the inner acrylic is measured with a precision of 0.4~mm, which allows for a small uncertainty of the IAV effect.
The Birks' constant of ionization quenching is tuned by benchmark data using a small sample of Daya Bay GdLS, and by comparing non-linearity from calibration sources in the ADs between data and MC.
The energy calibration and reconstruction process of MC data follows the same procedure as applied to the measured data.
Figure~\ref{fig:matrix} shows the detector response matrix which is constructed using the map of reconstructed energy and the input antineutrino energy in MC.

Both methods produced consistent response matrices for the prompt energy above $1.25$~MeV.
The uncertainty below 1.25~MeV was inflated to cover the difference of 10\% between the two methods.

\subsection{Spectral Comparison}
To quantify the discrepancy between the measured and predicted spectra, the uncertainties in both spectra were estimated.
Besides the statistical uncertainty, the systematic uncertainties include reactor-related uncertainty, detector-related uncertainty and background-related uncertainty.
The reactor related uncertainty  presented in Sec.~\ref{sec:flux} is propagated to the prompt energy spectrum when converting the antineutrino energy spectrum to the prompt energy spectrum.
The uncertainty of the detection efficiency is assumed to be independent of energy, and therefore does not impact the spectral shape.
The uncertainty of the IAV effect on the prompt energy spectrum is 4\% below 1.25~MeV and rapidly drops to 0.1\% above 1.25~MeV.
The uncertainty of non-linearity shown as the error band in Fig.~\ref{fig:non-linearity} is propagated to the prompt energy spectrum when applying the non-linearity effect to generate the predicted spectrum.
Five major sources of background are identified in the Daya Bay detectors. They are the accidental background, cosmogenic $^9$Li and $^8$He beta-decays, fast neutrons, Am-C neutron sources, and $^{13}$C($\alpha$, n)$^{16}$O reactions.
The background uncertainty is incorporated when subtracting the background from the measured spectrum.

To incorporate statistical, reactor-related, detector-related and background-related uncertainties, a covariance matrix $V$ was constructed as
\begin{equation}
  \label{eq:kinetic_positron}
  V = V^{stat} + V^{sys},
\end{equation}
where $V^{stat}$ is the statistical component, and $V^{sys}$ is the shape-only systematic component.
The statistical component has only diagonal terms and is calculated analytically.
Large samples of prompt spectra were generated to include the fluctuation due to various systematic uncertainties from the reactor, detector energy response, and background uncertainties. The elements in the covariance matrix of the systematic component were calculated as
\begin{equation}
  \label{eq:kinetic_positron}
  V^{sys}_{ij} = \frac{1}{N^{expts}}\sum^{N^{expts}}(N_i^{ran}-N_i^{nom})(N_j^{ran}-N_j^{nom}),
\end{equation}
where $N^{expts}$ is the number of toy MC samples, $N_i^{ran(nom)}$ is the random (nominal) predicted number of events at the prompt energy bin $i$.
Total number of events in the random predicted spectra are normalized to the nominal prediction.
Finally, the total covariance matrix was calculated by summing these two components, $V = V^{stat} + V^{sys}$.
Figure~\ref{fig:uncertainty} shows the elements of the correlation matrix, $V_{ij}/\sqrt{V_{ii}V_{jj}}$, and the fractional size of the diagonal elements of the covariance matrix, $V_{ii}/N_i^{pred}$, for each component.
The uncertainty is dominated by the reactor and the detector systematic uncertainties.

\begin{center}
\includegraphics[width=0.9\columnwidth]{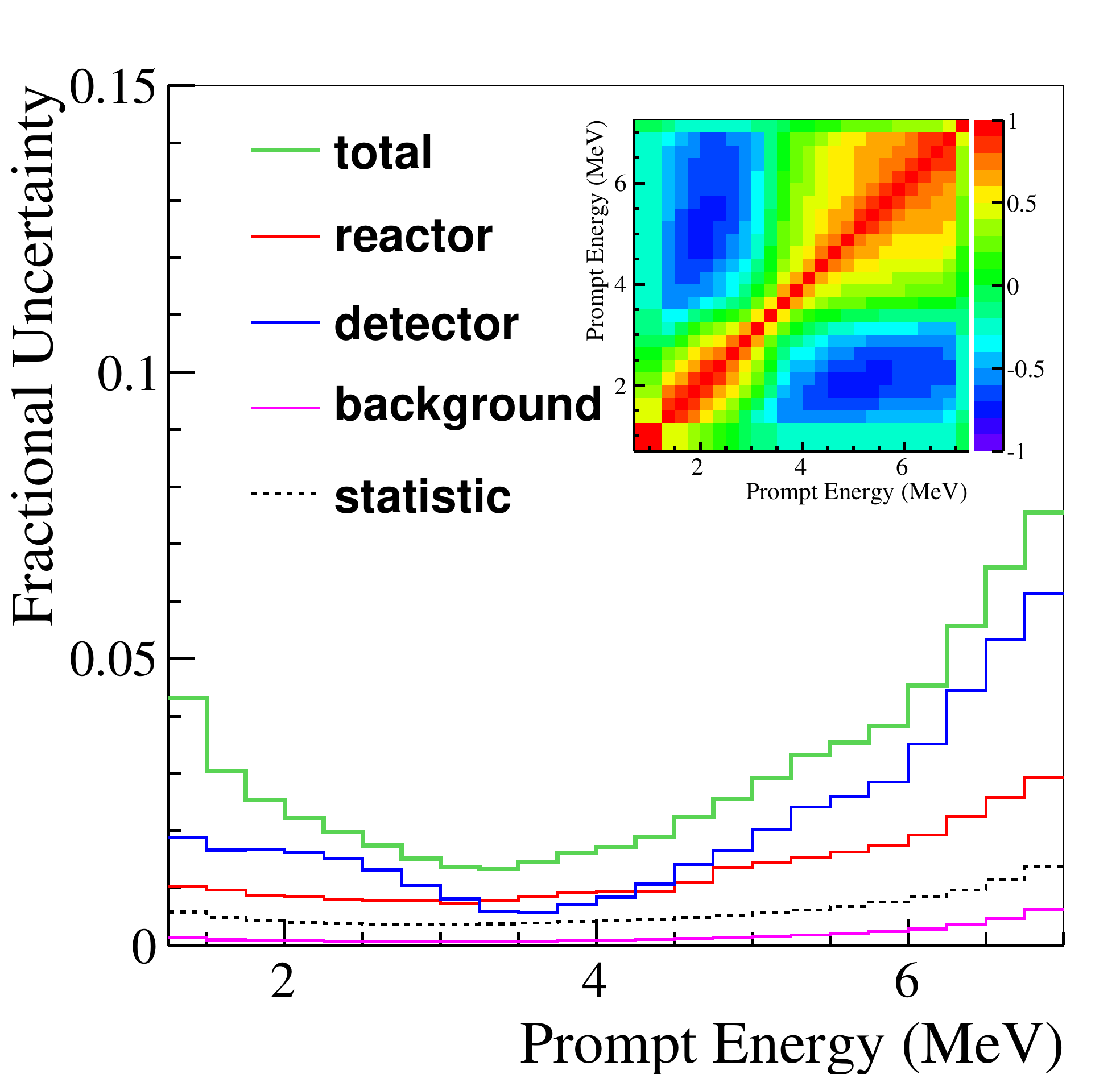}
\figcaption{The fractional size of the diagonal elements of the covariance matrix, $V_{ii}/N_i^{pred}$, for each component in each prompt energy bin.
Inset: the elements of the correlation matrix, $V_{ij}/\sqrt{V_{ii}V_{jj}}$ for the total uncertainty.
}
\label{fig:uncertainty}
\end{center}

A $\chi^2$ was defined to test the compatibility of the observed prompt energy spectrum with the predictions,
\begin{equation}
  \label{eq:chi2_prompt}
  \chi^2 = \sum_{i,j}(N^{\mathrm{obs}}_i - N^{\mathrm{pred}}_i)(V^{-1})_{ij}(N^{\mathrm{obs}}_j - N^{\mathrm{pred}}_j),
\end{equation}
where $N^{\mathrm{obs(pred)}}_i$ is the observed(predicted) number of events at the $i$-th prompt energy bin and $V$ is the covariance matrix that includes all the statistical and systematic uncertainties.
Figure~\ref{fig:prompt}A shows a comparison of the observed near-site prompt energy spectrum with the prediction.
The predicted spectrum was normalized to the measurement.
A clear discrepancy between the data and the prediction near 5 MeV is observed, while the agreement is reasonable in other energy regions.
A comparison to the Huber+Mueller model yields a $\chi^2/dof$ of 46.6/24 in the full energy range from 0.7 to 12 MeV, corresponding to a 2.9~$\sigma$ discrepancy.
The ILL+Vogel model shows a similar level of discrepancy from the data.

Another compatibility test was performed with a modified fitting algorithm.
In this method, $N$(=number of prompt energy bins) free-floating nuisance parameters are introduced to the oscillation parameter fit to adjust the normalization for each bin, as described in~\cite{bib:prl_shape}. The compatibility was tested by evaluating
\begin{equation}
\label{eq:mod_chi2}
\Delta \chi^2 = \chi^2(\mathrm{standard}) - \chi^2(N~\mathrm{\ extra\ parameters})
\end{equation}
for $N$ degrees of freedom.
We obtained $\Delta \chi^2/N = 50.1/25$, which is consistent with the results obtained by the first method using Eq.~\ref{eq:chi2_prompt}.

\begin{center}
\includegraphics[width=\columnwidth]{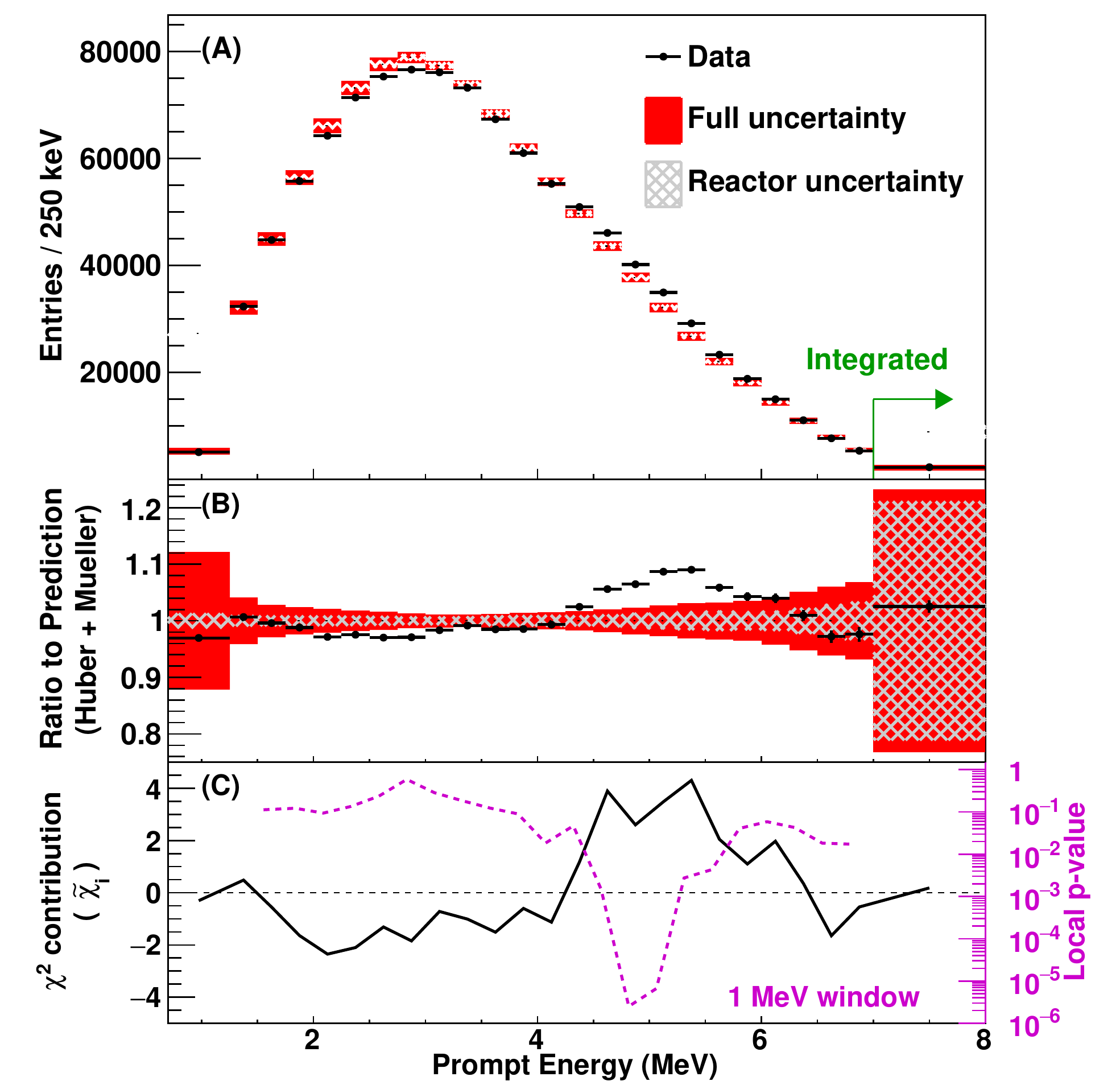}
\figcaption{(A) Comparison of predicted and measured prompt energy spectra.
  The prediction is based on the Huber+Mueller model and normalized to the number of measured events.
  The error bars on the data points represent the statistical uncertainty.
  The hatched and red filled bands represent the square-root of diagonal elements of the covariance matrix ($\sqrt{(V_{ii})}$) for the reactor related and the full systematic uncertainties, respectively.
  (B) Ratio of the measured prompt energy spectrum to the predicted spectrum   (Huber+Mueller model).
  (C) The defined $\chi^2$ distribution ($\widetilde{\chi}_i$) of each bin (black solid curve) and local p-values for 1~MeV energy windows (magenta dashed curve).
  See Eq.~\ref{eq:chi2_prompt2} and relevant text for the definitions.}
\label{fig:prompt}
\end{center}

\subsection{Quantification of the Local Deviation}
The ratio of the measured to predicted energy spectra is shown in Fig.~\ref{fig:prompt}B.
The spectral discrepancy around 5 MeV prompt energy is clearly visible.
Two approaches are adopted to evaluate the significance of this discrepancy.
The first method evaluates the $\chi^{2}$ contribution of each energy bin,
\begin{align}
\begin{split}
  \label{eq:chi2_prompt2}
\widetilde{\chi}_i & = \frac{N^{\rm obs}_i - N^{\rm pred}_i}{|N^{\rm obs}_i - N^{\rm pred}_i|} \sqrt {\sum_{j}\chi^2_{ij}}, \\
\chi^2_{ij} &= (N^{\rm obs}_i - N^{\rm pred}_i)(V^{-1})_{ij}(N^{\rm obs}_j - N^{\rm pred}_j).
\end{split}
\end{align}
By definition, $\sum_i \widetilde{\chi}_i^2$ is equal to the value of $\chi^2$ defined in Eq.~\ref{eq:chi2_prompt}.
As shown in Fig.~\ref{fig:prompt}C, an enhanced contribution is visible around 5 MeV.

In the second approach, the significance of the deviation is evaluated based on the modified oscillation analysis similar to Eq.~\ref{eq:mod_chi2}.
Instead of allowing all the $N$ nuisance parameters to be free floating, only parameters within a selected energy window are varied in the fit.
The difference between minimum $\chi^2$s before and after introducing these nuisance parameters within the selected energy window was used to evaluate the p-value of the local variation from the predictions.
The p-values with 1~MeV sliding energy window are shown in Fig.~\ref{fig:prompt}C.
The local significance for a discrepancy is greater than $4~\sigma$ at the highest point around 5~MeV.
In addition, the local significance for the 2~MeV window between 4 and 6 MeV were evaluated.
We obtained a $\Delta \chi^2/N$ value of 37.4/8, which corresponds to the p-value of $9.7\times 10^{-6} (4.4~\sigma)$.
Comparing with the ILL+Vogel model shows a similar level of local discrepancy between 4 and 6 MeV.

The excess between 4 and 6 MeV was $\sim$1.5\% of the total observed IBD candidates.
An excess of events in a same energy range was not observed in the spallation $^{12}$B beta decay spectrum, ruling out detector effects as an explanation.
Adding a simple beta-decay branch or a mono-energetic peak cannot reproduce the observed excess, indicating that it cannot be explained by a simple background contribution.
Contributions from other interaction channels (e.g. $\bar{\nu}_{e}$+$^{13}$C) were investigated and were found to be too small to account for the excess.
The events in the energy region around 5~MeV are carefully examined: the neutron capture time, the delayed energy spectrum, and the distance distribution for the delayed neutron capture signal were found to match IBD event characteristics.
The vertex distribution of the prompt signal was found to be uniform and consistent with IBD events.

Figure~\ref{fig:4-6MeVexcess} shows the event rate versus time in the energy window of 4.5-5.5~MeV and other windows.
The strong correlation indicates that the excess around 5~MeV is proportional to the reactor antineutrino flux.
Therefore, it strongly suggests that the deviation is due to the imperfect modelling of the reactor antineutrino spectrum.
A recent ab initio calculation of the antineutrino spectrum showed a similar deviation from previous predictions in the 4-6 MeV energy region~\cite{bib:dan}, and identified prominent fission daughter isotopes as a potential explanation.
Similar discussions can be found in Ref.~\cite{bib:sonzogni}.
Furthermore, a recent evaluation of uncertainties in forbidden decays suggests an additional $\sim$5\% uncertainty in both the rate and spectral shape of reactor antineutrino flux models using beta-to-antineutrino conversions~\cite{bib:hayes}, which may be another source of the discrepancy.

\begin{center}
\includegraphics[width=\columnwidth]{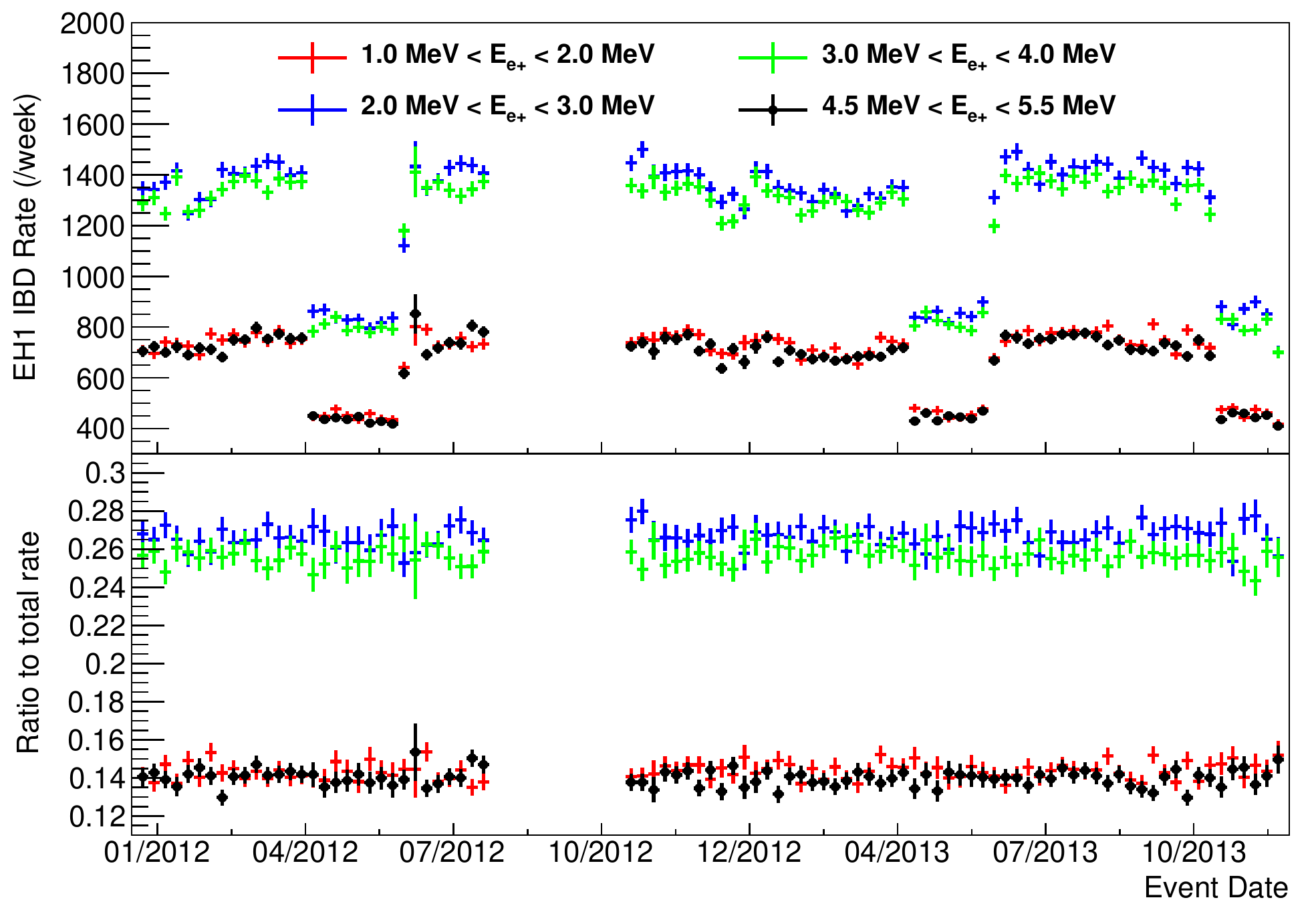}
\figcaption{History of the event rates in the region of the excess (4.5 MeV$<E_{e^+}<5.5$ MeV, black) and outside this region are shown in the top panel.
The fractional rates of the various energy windows are shown in the lower panel.
The flat distribution indicates that the event rate in the region of 4.5-5.5~MeV is proportional to the reactor antineutrino flux.
}
\label{fig:4-6MeVexcess}
\end{center}

\section{Generic Antineutrino Spectrum of IBD Reactions}\label{sec:generic_spectrum}
Since the predicted and measured prompt energy spectra have some discrepancies, it can be useful to extract an antineutrino spectrum weighted by the IBD cross section using the measured prompt energy spectrum at Daya Bay.
The unfolded antineutrino spectrum can be used as a model-independent input for reactor antineutrino flux and spectra prediction for future reactor antineutrino experiments.

\subsection{Solutions of Linear Inverse Problems}
A measured quantity (such as the energy spectrum) usually includes various detector effects, such as finite energy resolution and the limited acceptance of the detector.
Therefore, a correction or a transformation of the measured spectrum is necessary to deduce the true energy spectrum without the specific detector effects.
The result allows for a direct comparison with other experiments and theoretical predictions.
The transformation from measured to true spectrum is called unfolding and belongs to the class of linear inverse problems.
It requires a thorough understanding of the detector physics response.
Due to the properties of the detector response, e.g. finite energy resolution, the inverse problem is ill-posed: a small fluctuation of the measured spectrum can result in a large change of the unfolded result.
Non-proper solution by simple inversion yields unstable results.
The singular value decomposition (SVD) regularization method~\cite{bib:svd}, i.e.  the standard unfolding method for the linear inverse problem, discussed below, is used for obtaining the antineutrino spectrum at Daya Bay.
Alternative methods are also discussed and the results are cross-checked.

\subsection{Unfolding with Different Methods}

\subsubsection{SVD and Generalized Inverse of Response Matrix}
The solution of a linear inverse problem of the type $ A x = y $ (where $A$ is the detector response matrix, $x$ is the true distribution vector of $n$ dimensions, $y$ is the measured distribution vector of $m$ dimensions) requires the construction of the generalized inverse matrix, $A^\#$ ($A^{-1}$ in the case of square matrix).
The case of $m=n$ can be solved simply by matrix inversion.
In practice, $y$ has statistical fluctuations, and a simple matrix inversion results in large fluctuations in $x$ and negative correlations between bins of $x$.
One method, along the idea of a least squares fit, is to minimize
\begin{equation}\label{equ_ls}
||Ax - y||^2  = (Ax - y)^T V_y^{-1} (Ax - y),
\end{equation}
where $V_y$ is the covariance matrix for the measurement of $y$.
A larger dimension for the measurement, i.e. $m>n$, would lead to a more precise solution.

The SVD method is an orthogonalization method applied to the $m$-by-$n$ matrix $A$.
As a prerequisite, a scaling process is applied to the equations $A x = y$: the rows of $A$ and $y$ are both divided by the error vector of $y$. (The scaled matrix and vector are still written as $A$ and $y$ for convenience.)
The SVD of the $m$-by-$n$ matrix A with $m>n$ is expressed:
\begin{equation}\label{equ_svd}
A = U\Sigma V^T = \sum^n_{i=1} u_i \sigma_i v_i^T,
\end{equation}
where $U$ and $V$ are $m \times m$ and $ n \times n$ orthogonal matrices ($U^T U = I$, $V^T V= I$), $u_i$ and $v_i$ are the corresponding vectors, while $\Sigma$ is an $m\times n$ diagonal matrix with non-negative diagonal elements. The singular values $\sigma_i$ of $\Sigma$ are ordered and positive. 
Using the SVD matrices $U$, $\Sigma$ and $V$, a generalized inverse $A^\#$ of the matrix A can be defined by
\begin{equation}\label{equ_svd_2}
A^\# = V\Sigma^{-1} U^T
\end{equation}
with $A^\#A =  V\Sigma^{-1} U^T \dot U\Sigma V^T = I$.
Multiplying the equation $Ax =y$ by the generalized inverse $A^\#$, the vector $x$ is obtained by
\begin{equation}\label{equ_svd_solu}
x = (V\Sigma^{-1}U^T) y = \sum_{j=1}^n \frac{1}{\sigma_j} c_j v_j
\end{equation}
with coefficient $c_j = y^T u_j$. 
The covariance matrix for the estimate of $x$ is given by
\begin{equation}\label{equ_svd_solu_2}
V_x = \sum_{j=1}^n \frac{1}{\sigma_j^2} v_j v_j^T.
\end{equation}

For the case of an $n$-by-$n$ matrix A, the orthogonalization is a simple diagonalization:
\begin{eqnarray}\label{equ_square}
(A^T V_y^{-1}A)x&=&(A^T V_y^{-1}y), \\ \nonumber
Cx&=&b,
\end{eqnarray}
where $C = (A^T V_y^{-1}A)$ and $b=(A^TV_y^{-1})y$.
$C$ is then diagonalized
\begin{equation}\label{equ_square_c}
C = U\Lambda U^T,
\end{equation}
where $\Lambda$ is a diagonal matrix with eigenvalues $\lambda_j$.
The final solution after transformation is
\begin{equation}\label{equ_svd_square_x}
x = U\Lambda^{-1/2}(\Lambda^{-1/2} U^T) b = \sum_{j=1}^n \frac{1}{\sqrt {\lambda_j}} c_j u_j
\end{equation}
with coefficient $c_j =1/\sqrt{\lambda_j}(b^T u_j)$.
The corresponding expression for the covariance matrix $V_x$ is given by
\begin{equation}\label{equ_svd_square_vx}
V_x =  \sum_{j=1}^n \frac{1}{\lambda_j} u_j u_j^T.
\end{equation}

However, due to the properties of the response matrix $A$, the singular values (or eigenvalues) typically span many orders of magnitude.
The larger singular values represent the dominant components of the detector response matrix, however the small singular values would dominate the result if included in the solution.
The standard technique to reduce or suppress the contribution from the smaller eigenvalues is the regularization method, which does not introduce a bias if the regularization parameter is well defined and the response matrix $A$ is known.

For the Daya Bay experiment, the inputs of the linear inverse problem are the measured prompt energy spectra and the detector response matrix.
The measurement vector $y$ is the sum of the prompt energy spectra of the four near-site ADs weighted by their target mass relative to average target mass ($M_n$) of all ADs:
\begin{equation}
 S_\textrm{combined}(E_{prompt})  = \sum_{i=1}^4 S_i(E_{prompt})M_n/M_i, \label{eq:S_combined}
\end{equation}
where $E_{prompt}$ is the bin center of the prompt energy spectra at each bin.
The covariance matrix $V_y$ is composed of the statistical, systematic and background uncertainties described in Sec. \ref{sec:abs_spectrum}.
The response matrix $A$ is constructed by either of the two methods as described in Sec. \ref{sec:abs_spectrum}.

With the SVD method, the linear inverse problem is solved by a linear transformation of the measured prompt energy spectra.
This transformation is realized by the generalized inverse $A^\#$ of the response matrix $A$.
The construction of this generalized inverse allows the use of the standard method for propagating uncertainties.
The resulting covariance matrix of the unfolded result necessarily describes the correlations between bins of the unfolded spectrum.

\subsubsection{Regularization Method}
The exact solution of the linear system is equivalent to the minimization of
\begin{equation}\label{equ_reg_svd}
\chi^2(x) = (Ax - y)^T V_y^{-1} (Ax - y).
\end{equation}
A simple method of regularization is the truncation of the diagonalized matrix to exclude huge $1/\sigma_i$ components in solution $x$, which is equivalent to ignoring the insignificant components of the detector response matrix.
Though simple truncation is better than keeping all $j$, this introduces biases which are difficult to control.
One of the usual choices, used in high energy physics, is requiring that the regularized solution be smooth.
Technically, this requirement is introduced into the $\chi^2$ minimization condition by adding an extra term\cite{bib:svd,bib:blobel}:
\begin{equation}\label{equ_reg_svd}
\chi^2(x) = (Ax - y)^T V_y^{-1} (Ax - y) + \tau (C x)^T Cx.
\end{equation}

The parameter $\tau$ plays the role of the Lagrange multiplier in the new conditional minimization problem.
A small value of $\tau$ has a weak regularization effect, the correlations between bins remain mainly negative, and the result is still dominated by a large statistical fluctuation.
A very large value of $\tau$ reduces the statistical fluctuation but introduces positive correlations between bins of the solution.

For the regularization method, it is important to choose a proper regularization parameter $\tau$.
Usually Monte Carlo samples with different statistical and systematic uncertainties are analyzed to obtain the optimal regularization parameter.
For unfolding the measured prompt energy spectrum with the same statistics as our data set ($\sim$ 1 million IBD events), an unfolding package based on RooUnfold\cite{bib:rooUnfold}, and TSVD in ROOT is developed and used.
In RooUnfold, a positive integer, $k$, acts as the regularization parameter.
To determine a proper $k$ value, different toy MC samples of prompt energy spectra were generated by folding different true antineutrino energy spectra with one detector response matrix.
Two nominal true antineutrino energy spectra were constructed: one with the Huber+Mueller model and the other with the `model' of a recent ab initio calculations~\cite{bib:dan}.
The true antineutrino spectrum samples were generated by varying the nominal spectrum according to its systematic uncertainties.
Each sampled true spectrum was folded with the detector response matrix to become a true prompt energy spectrum.
Then, each bin was varied with its statistical uncertainty.
The energy range and number of bins of the sampled prompt energy spectra were the same as the measured spectra. A least squares method was defined to determine the regularization parameter $k$:
\begin{equation}\label{equ_reg_svd}
\chi^2 = \sum_{i=1}^n\frac{(x^i_{true} - x^i_{unfold})^2} {V_x^{ii}},
\end{equation}
where $x^i_{true}$ is the bin content of the true antineutrino spectrum, $x^i_{unfold}$ is the bin content of the unfolded spectrum, which was generated from the corresponding $x_{true}$, and $V_x^{ii}$ is the $i$-th diagonal element of the covariance matrix given by the unfolding.
A $k$ value scan for $x_{unfold}$ was carried out during unfolding to find the minimum $\chi^2 $.
The best $k$ value was found to be 15, for the $\sim$ 1 million IBD candidates, and when the nominal true antineutrino distribution was constructed with the Huber+Mueller model.
The best $k$ value is not sensitive to the choice of the reactor antineutrino spectrum model, but becomes smaller as the size of the sample increases.

Once the regularization parameter $k$ was determined with the input of the measured prompt spectrum of Eq.~\ref{eq:S_combined}, its full covariance matrix, and the detector response matrix from Geant4 MC simulation, the true antineutrino energy spectrum $S_\textrm{combined}(E)$ and its covariance matrix $V_\textrm{combined}(E)$ were obtained by unfolding with the SVD regularization method, shown in Fig~\ref{fig:unfolded_spectrum}.

\begin{center}
\includegraphics[width=\columnwidth]{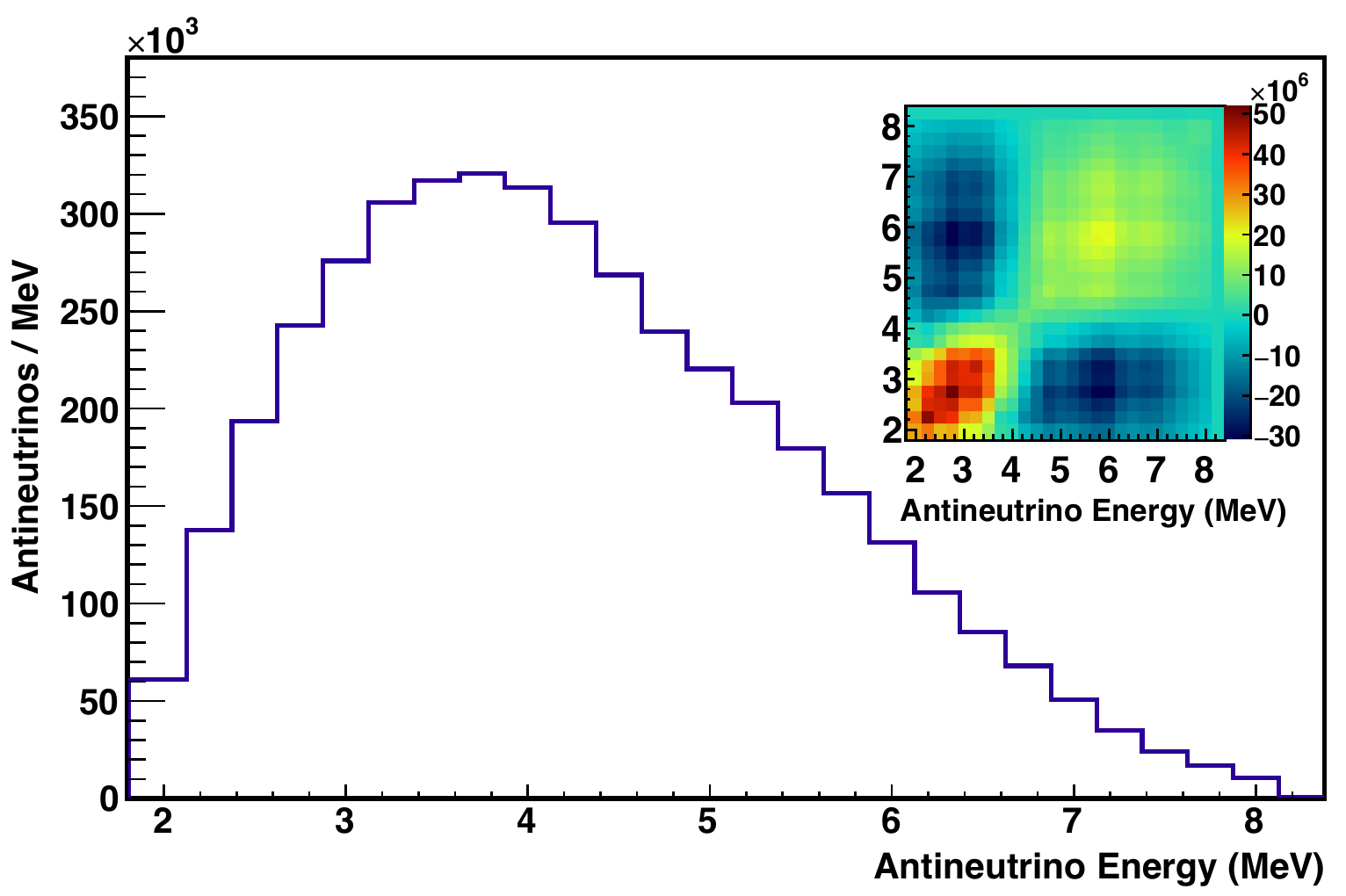}
\figcaption{The unfolded antineutrino energy spectrum using the combined prompt energy spectrum in the four near site ADs, and its covariance matrix (inset). The content of last bin is the integral up to 12 MeV.
      }
\label{fig:unfolded_spectrum}
\end{center}

\subsubsection{Bias Estimation}
An important requirement of unfolding is to minimize the bias.
The bias of each bin is defined as the average difference between the true and unfolded distribution sample pairs, written as $<|\bold{x_t}$ - $\bold{x_u}|>$.
True antineutrino spectrum samples were generated and used for the bias estimation, including different reactor antineutrino models, different statistics, different bin numbers, and different sample sizes.
Bias estimation was processed with the same bin width and statistics of the experimental data.
The bias of each bin is illustrated in Fig.~\ref{fig:bias}.
Between 2.75 and 6.5 MeV, the bias is $0.5\%$, which is comparable to the statistical uncertainty.
The bias increases outside this region due to the lower statistics.

\begin{center}
\includegraphics[width=\columnwidth]{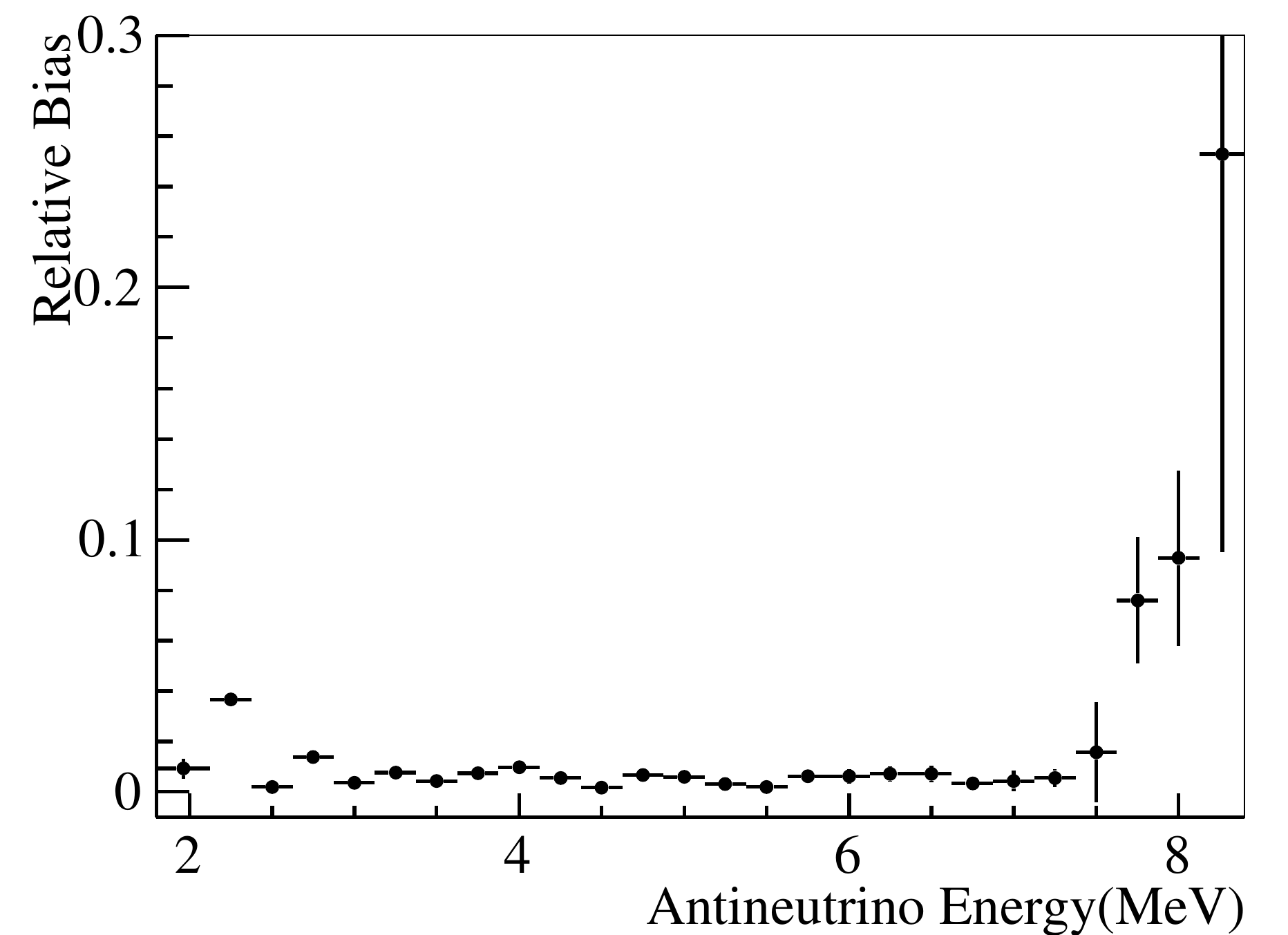}
\figcaption{Bias of unfolding in each bin with SVD regularization method.}
\label{fig:bias}
\end{center}

To have zero bias during unfolding, the condition is that the detector response matrix is exactly known. As mentioned in Sec.~\ref{sec:abs_spectrum}, there are two methods to construct the response matrix: analytical and full Geant4 MC simulation.
With one antineutrino spectrum as input, the two matrices generate two prompt spectra with $0.5\%$ bin-by-bin differences.
We used both matrices during bias estimation to obtain conservative results; i.e., using one matrix for the folding process to generate the true prompt energy spectrum samples, and the other matrix for the unfolding process to obtain the corresponding antineutrino energy spectra.
If the folding and unfolding processes use the same matrix, the bias is reduced when the statistics of the samples increase, and is negligible for 1 million events.



\subsubsection{Iterative Methods}

Another common way to unfold is using iterative methods.
Iterative methods have advantages for obtaining the true distributions of multiple dimensions, but require a starting value.
One typical example is the Bayesian iterative method~\cite{bib:bayes}.
An initial guess of the antineutrino spectrum can be set as the starting value, which is updated iteratively by a calculation that takes into account the response matrix and the observed prompt spectrum.
The iteration is stopped when the change in the antineutrino spectrum is small enough.
This method has an implicit regularization property, i.e. the number of iterations is similar to the regularization parameter in the SVD regularization method.
The summed prompt spectrum of the four ADs was also unfolded by the Bayesian iterative method, with the response matrix obtained by the Geant4 MC simulation method.
Figure~\ref{fig:svd_byes} compares the unfolded antineutrino spectrum obtained with the Bayesian method and the SVD regularization method.
The two methods yield consistent results and the difference below 8~MeV is negligible compared with the spectrum uncertainty.

\begin{center}
\includegraphics[width=\columnwidth]{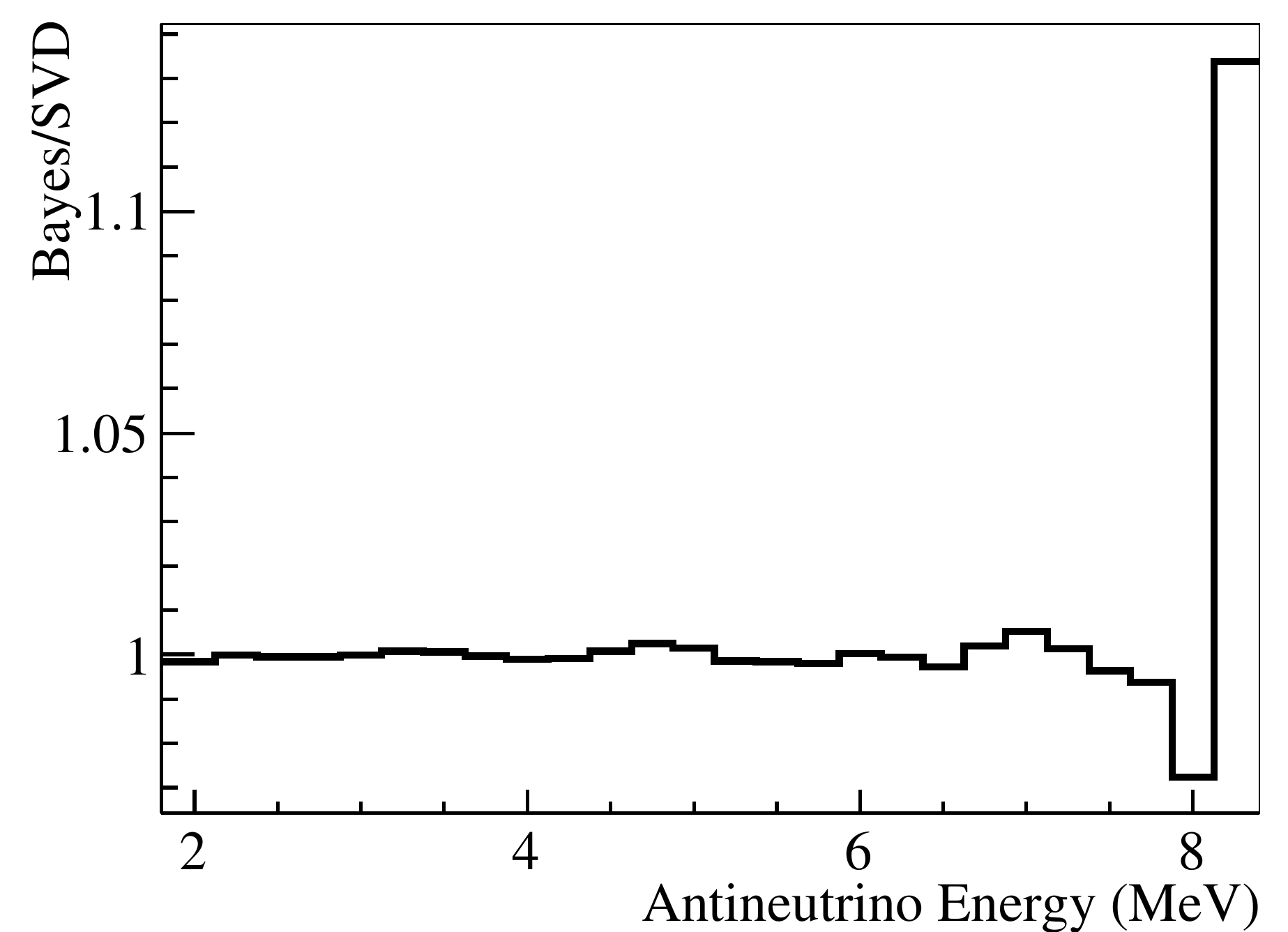}
\figcaption{Comparison of unfolding result with the SVD regularization method and the Bayesian iterative method.}
\label{fig:svd_byes}
\end{center}

\subsection{Antineutrino Spectrum Weighted by the IBD Cross Section}
\subsubsection{Normalization of Antineutrino Spectrum Weighted by the IBD Cross Section}
A generic reactor antineutrino spectrum for the IBD reaction was extracted from the measurement to provide a model-independent input for predicting reactor antineutrino flux and spectra.
The detector response effects were removed by unfolding the combined prompt spectrum $S_\textrm{combined}(E_{prompt})$ to an antineutrino spectrum $S_\textrm{combined}(E)$ for IBD reactions (Fig.~\ref{fig:unfolded_spectrum}).
Oscillation effects were also removed and each bin of the antineutrino spectrum was normalized to cm$^2$/fission/MeV using reactor information, which can be directly compared with the isotope spectra weighted by the IBD cross section.
The generic antineutrino spectrum is expressed as
\begin{equation}\label{equ_generic}
S_\textrm{generic} (E) = \frac{S_\textrm{combined}(E)} {P_\textrm{sur}(E,L) N_P F_\textrm{total}},
\end{equation}
where $P_\textrm{sur}(E,L )$ is the average survival probability of the $\bar\nu_e$ calculated with the fluxes from the six reactors to the four detectors,
$N_{P}$ is the number of target protons in the average target mass $M_n$,
and $F_\textrm{total}$ is a normalization factor based on the baseline-weighted total number of fissions.

The average survival probability $P_\textrm{sur}(E,L )$ is obtained by weighting the antineutrino contributions $B_{dr}$ from different reactors $r$
to each detector $d$,
\begin{equation}\label{equ_B_r}
B_{dr} = S_{dr}(E_{\nu})/S_d(E_{\nu}),
\end{equation}
where $S_{dr}(E_{\nu})$ and $S_d(E_{\nu})$ are the expected antineutrino spectrum with contributions from each reactor and from all reactors, calculated using Eq.~\ref{equ:singleReactor_prediction} and Eq.~\ref{equ:single_Reactor_prediction}.
The average survival probability $P_\textrm{sur}(E,L )$ is then
\begin{equation}\label{equ_effective_P}
P_\textrm{sur}(E,L ) = \frac{\sum_{d}S_d(E_{\nu})\sum_{r}B_{dr} P_{sur}^{dr}(E, L_{dr})} {\sum_d S_d(E_{\nu})},
\end{equation}
where $L_{dr}$ is the baseline between the $r$-th reactor to $d$-th detector;
$P_{sur}^{dr}(E, L_{dr})$ is the survival probability of antineutrinos after travelling from the $r$-th reactor to the $d$-th detector.
The oscillation probability is calculated using the oscillation parameters in the oscillation analysis of the same data set~\cite{bib:prl_shape2}.


The normalization factor $F_\textrm{total}$ for all four ADs contributed from six reactors is calculated as
\begin{equation}\label{equ_Ftotal}
F_\textrm{total} = \sum_d \sum_r \sum_t \frac{1}{4\pi L^2_{dr}}\frac{W^t_r}{\sum_i \alpha^t_{ir} e_i} \epsilon_d,
\end{equation}
where $W_r^t$ is the average power of the $t$-th week; $\alpha_{ir}^t$ is the weekly fission fraction of $i$-th isotope; $e_i$ is the fission energy; $\epsilon_d$ is the detection efficiency of each AD, which is the multiplication of the detection efficiency of $d$-th detector, which is the product of the IBD detection efficiency of all ADs, $\epsilon_0$= 80.6\%, the weekly multiplicity cut and muon veto efficiencies ($\epsilon_m$ and $\epsilon_\mu$), and the weekly live time.
The uncertainty of $N_P F_\textrm{total}$ is not dependent on the antineutrino energy and only contributes to the rate uncertainty of the generic spectrum. 
The rate uncertainty is $2.0\%$, which is contributed from the uncertainties of the efficiencies ($1.93\%$) listed in Table~\ref{tab:AllEffs}, the fission energy ($0.2\%$), and the reactor power and fission fractions (0.5\%).


From Eqs.~\ref{equ_generic}--\ref{equ_Ftotal}, the normalized reactor antineutrino spectrum measured at the two near sites is obtained.
The obtained generic antineutrino spectrum is shown in the top panel of Fig.~\ref{fig:generic}.
The values of the spectrum and the covariance matrix are shown in Tables~\ref{tab:generic1} and \ref{tab:generic2} in the appendix.
The middle panel of Fig.~\ref{fig:generic} is the ratio of the generic reactor antineutrino spectrum to the prediction using the isotope spectra of the Huber+Mueller model and the effective fission fractions listed in Table~\ref{tab:norm3m}.
The bottom panel of Fig.~\ref{fig:generic} shows the ratio of the spectrum obtained in the 6+8 AD period to that in the 6 AD period~\cite{prl_abs}.
The deviation of the ratio from one is due to the difference of fission fractions in the two data period and the statistic fluctuation.
The average deficit is equal to the overall flux deficit reported in Sec.~\ref{sec:abs_flux}.
The bump in the 5-7~MeV antineutrino energy corresponds to that in the 4-6~MeV prompt energy in Fig.~\ref{fig:prompt}.
The correlation matrix of the generic spectrum is obtained from its covariance matrix, which is calculated by both toy MC sampling method, and standard error propagation with matrices.
Figure~\ref{fig:covmatrix} shows the correlation matrix of the generic spectrum and its components for the energy-dependent uncertainties.

\begin{center}
\includegraphics[width=\columnwidth]{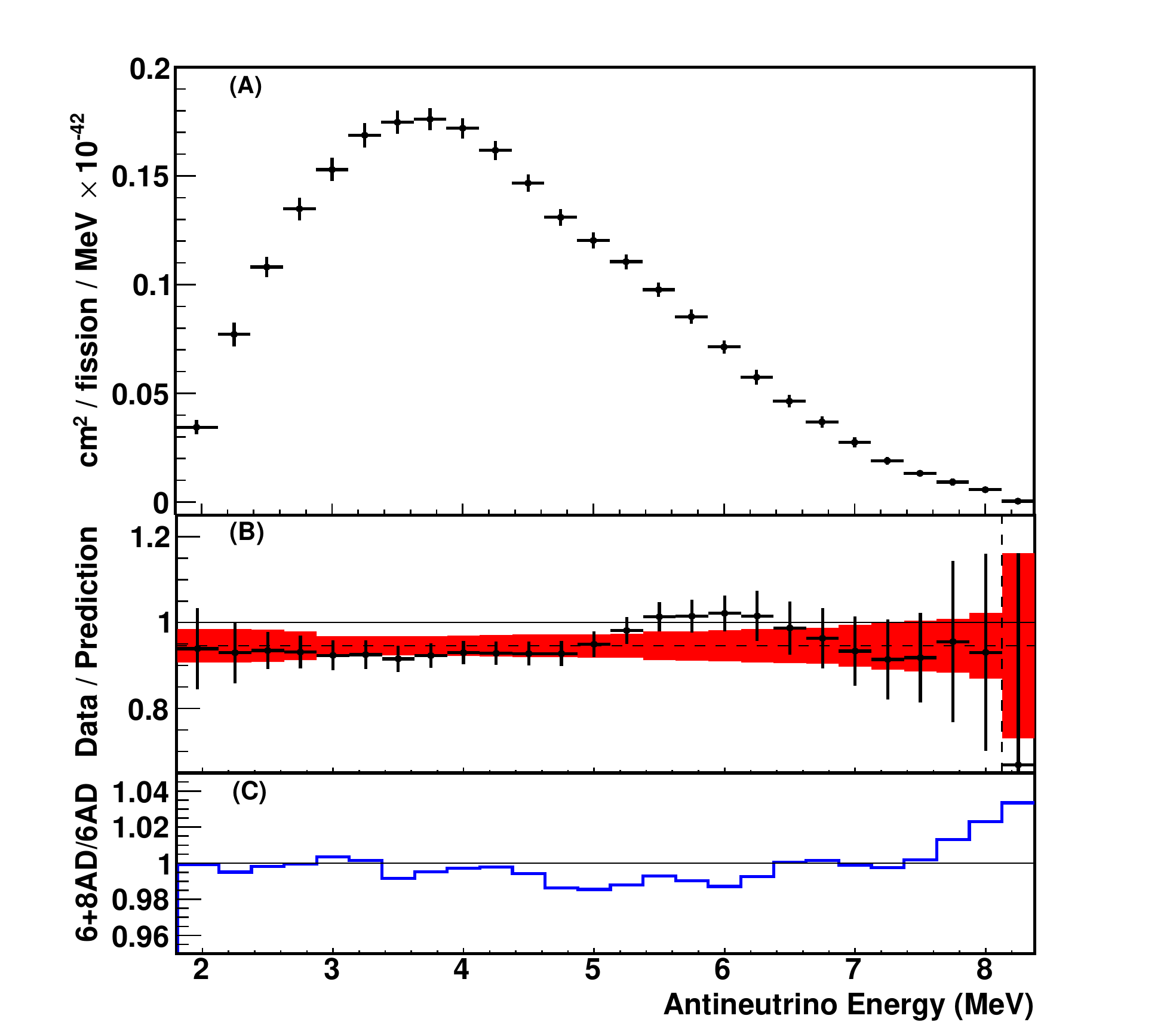}
\figcaption{(A) The antineutrino spectrum weighted by the IBD cross section. The last bin is integrated up to 12MeV. (B) Ratio of the extracted reactor antineutrino spectrum to the Huber+Mueller prediction. The error bars of the data points are the square-roots of the diagonal elements of the antineutrino spectrum covariance matrix. The solid red band represents the square-roots of the diagonal elements of the prediction covariance matrix, including both reactor and Huber+Mueller model uncertainties. (C) the ratio of the spectra from the 6+8 AD periods used in this analysis and the 6 AD period used in the previous
analysis~\cite{prl_abs}.}
\label{fig:generic}
\end{center}


\subsubsection{Possible Application of Generic Antineutrino Spectrum}
The generic antineutrino spectrum has been weighted by the IBD cross sections.
Other reactor neutrino experiments not utilizing the IBD reaction can remove the IBD weighting factor to obtain the antineutrino spectrum from the reactor.
IBD reaction experiments could directly use the generic spectrum to predict the antineutrino spectrum with IBD cross section $S_A$ in their experiment. A simplified example is:
\begin{equation}\label{equ_generic_app}
S_A = S_{dyb} + \sum_i { (f_{A_i} - f_{dyb_i}) S_{mod_i} },
\end{equation}
where $S_{dyb}$ is the generic spectrum from the Daya Bay, i.e. $S_{generic}(E)$, $f_{dyb}$  and $f_{A}$
are the effective fission fractions of the Daya Bay experiment and the reactor antineutrino experiment $A$;
and $S_{mod}$ are the isotope antineutrino spectra from models, such as ILL+Vogel, Huber+Mueller, etc.
$S_A$ could then replace the isotope spectra related part $\sum_i {S_{mod_i} e_i}$ in the calculation of the spectrum prediction presented in Sec.~\ref{sec:flux}.
The idea of this application depends on the condition that the effective fission fractions of different reactor antineutrino experiments, i.e.
$f_{dyb}$ and $f_{A}$ are small; therefore, corrections from $\sum_i { (f_{A_i} - f_{dyb_i}) S_{mod_i} }$ would be relatively small, and
$S_A$ will be dominated by the measurement result $S_{dyb}$ rather than reactor models.

\begin{center}
\includegraphics[width=\columnwidth]{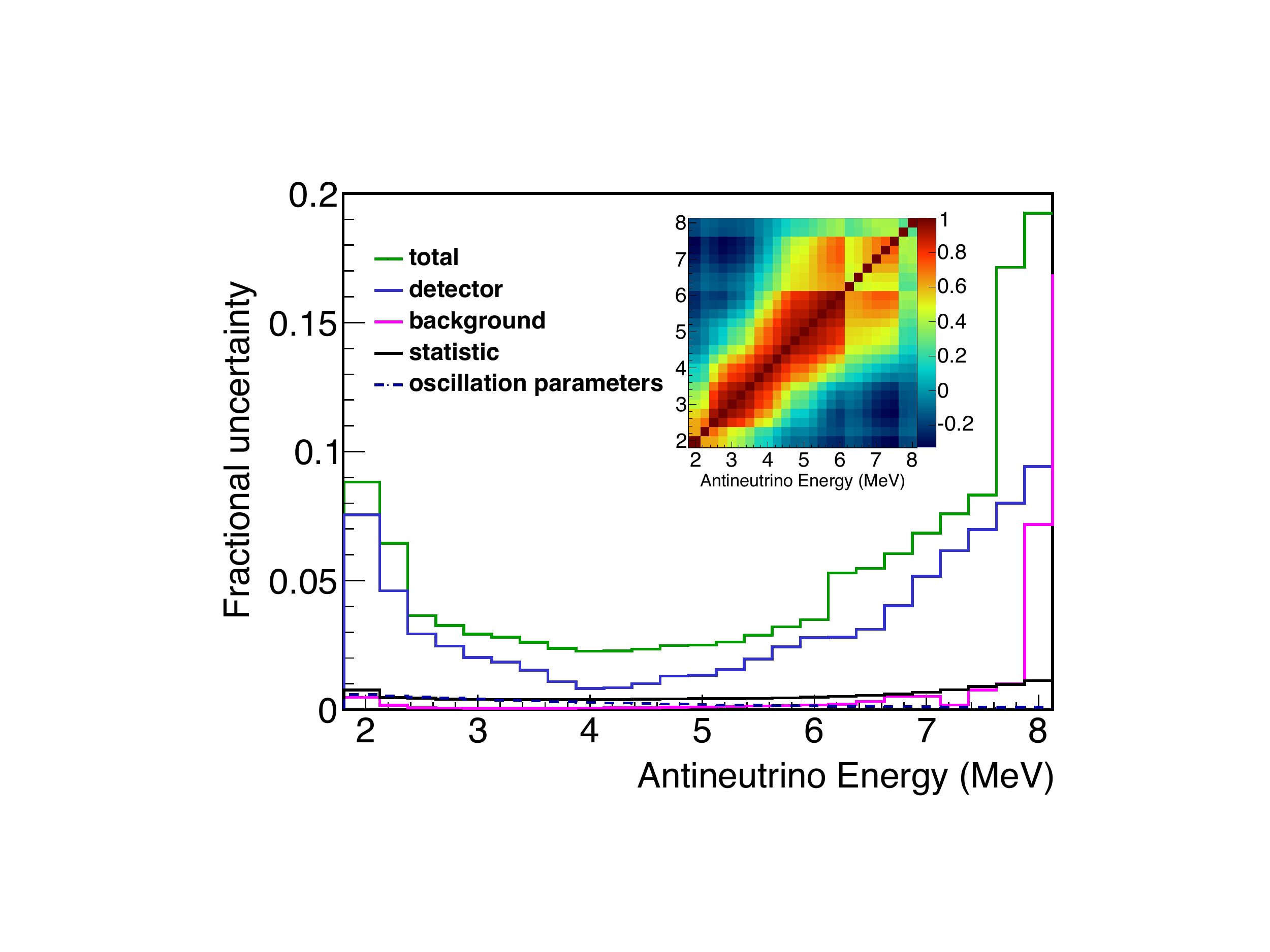}
\figcaption{Each component of the energy-dependent uncertainties for the generic spectrum. The inner plot shows the correlation matrix of the generic spectrum.}
\label{fig:covmatrix}
\end{center}

\section{Summary}\label{sec:summary}
After the final two detectors were installed in the Daya Bay experiment, an additional 404 days of data had been taken.
Including the previous 217 days of data taken by six ADs, more than 1.2 million IBDs were detected by the Daya Bay experiment.
The inverse beta decay (IBD) selection efficiency was found to be 80.6\% with a relative uncertainty of 1.93\% based on a detailed study of the detector performance.
The measured IBD yield is $(1.53 \pm 0.03) \times 10^{-18}$~cm$^2$/GW/day or $(5.91 \pm 0.12) \times 10^{-43}$~cm$^2$/fission.
The ratio of measured flux to the predictions is $0.946\pm0.020$ ($0.992\pm0.021$) for the Huber+Mueller (ILL+Vogel) model, which is consistent with the global average of previous short baseline experiments.
In addition, the predicted and measured spectra were compared, and a deviation of 2.9~$\sigma$ was found.
Particularly, an excess of events was found in the region of 4-6~MeV with a local significance of 4.4~$\sigma$.
Further investigation on the excess of events reveals possible problems in the reactor antineutrino flux predictions.
A reactor antineutrino spectrum weighted by the IBD cross section was extracted from the measurement at Daya Bay, providing a model-independent input for future reactor antineutrino experiments.

\section{Acknowledgement}
The Daya Bay Experiment is supported in part by
the Ministry of Science and Technology of China,
the United States Department of Energy,
the Chinese Academy of Sciences,
the CAS Center for Excellence in Particle Phsyics,
the National Natural Science Foundation of China,
the Guangdong provincial government,
the Shenzhen municipal government,
the China General Nuclear Power Group,
the Research Grants Council of the Hong Kong Special Administrative Region of China,
the MOST and MOE in Taiwan,
the U.S. National Science Foundation,
the Ministry of Education, Youth and Sports of the Czech Republic,
the Joint Institute of Nuclear Research in Dubna, Russia,
the NSFC-RFBR joint research program,
the National Commission for Scientific and Technological Research of Chile.
We acknowledge Yellow River Engineering Consulting Co., Ltd.\ and China Railway 15th Bureau Group Co., Ltd.\ for building the underground laboratory.
We are grateful for the ongoing cooperation from the China Guangdong Nuclear Power Group and China Light~\&~Power Company.

\section*{Appendix}
\end{multicols}
\newcommand{\tabincell}[2]{\begin{tabular}{@{}#1@{}}#2\end{tabular}}
\centering

\tabcaption{Generic antineutrino spectrum weighted by the IBD cross section. The spectrum is plotted in Panel A in Fig. \ref{fig:generic} .}
\footnotesize
\begin{tabular}{c|c}
\toprule
$\bar{\nu}_{e}$ Energy (MeV) & $cm^{2}/fission/MeV \times 10^{-46} $ \\\hline
1.8--2.125    & 344.19\\
2.125--2.375  & 770.96\\
2.375--2.625  & 1080.9\\
2.625--2.875  & 1348.4\\
2.875--3.125  & 1528.8\\
3.125--3.375  & 1687.0\\
3.375--3.625  & 1746.6\\
3.625--3.875  & 1760.6\\
3.875--4.125  & 1719.3\\
4.125--4.375  & 1617.6\\
4.375--4.625  & 1466.5\\
4.625--4.875  & 1309.3\\
4.875--5.125  & 1203.0\\
5.125--5.375  & 1105.4\\
5.375--5.625  & 976.50\\
5.625--5.875  & 852.31\\
5.875--6.125  & 713.19\\
6.125--6.375  & 573.90\\
6.375--6.625  & 463.54\\
6.625--6.875  & 368.70\\
6.875--7.125  & 274.56\\
7.125--7.375  & 190.00\\
7.375--7.625  & 132.08\\
7.625--7.875  & 92.114\\
7.875--8.125  & 56.689\\
8.125--12     & 4.0214\\

\bottomrule
\end{tabular}
\label{tab:generic1}

~\\
~\\
~\\
~\\
~\\
~\\
~\\
~\\
~\\
~\\
~\\
~\\
~\\
~\\
~\\

\centering
\tabcaption{Covariance matrix of antineutrino spectrum. Unit: $[cm^{2}/fission/MeV]^{2} \times 10^{-92}$.\protect\\
\ \ \ \ \ \ \ \ \ \ \ \ The square-roots of the diagonal elements are plotted in Panel A as error bars in Fig. \ref{fig:generic} .}
\scriptsize
\begin{tabular}{|l|p{0.8cm}|l|l|l|l|l|l|l|l|p{0.8cm}|l|l|p{0.8cm}|}
\hline
$\bar{\nu}_{e}$ Energy (MeV) & \tabincell{c}{1.8--\\2.125} & \tabincell{c}{2.125--\\2.375} & \tabincell{c}{2.375--\\2.625} & \tabincell{c}{2.625--\\2.875} & \tabincell{c}{2.875--\\3.125} & \tabincell{c}{3.125--\\3.375} & \tabincell{c}{3.375--\\3.625} & \tabincell{c}{3.625--\\3.875} & \tabincell{c}{3.875--\\4.125} & \tabincell{c}{4.125--\\4.375} & \tabincell{c}{4.375--\\4.625} & \tabincell{c}{4.625--\\4.875} & \tabincell{c}{4.875--\\5.125} \\
\hline
 1.8--2.125     & 920.6 & 933.9 & 731.6 & 725.7 & 584.5 & 543.9 & 412.8 & 216.3 & 171.3 & 35.29 & -49.44 & -97.78 & -85.43  \\\hline
 2.125--2.375   & 933.9 & 2471 & 1377 & 1317 & 1095 & 1049 & 808.1 & 473.2 & 393.9 & 166.2 & 23.04 & -66.35 & -56.99  \\\hline
 2.375--2.625   & 731.6 & 1377 & 1550 & 1601 & 1467 & 1483 & 1249 & 874.5 & 726.4 & 445.6 & 225.8 & 71.83 & 56.24  \\\hline
 2.625--2.875   & 725.7 & 1317 & 1601 & 1931 & 1838 & 1880 & 1673 & 1263 & 1048 & 711.3 & 409.6 & 191.3 & 154.8  \\\hline
 2.875--3.125   & 584.5 & 1095 & 1467 & 1838 & 2003 & 2016 & 1848 & 1504 & 1278 & 950.4 & 625.6 & 382.4 & 324.6  \\\hline
 3.125--3.375   & 543.9 & 1049 & 1483 & 1880 & 2016 & 2248 & 2013 & 1647 & 1423 & 1096 & 749.4 & 486.0 & 419.2  \\\hline
 3.375--3.625   & 412.8 & 808.1 & 1249 & 1673 & 1848 & 2013 & 2077 & 1718 & 1487 & 1227 & 913.3 & 658.6 & 578.4  \\\hline
 3.625--3.875   & 216.3 & 473.2 & 874.5 & 1263 & 1504 & 1647 & 1718 & 1750 & 1496 & 1305 & 1081 & 881.9 & 785.9  \\\hline
 3.875--4.125   & 171.3 & 393.9 & 726.4 & 1048 & 1278 & 1423 & 1487 & 1496 & 1512 & 1293 & 1088 & 931.8 & 847.8  \\\hline
 4.125--4.375   & 35.29 & 166.2 & 445.6 & 711.3 & 950.4 & 1096 & 1227 & 1305 & 1293 & 1354 & 1144 & 999.9 & 918.9  \\\hline
 4.375--4.625   & -49.44 & 23.04 & 225.8 & 409.6 & 625.6 & 749.4 & 913.3 & 1081 & 1088 & 1144 & 1179 & 1029 & 926.6  \\\hline
 4.625--4.875   & -97.78 & -66.35 & 71.83 & 191.3 & 382.4 & 486.0 & 658.6 & 881.9 & 931.8 & 999.9 & 1029 & 1058 & 919.5  \\\hline
 4.875--5.125   & -85.43 & -56.99 & 56.24 & 154.8 & 324.6 & 419.2 & 578.4 & 785.9 & 847.8 & 918.9 & 926.6 & 919.5 & 903.5  \\\hline
 5.125--5.375   & -118.8 & -93.59 & -4.512 & 53.22 & 200.6 & 284.5 & 439.7 & 659.6 & 737.0 & 835.8 & 870.2 & 860.6 & 815.7  \\\hline
 5.375--5.625   & -165.6 & -161.1 & -99.98 & -82.66 & 44.24 & 115.1 & 269.7 & 511.3 & 606.8 & 729.0 & 802.2 & 815.2 & 758.1  \\\hline
 5.625--5.875   & -190.7 & -207.5 & -175.5 & -190.5 & -85.11 & -26.42 & 123.0 & 375.8 & 486.9 & 624.3 & 721.6 & 760.3 & 707.6  \\\hline
 5.875--6.125   & -184.6 & -206.7 & -192.7 & -226.1 & -145.0 & -98.79 & 31.72 & 265.6 & 375.1 & 509.1 & 610.2 & 657.4 & 618.7  \\\hline
 6.125--6.375   & -143.5 & -153.2 & -144.9 & -177.9 & -119.3 & -83.83 & 17.02 & 201.7 & 291.3 & 399.5 & 482.9 & 521.4 & 493.2  \\\hline
 6.375--6.625   & -134.1 & -140.6 & -134.7 & -172.4 & -130.1 & -102.7 & -21.56 & 136.0 & 216.1 & 312.0 & 390.2 & 428.1 & 405.3  \\\hline
 6.625--6.875   & -151.2 & -166.4 & -164.3 & -213.0 & -183.7 & -163.0 & -92.45 & 58.06 & 138.7 & 236.0 & 321.2 & 366.3 & 348.1  \\\hline
 6.875--7.125   & -156.4 & -179.1 & -178.4 & -230.1 & -210.7 & -195.5 & -136.0 & -0.5791 & 74.43 & 166.3 & 250.1 & 296.6 & 283.5  \\\hline
 7.125--7.375   & -135.2 & -156.4 & -154.7 & -199.4 & -188.0 & -177.4 & -132.8 & -27.01 & 32.84 & 107.0 & 176.5 & 215.8 & 207.3  \\\hline
 7.375--7.625   & -109.3 & -125.4 & -121.1 & -156.0 & -149.9 & -142.4 & -111.2 & -33.25 & 11.71 & 67.80 & 121.6 & 152.4 & 147.0  \\\hline
 7.625--7.875   & -85.42 & -96.26 & -89.62 & -116.0 & -113.8 & -108.7 & -88.67 & -32.88 & 0.6093 & 42.19 & 83.63 & 107.3 & 104.2  \\\hline
 7.875--8.125   & -55.62 & -60.71 & -54.14 & -71.54 & -72.52 & -69.60 & -59.96 & -26.14 & -3.979 & 22.94 & 51.16 & 67.04 & 66.00  \\\hline
 8.125--12      & -3.999 & -4.169 & -3.533 & -4.848 & -5.135 & -4.954 & -4.530 & -2.244 & -0.5493 & 1.444 & 3.647 & 4.856 & 4.865  \\\hline

\end{tabular}
\begin{multicols}{2}
\end{multicols}

\centering
\scriptsize
\begin{tabular}{|l|p{0.8cm}|l|l|l|l|l|l|l|l|p{0.8cm}|l|l|p{0.8cm}|}
\hline
$\bar{\nu}_{e}$ Energy (MeV) & \tabincell{c}{5.125--\\5.375} & \tabincell{c}{5.375--\\5.625} & \tabincell{c}{5.625--\\5.875} & \tabincell{c}{5.875--\\6.125} & \tabincell{c}{6.125--\\6.375} & \tabincell{c}{6.375--\\6.625} & \tabincell{c}{6.625--\\6.875} & \tabincell{c}{6.875--\\7.125} & \tabincell{c}{7.125--\\7.375} & \tabincell{c}{7.375--\\7.625} & \tabincell{c}{7.625--\\7.875} & \tabincell{c}{7.875--\\8.125} & \tabincell{c}{8.125--\\12} \\\hline
 1.8--2.125   & -118.8 & -165.6 & -190.7 & -184.6 & -143.5 & -134.1 & -151.2 & -156.4 & -135.2 & -109.3 & -85.42 & -55.62 & -3.999  \\\hline
 2.125--2.375 & -93.59 & -161.1 & -207.5 & -206.7 & -153.2 & -140.6 & -166.4 & -179.1 & -156.4 & -125.4 & -96.26 & -60.71 & -4.169  \\\hline
 2.375--2.625 & -4.512 & -99.98 & -175.5 & -192.7 & -144.9 & -134.7 & -164.3 & -178.4 & -154.7 & -121.1 & -89.62 & -54.14 & -3.533  \\\hline
 2.625--2.875 & 53.22 & -82.66 & -190.5 & -226.1 & -177.9 & -172.4 & -213.0 & -230.1 & -199.4 & -156.0 & -116.0 & -71.54 & -4.848  \\\hline
 2.875--3.125 & 200.6 & 44.24 & -85.11 & -145.0 & -119.3 & -130.1 & -183.7 & -210.7 & -188.0 & -149.9 & -113.8 & -72.52 & -5.135  \\\hline
 3.125--3.375 & 284.5 & 115.1 & -26.42 & -98.79 & -83.83 & -102.7 & -163.0 & -195.5 & -177.4 & -142.4 & -108.7 & -69.60 & -4.954  \\\hline
 3.375--3.625 & 439.7 & 269.7 & 123.0 & 31.72 & 17.02 & -21.56 & -92.45 & -136.0 & -132.8 & -111.2 & -88.67 & -59.96 & -4.530  \\\hline
 3.625--3.875 & 659.6 & 511.3 & 375.8 & 265.6 & 201.7 & 136.0 & 58.06 & -0.5791 & -27.01 & -33.25 & -32.88 & -26.14 & -2.244  \\\hline
 3.875--4.125 & 737.0 & 606.8 & 486.9 & 375.1 & 291.3 & 216.1 & 138.7 & 74.43 & 32.84 & 11.71 & 0.6093 & -3.979 & -0.549  \\\hline
 4.125--4.375 & 835.8 & 729.0 & 624.3 & 509.1 & 399.5 & 312.0 & 236.0 & 166.3 & 107.0 & 67.80 & 42.19 & 22.94 & 1.444  \\\hline
 4.375--4.625 & 870.2 & 802.2 & 721.6 & 610.2 & 482.9 & 390.2 & 321.2 & 250.1 & 176.5 & 121.6 & 83.63 & 51.16 & 3.647  \\\hline
 4.625--4.875 & 860.6 & 815.2 & 760.3 & 657.4 & 521.4 & 428.1 & 366.3 & 296.6 & 215.8 & 152.4 & 107.3 & 67.04 & 4.856  \\\hline
 4.875--5.125 & 815.7 & 758.1 & 707.6 & 618.7 & 493.2 & 405.3 & 348.1 & 283.5 & 207.3 & 147.0 & 104.2 & 66.00 & 4.865  \\\hline
 5.125--5.375 & 837.3 & 766.6 & 714.6 & 629.1 & 506.2 & 423.0 & 372.3 & 310.1 & 230.8 & 166.4 & 120.6 & 78.65 & 5.983  \\\hline
 5.375--5.625 & 766.6 & 790.9 & 733.1 & 644.7 & 516.6 & 439.2 & 398.6 & 339.9 & 257.0 & 187.8 & 138.5 & 92.10 & 7.131  \\\hline
 5.625--5.875 & 714.6 & 733.1 & 748.5 & 654.2 & 519.9 & 443.6 & 412.2 & 359.2 & 275.4 & 203.4 & 152.0 & 102.7 & 8.074  \\\hline
 5.875--6.125 & 629.1 & 644.7 & 654.2 & 616.3 & 484.4 & 411.8 & 384.9 & 340.0 & 264.0 & 197.0 & 148.8 & 101.9 & 8.114  \\\hline
 6.125--6.375 & 506.2 & 516.6 & 519.9 & 484.4 & 923.8 & 339.8 & 315.4 & 277.6 & 216.2 & 162.3 & 123.7 & 85.79 & 6.924  \\\hline
 6.375--6.625 & 423.0 & 439.2 & 443.6 & 411.8 & 339.8 & 642.7 & 282.9 & 249.4 & 193.7 & 145.7 & 113.0 & 80.87 & 6.748  \\\hline
 6.625--6.875 & 372.3 & 398.6 & 412.2 & 384.9 & 315.4 & 282.9 & 495.8 & 251.0 & 195.6 & 147.4 & 116.3 & 86.13 & 7.439  \\\hline
 6.875--7.125 & 310.1 & 339.9 & 359.2 & 340.0 & 277.6 & 249.4 & 251.0 & 352.8 & 184.1 & 140.0 & 110.4 & 81.07 & 6.924  \\\hline
 7.125--7.375 & 230.8 & 257.0 & 275.4 & 264.0 & 216.2 & 193.7 & 195.6 & 184.1 & 208.0 & 116.7 & 90.36 & 61.83 & 4.838  \\\hline
 7.375--7.625 & 166.4 & 187.8 & 203.4 & 197.0 & 162.3 & 145.7 & 147.4 & 140.0 & 116.7 & 120.5 & 71.22 & 46.29 & 3.350  \\\hline
 7.625--7.875 & 120.6 & 138.5 & 152.0 & 148.8 & 123.7 & 113.0 & 116.3 & 110.4 & 90.36 & 71.22 & 249.4 & 44.56 & 3.923  \\\hline
 7.875--8.125 & 78.65 & 92.10 & 102.7 & 101.9 & 85.79 & 80.87 & 86.13 & 81.07 & 61.83 & 46.29 & 44.56 & 118.9 & 5.278  \\\hline
 8.125--12    & 5.983 & 7.131 & 8.074 & 8.114 & 6.924 & 6.748 & 7.439 & 6.924 & 4.838 & 3.350 & 3.923 & 5.278 & 3.266  \\\hline

\end{tabular}
\label{tab:generic2}
\begin{multicols}{2}

\end{multicols}

\vspace{-1mm}
\centerline{\rule{80mm}{0.1pt}}
\vspace{2mm}
\begin{multicols}{2}

\bibliographystyle{apsrev4-1}
\bibliography{Reactor_8AD_CPC}{}

\end{multicols}

\end{document}